\documentclass[10pt,aps,prd,reprint,superscriptaddress,amsmath,amssymb,amsfonts,showkeys,noshowpacs,nofootinbib,floatfix]{revtex4-1}
\usepackage{graphicx}
\usepackage{bm}
\usepackage{calc}
\usepackage{color}
\usepackage{url}
\usepackage{array}
\usepackage{tabularx}

\newif\ifAMStwofonts
\AMStwofontstrue

\def\gsim{~\rlap{$>$}{\lower 1.0ex\hbox{$\sim$}}}

\def\simpropto{\lower.2ex\hbox{$\; \buildrel \propto \over \sim \;$}}
\def\ltsim{\lower.5ex\hbox{$\; \buildrel < \over \sim \;$}}
\def\gtsim{\lower.5ex\hbox{$\; \buildrel > \over \sim \;$}}
\def\ltsim{\lower.5ex\hbox{$\; \buildrel < \over \sim \;$}}
\def\gtsim{\lower.5ex\hbox{$\; \buildrel > \over \sim \;$}}

\def\prd{Physical Review D.}
\def\nat{Nature}
\def\astjo{Astronomical Journal}

\def\dd{\,{\rm d}}

\def\dd{{\rm d}}

\def\pmb#1{\setbox0=\hbox{#1}%
\kern-.025em\copy0\kern-\wd0
\kern.05em\copy0\kern-\wd0
\kern-.025em\raise.0433em\box0}

\def\vv{\boldsymbol{v}}

\DeclareMathOperator{\tr}{Tr}
\renewcommand{\theenumi}{(\alph{enumi})}

\def\vx{\boldsymbol{x}}

\def\vr{\boldsymbol{r}}

\def\vd{\boldsymbol{d}}
\def\va{\boldsymbol{a}}
\def\hvr{\hat{\vr}}

\def\vk{\boldsymbol{k}}

\def\aj{AJ}

\def\simlt{\lower.5ex\hbox{$\; \buildrel < \over \sim \;$}}
\def\simgt{\lower.5ex\hbox{$\; \buildrel > \over \sim \;$}}

\newcommand{\beq}{\begin{equation}}
\newcommand{\eeq}{\end{equation}}
\def\beqa{\begin{eqnarray}}
\def\eeqa{\end{eqnarray}}
\def\fixit#1{}

\def\dd{{\rm d}}
\def\apj{ApJ}
\def\jcap{JCAP}

\def\apjs{ApJ. S}

\def\physrep{Physics Reports}
\def\apjl{ApJL}
\def\mnras{MNRAS}
\def\aap{A\&A}

\begin{document}
\title{Tracing the cosmic velocity field at \boldmath{$z\sim 0.1$} from galaxy luminosities in the SDSS DR7}
%\title{Extracting peculiar velocity information from the SDSS DR7 galaxy luminosity function at \boldmath{$z\sim 0.1$}}
%\title{Measuring the peculiar velocity field through the galaxy luminosity function at \boldmath{$z=0.1$}}
\author{Martin Feix}
\email[Electronic address: ]{mfeix@physics.technion.ac.il}
\affiliation{Department of Physics, Israel Institute of Technology - Technion, Haifa 32000, Israel}
\author{Adi Nusser}
\email[Electronic address: ]{adi@physics.technion.ac.il}
\affiliation{Department of Physics, Israel Institute of Technology - Technion, Haifa 32000, Israel}
\affiliation{Asher Space Science Institute, Israel Institute of Technology - Technion, Haifa 32000, Israel}
\author{Enzo Branchini}
\email[Electronic address: ]{branchin@fis.uniroma3.it}
\affiliation{Department of Physics, Universit\`a Roma Tre, Via della Vasca Navale 84, Rome 00146, Italy}
\affiliation{INFN Sezione di Roma 3, Via della Vasca Navale 84, Rome 00146, Italy}
\affiliation{INAF, Osservatorio Astronomico di Roma, Monte Porzio Catone, Italy}

\begin{abstract}
Spatial modulations in the distribution of observed luminosities (computed using redshifts) of $\sim 5\times 10^5$
galaxies from the SDSS Data Release $7$, probe the cosmic peculiar velocity field out to $z\sim 0.1$. Allowing for luminosity
evolution, the $r$-band luminosity function, determined via a spline-based estimator, is well represented by a Schechter form
with $ M^{\star}(z)-5{\rm log_{10}} h=-20.52-1.6(z-0.1)\pm 0.05$ and $\alpha^{\star}=-1.1\pm 0.03$. Bulk flows and higher
velocity moments in two redshift bins, $0.02 < z < 0.07$ and $0.07 < z < 0.22$, agree with the predictions of the $\Lambda$CDM
model, as obtained from mock galaxy catalogs designed to match the observations. Assuming a $\Lambda$CDM model, we estimate
$\sigma_{8}\approx 1.1\pm 0.4$ for the amplitude of the linear matter power spectrum, where the low accuracy is due to the
limited number of galaxies. While the low-$z$ bin is robust against coherent photometric uncertainties, the bias of results
from the second bin is consistent with the $\sim1$\% magnitude tilt reported by the SDSS collaboration. The systematics are
expected to have a significantly lower impact in future datasets with larger sky coverage and better photometric calibration.
\end{abstract}

\keywords{Cosmology: theory, observations, large-scale structure of the universe, dark matter, redshift surveys}
%\arxivnumber{}
% $0.02 < z_{1} < 0.07 < z_{2} < 0.22$,
\maketitle

\section{Introduction}
\label{sec:int}
In recent years, the amount of available extragalactic data has helped to establish a comprehensive  picture of our Universe
and its evolution \cite[e.g.][]{Percival2010, Riess2011, Hinshaw2013, Planck2013}. These data, by and large, have enforced
the standard cosmological paradigm where initial perturbations in the mass density field grow via gravitational instability
and eventually form the cosmic structure we observe today. The clustering process is inevitably associated with peculiar
motions of matter, namely deviations from a pure Hubble flow. On large scales, these motions exhibit a coherent pattern, with
matter generally flowing from underdense to overdense regions. If galaxies indeed move much like test particles, they should
appropriately reflect the underlying peculiar velocity field which contains valuable information and, in principle, could be
used to constrain and discriminate between different cosmological models.

Usually relying on galaxy peculiar velocities estimated from measured redshifts and distance indicators, most approaches in
the literature have focused on extracting this information within local volumes of up to $100h^{-1}$ Mpc and larger centered
on the Milky Way \cite[e.g.,][]{Riess1995, Dekel1999, Zaroubi2001, Hudson2004, Sarkar2007, Lavaux2010, feldwh10, ND11,
Turnbull2012, Feindt2013}. Common distance indicators are based on well-established relations between observable intrinsic
properties of a given astronomical object, where one of them depends on the object's distance. A typical example is the
Tully-Fisher relation \cite{TF77} between rotational velocities of spiral galaxies and their absolute magnitudes. Due to
observational challenges, the number of galaxies in distance catalogs is relatively small compared to that of redshift catalogs,
limiting the possibility of exploring the cosmological peculiar velocity field to low redshifts $z\sim$ 0.02--0.03. Moreover,
all known distance indicators are potentially plagued by systematic errors \cite{lyn88, Strauss1995} which could give rise to
unwanted biases in the inferred velocities and thus renders their use for cosmological purposes less desirable.

To probe the flow of galaxies at deeper redshifts, one needs to resort to non-traditional distance indicators. One method,
for instance, exploits the kinetic Sunyaev-Zel'dovich effect to measure the cosmic bulk flow, i.e. the volume average of
the peculiar velocity field, out to depths of around 100--500$h^{-1}$ Mpc \cite[e.g,][]{Haehnelt1996, Osborne2011, Lavaux2013,
planck_bf}. Another strategy is based on the apparent anisotropic clustering of galaxies in redshift space which is commonly
described as redshift-space distortions. This effect is a direct consequence of the additional displacement from distances to
redshifts due to the peculiar motions of galaxies, and it yields reliable constraints on the amplitude of large-scale coherent
motions and the growth rate of density perturbations \cite[e.g.,][]{Hamilton1998, Peacock2001, Scoccimarro2004, Guz08}.

Galaxy peculiar motions also affect luminosity estimates based on measured redshifts, providing another way of tackling the
problem. Since the luminosity of a galaxy is independent of its velocity, systematic biases in the estimated luminosities
of galaxies can be used to explore the peculiar velocity field. The idea has a long history. It was first adopted to constrain
the velocity of the Virgo cluster relative to the Local Group by correlating the magnitudes of nearby galaxies with their
redshifts \cite{TYS}. Although in need of very large galaxy numbers to be effective, methods based on this idea use only
measured galaxy luminosities and their redshifts to derive bounds on the large-scale peculiar velocity field. Therefore, these
methods do not require the use of traditional distance indicators and they are also independent of galaxy bias. Using the
nearly full-sky 2MASS Redshift Survey (2MRS) \cite{Huchra2012}, for example, this approach has recently been adopted to
constrain bulk flows in the local Universe within $z\sim 0.01$ \cite{Nusser2011, Branchini2012}. Furthermore, it has been used
to determine the current growth rate of density fluctuations by reconstructing the full linear velocity field from the
clustering of galaxies \cite{Nusser1994, Nusser2012}.

Here we seek to apply this luminosity-based approach to obtain peculiar velocity information from galaxy redshifts and
apparent magnitudes of the Sloan Digital Sky Survey (SDSS) \cite{York2000}. The goals of our analysis are:
\begin{itemize}
\item A demonstration of the method's applicability to datasets with large galaxy numbers.
\item An updated estimate of the $r$-band luminosity function of SDSS galaxies at $z\sim 0.1$, accounting for evolution
in galaxy luminosities.
\item Novel bounds on bulk flows and higher-order moments of the peculiar velocity field at redshifts $z\sim 0.1$.
\item First constraints on the angular peculiar velocity power spectrum and cosmological parameters without additional
input such as galaxy clustering information.
\end{itemize}  
The paper is organized as follows: we begin with introducing the luminosity method and its basic equations in section \ref{section2}.
In section \ref{section3}, we then describe the SDSS galaxy sample used in our analysis, together with a suite of mock catalogs
which will allow us to assess uncertainties and known systematics inherent to the data. After a first test of the method, we attempt
to constrain peculiar motions in section \ref{section4}, assuming a redshift-binned model of the velocity field. Because of the mixing
between different velocity moments arising from the SDSS footprint, bulk flow measurements are interpreted with the help of
galaxy mocks. Including higher-order velocity moments, we proceed with discussing constraints on the angular velocity power in
different redshift bins and their implications. As an example of cosmological parameter estimation, we further infer the quantity
$\sigma_{8}$, i.e. the amplitude of the linear matter power spectrum on a scale of $8h^{-1}$ Mpc, and compare the result to the
findings from the corresponding mock analysis. Other potential issues and caveats related to our investigation are addressed at the
section's end. In section \ref{section5}, we finally summarize our conclusions and the method's prospects in the context of
next-generation surveys. For clarity, some of the technical material is separately given in an appendix. Throughout the paper, we
adopt the standard notation, and all redshifts are expressed in the rest frame of the cosmic microwave background (CMB) using the
dipole from ref. \cite{Fixsen1996}.

\section{Methodology}
\label{section2}
\subsection{Variation of observed galaxy luminosities}
\label{section2a}
In an inhomogeneous universe, the observed redshift $z$ of an object (a galaxy) is generally different from its cosmological
redshift $z_{c}$ defined for the unperturbed background. To linear order in perturbation theory, one finds the well-known
expression \cite{SW}
\begin{equation}
\begin{split}
\frac{z-z_{c}}{1+z} &= \frac{V(t,r)}{c} - \frac{\Phi(t,r)}{c^2}\\
&{ } - \frac{2}{c^2}\int_{t(r)}^{t_0}\dd t \frac{\partial\Phi\left\lbrack\hvr r(t),t\right\rbrack}{\partial t}\approx \frac{V(t,r)}{c},
\end{split}
\label{eq:sw}
\end{equation}
where $V$ is the physical radial peculiar velocity of of the object, $\Phi$ denotes the gravitational potential and $\hvr$
is a unit vector along the line of sight to the object. The last step explicitly assumes low redshifts where the velocity
$V$ makes the dominant contribution.\footnote{The first two terms on the right-hand side of eq. \eqref{eq:sw} describe
the Doppler effect and the gravitational redshift, respectively. The last one reflects the energy change of a photon
passing through a time-dependent potential well and is equivalent to the late-time integrated Sachs-Wolfe effect.} Note
that all fields are considered relative to their present-day values at a comoving radius of $r(t=t_{0})$ and that we have
substituted $z$ for $z_{c}$ in the denominator on the left-hand side of eq. \eqref{eq:sw}, which simplifies part of the
analysis presented below and is consistent at the linear level.

The observed absolute magnitude $M$, computed using the galaxy redshift $z$, rather than the (unknown) cosmological
redshift $z_{c}$, differs from the true value $M^{(t)}$ because of the shift ${\rm DM}(z)-{\rm DM}(z_c)$ in the distance
modulus ${\rm DM}=25+5\log_{\rm 10}\lbrack D_{L}/{\rm Mpc}\rbrack$, where $D_{L}$ is the luminosity distance. Hence, 
\begin{equation}
\begin{split}
M &= m - {\rm DM}(z) - K(z) + Q(z)\\
&= M^{(t)} + 5\log_{10}\dfrac{D_{L}(z_{c})}{D_{L}(z)},
\end{split}
\label{eq:magvar}
\end{equation}
where $m$ is the apparent magnitude, $K(z)$ is the $K$-correction \cite[e.g.,][]{Blanton2007}, and the function $Q(z)$
accounts for luminosity evolution. Since the variation $M-M^{(t)}$ of magnitudes distributed over the sky is systematic,
it can be used to gain information on the peculiar velocity field. In the following, we will discuss how this may be
achieved with the help of maximum-likelihood techniques.

%\subsection{Bulk flows from the luminosity distribution of galaxies}
%\label{section2b}

\subsection{Statistical description }% Higher-order moments and the angular velocity power spectrum}
\label{section2c}
%Building on the pioneering work of \cite{TYS}, the  approach has first been use by \cite{Nusser2011} for constraining  bulk flows by minimization of
%the variations in the observed galaxy magnitudes in the Two-Mass redshift Survey {\bf AN: give ref}. Characterizing the peculiar velocity
%field by a bulk flow $\vv_{B}$, one starts from the probability of observing a galaxy with magnitude $M$ given only its
%redshift and its angular position $\hvr$ on the sky, 
%\begin{equation}
%P\left (M\vert z,\vv_{B}\right ) = P\left (M\vert z,V\right ) = \frac{\phi(M)}{\eta\left (M^{+},M^{-}\right )},
%\label{eq:2b1}
%\end{equation}
%where the radial velocity is expressed as $V=\hvr\cdot\vv_{B}$ and the
%Compared to other bulk flow estimators proposed in the literature, the above method is independent of traditional distance
%indicators and the poorly understood galaxy bias, and thus avoids the potential systematics associated with either one of
%them.
\subsubsection{Inference of bulk flows and other velocity moments}
Before introducing our methodology, we need to specify a suitable model of the velocity field. Although a popular option
is to characterize peculiar velocities in terms of bulk flows, one could aim at a more complete description of the
peculiar velocity field. Given the current data, however, a full three-dimensional estimate of the velocity field would
be entirely dominated by the noise. A more promising approach is the following: first, we subdivide the galaxy data into
suitable redshift bins and consider the bin-averaged velocity $\tilde{V}$. Supposing for the moment that we are dealing
with a single bin, we then proceed to decompose $\tilde{V}(\hvr )$ (evaluated at the galaxy position $\hvr$) into
spherical harmonics, i.e.
\begin{equation}
\begin{split}
a_{lm} &= \int\dd\Omega\tilde{V}(\hvr )Y_{lm}(\hvr ),\\
\tilde{V}(\hvr ) &= \sum\limits_{l,m}a_{lm}Y_{lm}^{*}(\hvr ),\quad l>0,
\end{split}
\label{eq:2c1}
\end{equation}
where the sum over $l$ is cut at some maximum value $l_{\rm max}$. A bulk flow of the entire volume, denoted as $\vv_{B}$,
corresponds to the dipole term ($l=1$) in eq. \eqref{eq:2c1}. Building on the pioneering work of \cite{TYS}, for example,
the analysis presented in \cite{Nusser2011} has initially been restricted to a model with $l_{\rm max}=1$ when considering
galaxies from the 2MRS \cite{Huchra2012}.  

Assuming that redshift errors can be neglected \cite{Nusser2011}, we write the probability of observing a
galaxy with magnitude $M$, given only its redshift and angular position $\hvr$ on the sky, as
\begin{equation}
P\left (M\vert z,a_{lm}\right ) = P\left (M\vert z,\tilde{V}(\hvr )\right ) = \frac{\phi(M)}{\eta\left (M^{+},M^{-}\right )},
\label{eq:2c2}
\end{equation}
where $\phi(M)$ is the galaxy luminosity function (LF) and $\eta\left (M^{+},M^{-}\right )$ is defined as
\begin{equation}
\eta\left (M^{+},M^{-}\right ) = \int_{M^{+}}^{M^{-}}\phi(M)\dd M.
\label{eq:2b2}
\end{equation}
The corresponding limiting magnitudes $M^{\pm}$ are given by
\begin{equation}
\begin{split}
M^{+} &= \max\left\lbrack M_{\rm min}, m^{+} - {\rm DM}(z_{c}) - K(z) + Q(z)\right\rbrack,\\
M^{-} &= \min\left\lbrack M_{\rm max}, m^{-} - {\rm DM}(z_{c}) - K(z) + Q(z)\right\rbrack,
\end{split}
\label{eq:app1c}
\end{equation}
where $m^{\pm}$ are the sample's limiting apparent magnitudes and the cosmological redshift $z_{c}$ depends on
the velocity $\tilde{V}$ and the observed redshift $z$ because of eq. \eqref{eq:sw}. The velocity model enters the
expression for the limiting magnitudes $M^{\pm}$ since it induces a shift in the distance modulus.
 The coefficients
$a_{lm}$ of the flow modes can, therefore, be inferred by maximizing the total log-likelihood obtained from the sum
over all galaxies in a sample, i.e. $\log P_{\rm tot} = \sum\log P_{i}$. The rational for this is to find the set of
$a_{lm}$ which minimizes the spread in the observed magnitudes \cite{Nusser2011}. The spherical harmonics provide
an orthogonal basis only in the case of an all-sky survey, and the partial sky coverage of the SDSS implies that
the inferred moments will not be statistically independent. For example, a quadrupole velocity mode ($l=2$) would 
contaminate the estimate of a bulk flow $\vv_{B}$, which must be taken into account when interpreting any results. 
The monopole term ($l=0$) is completely degenerate with an overall shift of the magnitudes, and hence it is not
included. 

If the number of available galaxies is large enough, the central limit theorem implies that $P_{\rm tot}$ becomes
approximately normal, and we have
\begin{equation}
\log P_{\rm tot}\left (\vd\vert\vx\right ) = -\frac{1}{2}\left (\vx-\overline{\vx}\right )^{\rm T}\bm{\Sigma}^{-1}
\left (\vx-\overline{\vx}\right ),
\label{eq:2c3}
\end{equation}
where $\vx$ is a vector of all model parameters, $\overline{\vx}$ is the corresponding mean, $\vd$ denotes the data
(or Bayesian evidence), and $\bm{\Sigma}$ is the covariance matrix describing the expected error of our estimate. We
have numerically verified that this approximation is extremely accurate for the SDSS which comprises several hundred
thousands of galaxies. The distribution's mean in eq. \eqref{eq:2c3} simply corresponds to the maximum-likelihood
estimate $\hat{\vx}^{\rm ML}$ of the vector $\vx$, and $\bm{\Sigma}$ can be estimated either by inverting the
observed Fisher matrix $\mathbf{F}$ which is defined as
$\mathbf{F}_{\alpha\beta}=-\partial\log P_{\rm tot}/(\partial x_{\alpha}\partial x_{\beta})$ evaluated at the
maximum value $\hat{\vx}^{\rm ML}$ or from a realistic set of mock galaxy catalogs. The increasing number of
parameters associated with the higher-order moments of $\tilde{V}$ typically renders a full numerical evaluation of
$\log P_{\rm tot}$ unfeasible. A solution to this problem is based on approximating the total log-likelihood
function to second order (see section \ref{sectionnum} and appendix \ref{app1} for details). In the realistic
application to SDSS data, the model parameters $\vx$ include the coefficients $a_{lm}$ (for each redshift bin) as
well as the LF parameters. They will be determined simultaneously by maximizing $\log P_{\rm tot}$. 

\subsubsection{Inference of angular velocity power spectra}
Let us now focus on large scales where linear theory is applicable. Assuming Gaussian initial conditions, the
cosmological peculiar velocity field on these scales is then fully characterized by its power spectrum. The relevant
quantity here is the angular velocity power spectrum $C_{l}=\langle\lvert a_{lm}^{2}\rvert\rangle$. Under these
preliminaries, the problem of inferring the $C_{l}$ becomes equivalent to the more familiar estimation of the CMB
anisotropy power spectrum, and may thus be tackled with the same general techniques \cite{Tegmark1997, Bond1998}. To
estimate the power spectrum, one simply maximizes the probability of observing the data given the $C_{l}$, i.e.
\begin{equation}
P(C_{l})\equiv P\left (\vd\vert C_{l}\right )\propto\int\dd a_{lm}P\left (\vd\vert a_{lm}\right )P\left (a_{lm}\vert C_{l}\right ),
\label{eq:2c4}
\end{equation}
which is obtained by constructing the posterior likelihood according to Bayes' theorem and marginalizing over the
$a_{lm}$. Here the individual $a_{lm}$ are uncorrelated and taken to be normally distributed, i.e. one has
\begin{equation}
P\left (a_{lm}\vert C_{l}\right ) = \prod\limits_{l,m}\left (2\pi C_{l}\right )^{-1/2}\exp
\left (-\frac{\lvert a_{lm}\rvert^{2}}{2C_{l}}\right ),
\label{eq:2c5}
\end{equation}
and $P\left (\vd\vert a_{lm}\right )$ is derived from marginalizing $P_{\rm tot}\left (\vd\vert\vx\right )$ over the
remaining parameters in $\vx$. Within the Gaussian approximation, carrying out the integration in eq. \eqref{eq:2c4}
is straightforward and the resulting expressions are presented in appendix \ref{app3}.

Considering a particular model like, for example, the standard $\Lambda$CDM cosmology, the $C_{l}$ are fully specified
by a set of cosmological parameters $\zeta_{k}$. Therefore, accounting for this dependency in the prior probability,
the above technique may further be used to constrain cosmological key quantities such as $\sigma_{8}$ from the observed
peculiar velocity field alone. Given the characteristics of current galaxy redshift surveys, it is clear that these
constraints will be much less tight than those obtained by other means such as CMB analysis, but still valuable as a
complementary probe and consistency check.

A successful application of the method requires a large number of galaxies to beat the statistical (Poissonian)
errors. The method does not require accurate redshifts and can be used with photometric redshifts to recover signals
on scales larger than the spread of the redshift error. Other related maximum-likelihood approaches based on reduced
input (photometric redshifts or just magnitudes) \cite{Itoh2010,Abate2012} consider integrated quantities such as
number densities, resulting in less sensitive measurements of bulk flows and higher-order moments of the peculiar
velocity field.

\subsection{Estimating the galaxy luminosity function}
\label{section2d}
A reliable measurement of the galaxy LF represents a key step in our approach. A corresponding
estimator should be flexible enough to capture real features in the luminosity distribution, but also physical in
the sense of returning a smooth function over the range of interest. To meet these requirements, we shall adopt the
spline-based estimator introduced in \cite{Branchini2012} for our analysis.\footnote{Since the underlying principle
is a reduction in the spread of observed magnitudes, even unrealistic models of the luminosity distribution should
yield unbiased measurements of the velocity information \cite{Nusser2012}, albeit with larger statistical errors.}
In this case, the unknown LF is written as a piecewise-defined function, i.e.
\begin{equation}
\phi(M) = \varphi_{i}(M),\qquad M_{i-1}\leq M<M_{i},
\label{eq:2d1}
\end{equation}
where $\varphi_{i}$ is a third-order polynomial defined such that the second derivative of $\phi$ with respect to $M$ is
continuous on the interval $[M_{0},M_{N-1}]$ and vanishes at the boundaries.
The cubic spline in eq. \eqref{eq:2d1} may be regarded as a generalization of the stepwise estimator originally proposed
in \cite{efs88}, and the actual spline coefficients are determined employing the standard techniques summarized in
\cite{Press2002}. Since there occur only polynomial expressions, derivatives and integrals of $\phi(M)$ are of particularly
simple form, allowing quite an efficient evaluation of the previously defined likelihood functions. LFs which are obtained
according to this procedure might exhibit spurious wiggles, especially at the corresponding bright and faint ends.
As is already discussed in \cite{Branchini2012}, however, these wiggles can be sufficiently suppressed by adding an appropriate
penalty term to the total likelihood function or by enforcing (log-)linear behavior of $\phi(M)$ beyond suitable bright and
faint magnitude thresholds. Alternatively, it is also possible to simply choose magnitude cuts and the total number of spline
points in such a way that the number of galaxies in each magnitude interval is large enough to avoid this problem for all
practical purposes. In the present analysis, we will follow the latter approach when maximizing the total log-likelihood.

In addition to the spline-based estimator, which is most relevant when considering real observations, we shall also use a
parametric estimator that assumes a widely used Schechter form of the LF \cite{Sandage1979,schechter}, i.e.
\begin{equation}
\phi(M) \propto 10^{0.4(1+\alpha^{\star})(M^{\star}-M)}\exp{\left (-10^{0.4(M^{\star}-M)}\right )},
\label{eq:2d2}
\end{equation}
where $M^{\star}$ and $\alpha^{\star}$ are the usual Schechter parameters. The normalization of $\phi(M)$ cancels in the
likelihood function and does not concern us here. Although it does not provide a good fit to all datasets, the Schechter form
and its corresponding estimator turn out very useful for the analysis of both mock catalogs and the real galaxy sample
presented in section \ref{section4}.

\section{Datasets}
\label{section3}
\subsection{NYU Value-Added Galaxy Catalog}
\label{section3a}
We will use the SDSS galaxies from the latest publicly available NYU Value-Added Galaxy Catalog (NYU-VAGC)
\cite{Blanton2005}.\footnote{\url{http://sdss.physics.nyu.edu/vagc/}}. This catalog is based on the SDSS Data Release $7$
(DR7) \cite{abaz}, and contains galaxies with a median redshift of $z\approx 0.1$, observed in five different photometric
bands with magnitudes corrected for Galactic extinction according to \cite{Schlegel1998}. Using Petrosian magnitudes, we
decide to work with the $r$-band, mainly because it gives the largest spectroscopically complete galaxy sample
\cite{Blanton2001, Strauss2002}, which is an important factor for the statistical method we have introduced in section
\ref{section2}. To minimize incompleteness and to exclude galaxies with questionable photometry and redshifts, we choose
the subsample NYU-VAGC {\tt safe} which contains only galaxies whose apparent $r$-band magnitudes satisfy
$14.5 < m_{r} < 17.6$.\footnote{The SDSS photometry is known to exhibit small offsets from the AB magnitude system
\cite{Oke1983}. For the SDSS $r$-band, this amounts to a shift of around $0.01$ \cite{Eisenstein2006} which we will
take into account when calculating absolute magnitudes below. } The subsample accounts for fiber collisions following
the correction scheme {\tt nearest}, but this is expected to be of little relevance in our analysis which should be
insensitive to galaxy clustering. Also, since we are interested in minimizing
systematics due to uncertainties in $K$-corrections and luminosity evolution (see section \ref{section4}), we shall adopt
the $^{0.1}r$-bandpass when dealing with absolute magnitudes \cite{Blanton2003B}, and further impose cuts on redshifts
(expressed in the CMB frame) and observed absolute magnitudes $M_{r}$ such that only galaxies with $0.02 < z < 0.22$ and
$-22.5 < M_{r} - 5\log_{10}h < -17.0$ are selected. The number of galaxies contained in our final working sample is
approximately $5.4\times 10^{5}$ and may slightly vary, depending on the assumed background cosmology which enters the
calculation of $M_{r}$ through the luminosity distance. For realistic flat cosmologies with a total matter density
$\Omega_{m}\approx 0.3$, however, these variations are typically on the order of a few hundred galaxies and thus not very
significant. Since we are concerned with relatively low redshifts $z\ltsim 0.2$, we assume a linear dependence of the
luminosity evolution on redshift for simplicity, i.e. 
\begin{equation}
\label{eq:qz}
Q(z)= Q_{0}(z-z_{0}), 
\end{equation}
where we set the pivotal redshift $z_{0}=0.1$. Furthermore, $K$-corrections for individual galaxies are taken from the
NYU-VAGC and have been calculated with the software package {\tt kcorrect} {\tt v4{\_}1{\_}4} \cite{Blanton2007}. To
calculate the limiting absolute magnitudes $M^{\pm}$ in the $^{0.1}r$-bandpass at a given redshift $z$, however, we
resort to a mean $K$-correction of the form
\begin{equation}
\overline{K}(z) = -2.5\log_{10}(1.1) + \sum_{i=1}^{3}\gamma_{i}(z - 0.1)^{i},
\label{eq:4b}
\end{equation}
where $\gamma_{1}\approx 0.924$, $\gamma_{2}\approx 2.095$, and $\gamma_{3}\approx -0.184$ are determined by directly
fitting the individual $K$-corrections listed in the NYU-VAGC. When calculating the total likelihood function introduced
in section \ref{section2c}, all galaxies are weighted according to the angular (redshift) completeness. All remaining
details relevant to the analysis of the NYU-VAGC galaxy redshift data will be separately discussed in section \ref{sectionnum}.

\subsection{Mock galaxy catalogs}
\label{section3b}
To test the performance of our approach, we resort to two different suites of galaxy mock catalogs. The first set of mocks
is based on the LasDamas simulations \cite{McBride2009} while the second one is obtained from the real NYU-VAGC dataset that
we analyse in this work.

\subsubsection{LasDamas mock catalogs}
These mock galaxy catalogs are obtained by populating the LasDamas simulations \cite{McBride2009} with artificial galaxies,
using a halo occupation distribution model \cite[e.g.,][]{pesm,Seljak2000,Berlind2002} to match the observed clustering of
SDSS galaxies in a wide luminosity range. The goal of these catalogs is to benchmark our method and validate its implementation,
using a sample with overall characteristics (number density of objects, sky coverage, etc.) similar to that of the real catalog,
ignoring all sources of systematic biases.

Here we will consider a total of $60$ mocks from the public {\tt gamma} release, modeled after a volume-limited subsample of
SDSS DR7 cut at $M_{r}<-20$, which cover the full SDSS footprint (``North and South'') and a redshift range $0.02<z<0.106$ with
a median of $z\approx 0.08$.\footnote{\url{http://lss.phy.vanderbilt.edu/lasdamas/mocks.html}} The typical galaxy number in
these mocks is around $1-1.5\times 10^{5}$, and we shall use them as a basic test of bulk flow measurements. To this end, an
observed redshift is assigned to each galaxy according to
\begin{equation}
cz = cz_{c} + V + c\epsilon_{z},
\label{eq:3b1}
\end{equation}
where $z_{c}$ corresponds to the redshift entry in the mock catalog, the radial velocity $V=\hvr\cdot\vv_{B}$ is the
line-of-sight component of the bulk flow $\vv_{B}$, and $\epsilon_{z}$ is a random measurement error drawn from a
Gaussian distribution with $c\sigma_{z}=15$ km s$^{-1}$.\footnote{Although the redshifts listed in the LasDamas {\tt gamma}
mocks include distortions from peculiar velocities, we interpret them as cosmological redshifts $z_{c}$ for simplicity.
This has no adverse effect on testing our method's performance.} Similarly, observed $r$-band magnitudes are assigned
with the help of eq. \eqref{eq:magvar}, but without including the $K$-correction term. Assuming the linear luminosity
evolution in \eqref{eq:qz} with $Q_{0}=1.6$, the true galaxy magnitudes $M^{(t)}$ are randomly extracted from the
Schechter distribution given by eq. \eqref{eq:2d2} with the parameters $M^{\star}=-20.44+5\log_{10}h$ and
$\alpha^{\star}=-1.05$ \cite{Blanton2003}. Although it is irrelevant for the present purposes, this procedure ignores
the masses of dark matter halos, meaning that very massive halos may host very faint galaxies and vice versa. We also
add a Gaussian random error to $M_{r}$ with $\sigma_{M}=0.03$, and further trim the resulting mock catalogs by requiring
$M_{r}<-20.25$ to prevent problems related to Malmquist bias. Finally, our choice of the bulk flow $\vv_{B}$ used in the
benchmark runs will be described in section \ref{section4}. As the LasDamas simulations assume a flat $\Lambda$CDM model
with $\Omega_{m}=0.25$ and $h=0.7$, we adopt the same cosmology for the mocks. 

\begin{table*}
\caption{Summary of the $\Lambda$CDM cosmologies described in the text.}
%:Concerning the computation of the theoretical power spectrum $C_{l}$, note that all models assume a CMB temperature of $T_{\rm CMB}\approx 2.726$ K.}
\begin{tabular*}{0.95\linewidth}{@{\extracolsep{\fill}}lcccccc}
\noalign{\medskip}
\hline
\hline
\noalign{\smallskip}
Parameter set & $\Omega_{b}$ & $\Omega_{m}$ & $\Omega_{\Lambda}$ & $h$ & $n_{s}$ & $\sigma_{8}$
\tabularnewline
\noalign{\smallskip}
\hline
\noalign{\smallskip}
{\tt param{\_}mock} & 0.0455 & 0.272 & 0.728 & 0.702 & 0.961 & 0.8
\tabularnewline
{\tt param{\_}wmap} & 0.0442 & 0.2643 & 0.7357 & 0.714 & 0.969 & 0.814
\tabularnewline
{\tt param{\_}planck} & 0.049 & 0.3175 & 0.6825 & 0.671 & 0.962 & 0.834
\tabularnewline
\noalign{\smallskip}
\hline
\hline
\end{tabular*}
\label{table1}
\end{table*}

\subsubsection{NYU-VAGC mock catalogs}
Starting directly from the previously described NYU-VAGC dataset, we generate a second set of mock catalogs built from the
angular positions and spectroscopic redshifts of the observed galaxies. The goal of these mocks is to investigate the impact
of known observational biases, incompleteness, and cosmic variance while preserving the spatial distributions of the galaxies
in the real SDSS DR7 catalog.

Just as in the case of the LasDamas mock catalogs, we interpret the observed spectroscopic redshifts as the cosmological ones
and obtain the corresponding measured redshifts from eq. \eqref{eq:3b1}, where $V$ is now determined from the full linear
velocity field evaluated at redshift $z=0$. The velocity field is obtained from a random realization sampled on a cubic grid
with $1024^{3}$ points and a comoving mesh size of $4h^{-1}$ Gpc, assuming the linear power spectrum $P_{v}(k)$ of a flat
$\Lambda$CDM cosmology with total matter and baryonic density parameters $\Omega_{m}=0.272$ and $\Omega_{b}=0.0455$, respectively,
scalar spectral index $n_{s}=0.961$, $h=0.702$, and $\sigma_{8}=0.8$ (corresponding to the parameter set {\tt param{\_}mock}
which is listed in table \ref{table1}). To ensure a high level of (statistical) independence between the final mocks, we
perform appropriate translations and rotations of the survey data reference frame relative to the sampling grid before each
galaxy is assigned a velocity equal to that of the nearest grid point. Because of small-scale nonlinearities and the finite
grid sampling, we further add uncertainties to the line-of-sight components of these velocities which are generated from a
normal distribution with $\sigma_{V} = 250$ km s$^{-1}$. 

The luminosities are assigned exactly as for the LasDamas mocks using the appropriate cuts in apparent and absolute magnitudes
specified in section \ref{section3a}. In addition, we simulate two known systematic errors in the photometric calibration
of SDSS data \cite{Pad2008} that have a potential impact on our analysis. The first one arises from various magnitude offsets
between the individual SDSS stripes, and is modeled by considering another random error with $\sigma_{\rm stripe} = 0.01$. The
second, more serious error results from unmodeled atmospheric variations during the time of observation, ultimately causing an
overall zero-point photometric tilt of roughly $0.01$ in magnitudes over the survey region. To mimic this tilt, we include in
each mock a magnitude offset in the form of a randomly oriented dipole normalized such that its associated root mean square
(rms) over all galaxies is $\delta m_{\rm dipole}=0.01$ \cite{Pad2008}. 

With this procedure we obtain a total of $269$ galaxy mocks (both flux- and volume-limited), mimicking the characteristics of the
real NYU-VAGC sample. These mocks will be used to explore the distribution of measured bulk flow vectors and to study constraints
on the power spectrum $C_{l}$ or cosmological parameters for a realistic choice of the large-scale peculiar velocity field.

\section{Data analysis}
\label{section4}
We now proceed to apply our method to the SDSS data. To achieve our goals outlined in section \ref{sec:int}, we will begin with
a short description of some additional preliminaries and present the general line of action in section \ref{sectionnum}. After
this, we will estimate the $r$-band LF of SDSS galaxies at $z\sim 0.1$ in section \ref{sectionlfestimate}, which provides the
basis for our investigations. The results of our velocity analysis of the NYU-VAGC galaxy sample are then presented and discussed
in sections \ref{sec:bf} and \ref{sectionvcosmo}.

\subsection{General line of action}
\label{sectionnum}
A major obstacle in constraining the velocity field from the SDSS is the partial coverage (only about $20$\%) of the sky. Since
the $Y_{lm}$ no longer form an orthogonal basis on this limited mask, the maximum-likelihood approach yields a statistical mixing
between the estimated velocity moments, effectively probing a combination of different multipoles. In the case of the bulk flow,
for instance, this would correspond to a superposition of several terms up to even the hexadecapole of the peculiar velocity
field \cite{Tegmark2004}. Of course, one may resort to an orthogonal basis set for $l_{\rm max}$ in pixel space. Because we are
going through the full maximum-likelihood procedure, however, there is no gain in doing so, i.e. all the information is already
contained in the measured $a_{lm}$ and their covariance matrix. Also, the results expressed in such orthogonal bases typically
have a less obvious physical interpretation.

Additional difficulties arise from a too flexible LF model, i.e. oversampling issues related to the spline-based estimator, and
the linear evolution term $Q(z)$ which actually mimics the formally ignored monopole contribution in eq. \eqref{eq:2c1} over the
redshift range of interest. Both may contribute to the mode mixing and further complicate the interpretation of the corresponding
results. Similarly, the presence of systematic errors in the SDSS photometry (see section \ref{section3b}) can lead to spurious
flows which contaminate the velocity measurements and bias possible estimates of velocity power and cosmological parameters.

Despite these limitations, however, we show below that such measurements can still provide meaningful constraints if one
interprets them with the help of suitable mock catalogs sharing the same angular mask (see section \ref{section3b}). For
instance, estimates of different quantities can be directly compared to the corresponding distributions obtained from the
mocks where systematic effects are under control. As for the data (and mock) analysis presented below, we shall thus employ
the following basic strategy:
\begin{enumerate}
\item Assume a set of parameters that describe the background cosmology and select the galaxy sample according to the absolute
magnitudes and luminosity distances computed, respectively, from apparent magnitudes and redshifts (see section \ref{section3a}).
\item Assuming the linear luminosity evolution model specified in eq. \eqref{eq:qz}, determine the $^{0.1}r$-band LF parameters
including $Q_{0}$ for the case of a vanishing peculiar velocity field, i.e. $a_{lm} = 0$. The value of $Q_{0}$ is kept fixed in
the following steps while the LF parameters are free to vary, except when using the fixed LF estimator explored in section
\ref{sec:bf}.
\item Compute the maximum-likelihood estimate $\hat{\vx}^{\rm ML}$ of the parameter vector $\vx$ introduced in section \ref{section2c}
for a suitable $l_{\rm max}$. These parameters specify both the velocity model and the LF. Approximating $P_{\rm tot}$
locally by a Gaussian distribution and taking the previously found $\phi$ with $a_{lm} = 0$ as an initial guess, this is achieved
by iteratively solving for $\hat{\vx}^{\rm ML}$ until the (exact) likelihood peak is reached. The required derivatives of
$\log{P_{\rm tot}}$ can be calculated analytically and are summarized in appendix \ref{app1}. Convergence is reached after $3$--$5$
iterations for a relative accuracy $10^{-6}$--$10^{-10}$. The CPU time depends on the value of $l_{\rm max}$, but is typically
around a few tens of minutes for about half a million objects. The results are potentially prone to mask-induced degeneracies
related to the spline point separation $\Delta M$ when estimating the LF. We will describe below the various approaches used to
investigate this issue.
\item Estimate the random errors of $\hat{\vx}^{\rm ML}$ from the covariance matrix $\bm{\Sigma}$ which is computed by directly
inverting the observed Fisher matrix $\mathbf{F}$. The Fisher matrix
$\mathbf{F}_{\alpha\beta}=-\partial\log P_{\rm tot}/(\partial x_{\alpha}\partial x_{\beta})$ is evaluated at the maximum value
$\hat{\vx}^{\rm ML}$.\footnote{An immediate worry is that $\mathbf{F}$ could turn out singular or ill-conditioned.
Except for issues, which are related to the normalization of the spline-based LF and can be easily overcome with the
help of standard techniques \cite{James2006}, we have not encountered such problems in our study which considers only
the first few multipoles, i.e. $l_{\rm max}\ltsim 5$.} 
\item Marginalize the resulting distribution $P_{\rm tot}(\vd\vert\vx )$ over all LF parameters that are unrelated to the velocity
field and construct the posterior probability for the $a_{lm}$ and, subsequently, the probability $P(C_{l})$ according to the
prescription given in section \ref{section2c}. Then maximize the latter with respect to $C_{l}$ to estimate the angular power.
Given the characteristics of the SDSS data, such estimates are expected to be quite uncertain, and thus we will limit ourselves
to a proof of concept.
\item Alternatively, consider a spatially flat $\Lambda$CDM model where the $C_{l}$ are not free and independent, but
fully determined by the cosmological parameters $\zeta_{k}$, i.e. $C_{l}=C_{l}(\zeta_{k})$. Constraints are obtained
by sampling the probability $P[C_{l}(\zeta_{k})]$ as a function of $\zeta_{k}$ on a discrete grid. Although other
parameter choices are briefly discussed, we will focus on the quantity $\sigma_{8}$ which corresponds to the amplitude
of the linear matter power spectrum on a scale of $8h^{-1}$ Mpc. To ensure that linear theory remains a valid
description on the physical scales probed in the analysis, we further have to set $l_{\rm max}$ accordingly (see section
\ref{sectionvcosmo} for details).
\end{enumerate}

Except in the case of examining the LasDamas mocks, which contain substantially less galaxies than the other samples
(see section \ref{section3b} above) and cover a smaller redshift range, we will consider the peculiar velocity field
in two redshift bins with $0.02 < z_{1} < 0.07 < z_{2} < 0.22$, comprising about $N_{1}\sim 1.5\times 10^{5}$ and
$N_{2}\sim 3.5\times 10^{5}$ galaxies, respectively. This specific choice is mainly driven by the accuracy of the bulk
flow estimates presented in section \ref{sec:bf}. For two redshift bins, the uncertainties are typically around $100$ km
s$^{-1}$ which is also larger than the expected variation of the flow amplitude within the respective bins, yielding a good
compromise between accuracy and evolution. Further, the actual bin widths are determined by requiring comparable
signal-to-noise ratios which we roughly estimate from the expected variance of the velocity field within the corresponding
bin volumes and the Poisson noise due the finite number of objects. As for the study of the observed data sample, we adopt
the latest cosmological parameters based on the Wilkinson Microwave Anisotropy Probe (WMAP) combined with ground-based
experiments \cite{Calabrese2013} and the recent measurements by the Planck satellite \cite{Planck2013} which are summarized
in table \ref{table1} and denoted by {\tt param{\_}wmap} and {\tt param{\_}planck}, respectively.
%Since we are dealing with sufficiently low redshifts in this work, the computation of luminosity distances is simplified
%by using a third-order expansion in $z$. Depending on the assumed cosmological model, this yields a relative accuracy of
%$0.003$--$0.03$\%.

\begin{figure} 
\includegraphics[width=0.95\linewidth]{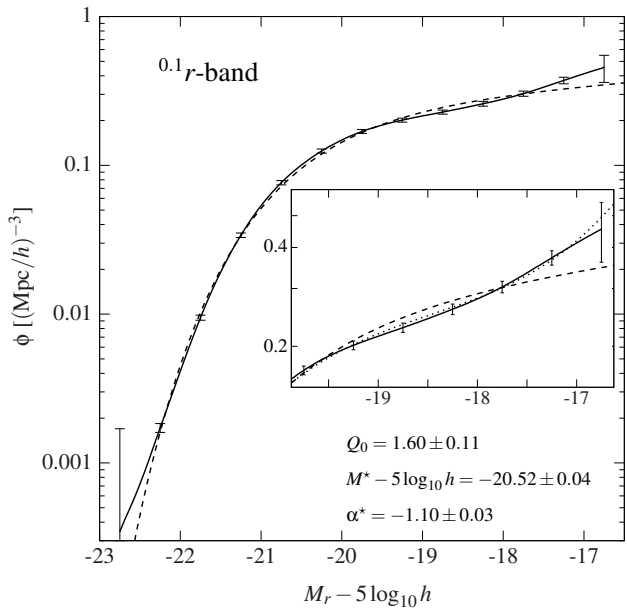}
\caption{The $^{0.1}r$-band LF as obtained from the NYU-VAGC sample: shown are the maximum-likelihood result adopting
the spline-based estimator with $\Delta M=0.5$ (solid line), and two fits based on the Schechter form (dashed line)
and its extension (dotted line; zoomed panel only) which is defined by eq. \eqref{eq:4a}.}
\label{fig2}
\end{figure}

\subsection{The \boldmath{$^{0.1}r$}-band luminosity function of NYU-VAGC galaxies}
\label{sectionlfestimate}

As described in section \ref{sectionnum}, we begin with estimating $\phi(M)$ in the $^{0.1}r$-band from the NYU-VAGC
data for a vanishing velocity field. Adopting the spline-based estimator with a separation of $\Delta M = 0.5$ between
individual spline points, the resulting $\phi(M)$ is shown as a solid line in figure \ref{fig2}, where the normalization
is chosen such that the integral of $\phi(M)$ over the considered absolute magnitude range becomes unity, i.e.
$\eta(-22.5+5\log_{10}h,-17+5\log_{10}h)=1$, and error bars are computed from the ``constrained'' covariance matrix
obtained by enforcing the LF normalization to guarantee a non-singular Fisher matrix. The shape of $\phi(M)$ and the
found evolution parameter, $Q_{0} = 1.6\pm 0.11$, are in good agreement with previous studies based on earlier data
releases \cite{Blanton2003,Montero2009}. While the simple Schechter form with $M^{\star}-5\log_{10}h=-20.52\pm 0.04$
and $\alpha^{\star}=-1.10\pm 0.03$ (dashed line) describes the estimated $\phi$ reasonably well, it does not capture
the visible feature at the faint end.\footnote{We emphasize that the variation in the faint-end slope is robust with
respect to different choices of $\Delta M$ and not an artifact caused by the spline estimator.} Therefore, we consider
an extension to eq. \eqref{eq:2d2} which, after using the relation $L/L^{\star}=10^{0.4(M^{\star}-M)}$, takes the form
\begin{equation}
\phi \propto \left (\frac{L}{L^{\star}}\right )^{\beta^{\star}_{1}}\left\lbrack 1 + 10^{-2.5}
\left (\frac{L}{L^{\star}}\right )^{\beta^{\star}_{2}}\right\rbrack\exp{\left (-\frac{L}{L^{\star}}\right )}
\label{eq:4a}
\end{equation}
and is equivalent to the sum of two Schechter functions with different choices of normalization and $\alpha^{\star}$.
Fitting the above to the spline estimate yields the parameters $M^{\star} = -20.46\pm 0.03$,
$\beta^{\star}_{1} = -1.01\pm 0.03$, and $\beta^{\star}_{2} = -1.64\pm 0.11$, giving a much better representation of
the observed trend. This is illustrated in the zoomed panel of figure \ref{fig2}, where the new result (dotted line) is
compared to both the spline (solid line) and the previous Schechter fit (dashed line).

To further assess our result, we
also calculate the predicted redshift distribution $\dd N/\dd z$ of galaxies which is directly proportional to the radial
selection function $S(z)$, i.e. the fraction of galaxies included in the sample at redshift $z$. The selection function
is easily obtained as an integral of the LF over the magnitude range defined by the redshift-dependent limiting
absolute magnitudes.
\begin{figure} 
\includegraphics[width=0.95\linewidth]{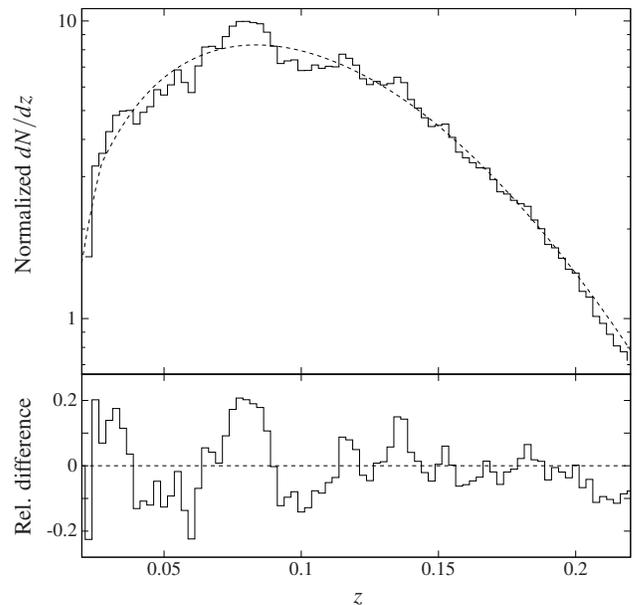}
\caption{Redshift distribution of SDSS galaxies from the NYU-VAGC sample: the histogram (solid line) represents the
observed distribution normalized to unity over the considered redshift range for bins with $\Delta z = 2.5\times 10^{-3}$.
The predicted distribution (dashed line) assumes the spline-based estimate of the LF. }
\label{fig3}
\end{figure}
Figure \ref{fig3} shows that the predicted and observed redshift distributions match quite well, except for a slight
disagreement on the order of a few percent near the high-$z$ cut. This small discrepancy is most likely caused by a
combination of both the limited linear evolution model and the use of different $K$-corrections (individual and mean)
when estimating $\phi(M)$ and the selection function (see section \ref{sectioncaveats} for a discussion of how
luminosity evolution and $K$-corrections impact our peculiar velocity results). Note that all of the above assumes
the set {\tt param{\_}wmap}. Repeating the analysis with the parameters from {\tt param{\_}planck}, however, does not
yield any significant changes.

\begin{figure} 
\includegraphics[width=0.95\linewidth]{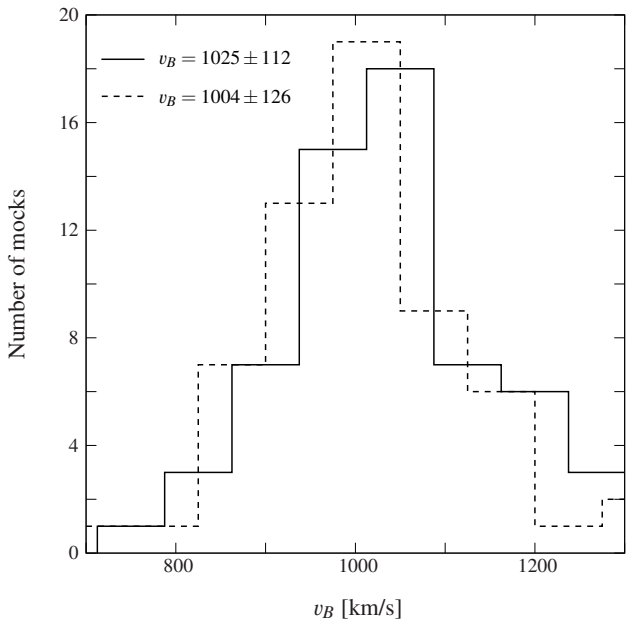}
\caption{Histograms of bulk flow measurements obtained from the customized LasDamas mocks: shown are
the recovered distributions for both a known (dashed line) and unknown (solid line) flow direction.}
\label{fig1}
\end{figure}

\subsection{Constraining bulk flows}
\label{sec:bf}
As a first application of our method, we address how it may be used to constrain the bulk flow, $\vv_{B}$, in the
NYU-VAGC data. We begin with the LasDamas mocks for testing the ability of the method to detect large, anomalous
bulk flows in a SDSS-like catalog. Then we apply the method to the real NYU-VAGC and discuss whether our results
are consistent with the standard $\Lambda$CDM cosmology.

\subsubsection{LasDamas benchmark}
\label{benchmark}
We make use of the LasDamas mocks introduced in section \ref{section3b} and assume
a constant $\vv_{B}$ of $1000$ km s$^{-1}$ pointing toward the direction $(l, b) \approx (266^{\circ}, 33^{\circ})$
expressed in Galactic coordinates. Note that both the magnitude and the direction of $\vv_{B}$ are chosen in accordance
with the recent controversial claim of a ``dark flow'' out to depths of around 300--600$h^{-1}$ Mpc \cite{Kash2008,
Keisler2009, Kash2010, Osborne2011}.

Using the Schechter estimator for the LF and setting $l_{\rm max}=1$, we follow
the procedure outlined in section \ref{sectionnum} to recover the flow $\vv_{B}$ from the customized LasDamas mocks.
The histogram in figure \ref{fig1} shows the resulting component along the input direction for the cases that it is
known (dashed line) and unknown (solid line), i.e. the direction is allowed to vary freely. Clearly, the magnitude of
$\vv_{B}$ is successfully extracted in both cases, and the corresponding rms values of $111$ and $125$ km s$^{-1}$ are
fully consistent with each other as is expected from Gaussian statistics. Although not presented here, the found
distributions along the other (perpendicular) directions for a freely varying $\vv_{B}$ are consistent with a zero
velocity and exhibit a similar scatter. Of course, the current setup neglects any contamination due to leakage from
other multipoles or systematic errors in the data. If these effects remain subdominant in the sense that their
combination leads to changes comparable to or less than the estimated random errors, our result suggests that the
method is capable of constraining large coherent bulk flows using the available galaxy data from the NYU-VAGC. As we
will show below, this condition seems reasonably satisfied, at least for the results in the low-redshift bin with
$0.02<z<0.07$.

\begin{table*}
\caption{Summary of ``bulk flow'' measurements ($l_{\rm max}=1$) in two redshift bins for the parameter set
{\tt param{\_}wmap} and the different models of the LF described in the text.}
\centering
\begin{tabular*}{0.95\linewidth}{@{\extracolsep{\fill}}lcccccc}
\noalign{\medskip}
\hline
\hline
\noalign{\smallskip}
 & \multicolumn{3}{c}{$0.02<z<0.07$} & \multicolumn{3}{c}{$0.07<z<0.22$}
\tabularnewline
\noalign{\smallskip}
$\phi(M)$ & $v_{x}$ [km/s] & $v_{y}$ [km/s] & $v_{z}$ [km/s] & $v_{x}$ [km/s] & $v_{y}$ [km/s] & $v_{z}$ [km/s]
\tabularnewline
\noalign{\smallskip}
\hline
\noalign{\smallskip}
Hybrid & $-227\pm 128$ & $-326\pm 113$ & $-239\pm 73$ & $-367\pm 92$ & $-439\pm 85$ & $-25\pm 71$
\tabularnewline
Fixed & $-175\pm 126$ & $-278\pm 111$ & $-147\pm 58$ & $-340\pm 90$ & $-409\pm 81$ & $-45\pm 43$
\tabularnewline
Schechter & $-151\pm 130$ & $-277\pm 116$ & $-102\pm 78$ & $-422\pm 93$ & $-492\pm 86$ & $-150\pm 74$
\tabularnewline
\noalign{\smallskip}
\hline
\hline
\end{tabular*}
\label{table2}
\end{table*}

\subsubsection{Constraints from the NYU-VAGC}
In the next step, we seek constraints on the velocity field for the case $l_{\rm max}=1$, now using the real NYU-VAGC
galaxy data. As we have argued above, the angular mask of our sample causes such ``bulk flow'' estimates to suffer
from multipole mixing and their interpretation requires the use of mock catalogs. Another mask-related problem arises
from additional degeneracies between the velocity multipoles and the LF, depending on the assumed spline point
separation. For $\Delta M=0.5$, this already becomes an issue, and the straightforward remedy is to increase the
separation to an adequate value. To account for alternative solutions and to further judge our method's robustness,
however, we will consider the following three representative approaches in our analysis:
\begin{enumerate}
\item Fix the LF to its estimate for a vanishing velocity field, i.e. use a predetermined shape of $\phi(M)$ for the
analysis. The rational of this fixed estimator is to evaluate the impact of adding degrees of freedom in the LF model.
\item Adopt a hybrid model by fitting a Schechter form to the spline-based LF estimate for a vanishing velocity field
and expressing the LF as the sum of a Schechter function and the corresponding (fixed) residual.
\item Work exclusively with the Schechter parameterization of the LF.
\end{enumerate}
Featuring the highest flexibility among the above, we expect estimates based on the hybrid LF model to be the most
reliable ones. The corresponding flow measurements will be expressed in a specific Cartesian coordinate system defined
by its $x$-, $y$-, and $z$-axes pointing toward Galactic coordinates $(l,b)\approx (81^{\circ},-7^{\circ})$,
$(172^{\circ},-1^{\circ})$, and $(90^{\circ},83^{\circ})$, respectively. In particular, the system's $z$-axis is
chosen such that it approximately penetrates the central patch of galaxies observed in the northern hemisphere, and thus
it is expected to give the tightest constraints. Using these coordinates, for example, the anomalous bulk flow incorporated
in the LasDamas mocks (see the first paragraph of section \ref{benchmark}) can be written as
$\vv_{B}^{\rm T}\approx (-894,-80,442)$ in units of km s$^{-1}$. To aid the following discussion of our results, let us
further introduce $v_{K}$ as the component along the direction of this anomalous flow which points toward
$(l, b)\approx (266^{\circ}, 33^{\circ})$.

\begin{figure} 
\centering
\includegraphics[width=0.95\linewidth]{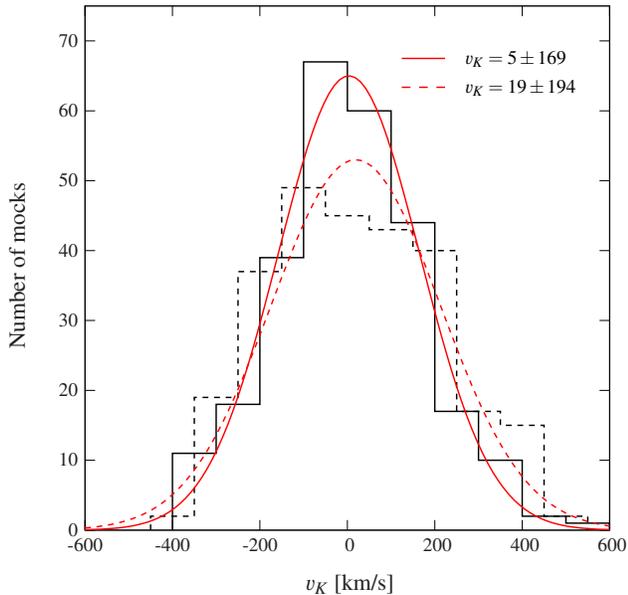}
%\vspace{0pt}
\caption{Histograms of ``bulk flow'' measurements obtained from the simple NYU-VAGC mocks: shown are the recovered
distributions (black lines) and corresponding Gaussian estimates (red lines) for the two redshift bins with
$0.02<z<0.07$ (solid lines) and $0.07<z<0.22$ (dashed lines) along the direction of the anomalous flow assumed in
the first paragraph of section \ref{benchmark}.}
\label{fig4}
\end{figure}

Regarding the real NYU-VAGC galaxy sample, the inferred ``bulk flows'' for the cosmology defined by {\tt param{\_}wmap}
are summarized in table \ref{table2}. A comparison between the estimated flow components shows that the results based on
the various LF models are different, but consistent within their $1\sigma$-values, where the quoted errors are derived
from the observed Fisher information. All measurements are in very good agreement with $v_{K}\approx 120\pm 115$ and
$355\pm 80$ km s$^{-1}$ for the first and second redshift bin, respectively. 
To make sense of these numbers, we compare them to the distribution of $v_{K}$ found with the help of the simple NYU-VAGC
mocks (see section \ref{section3b}) which is presented in figure \ref{fig4}. Note that the mock analysis leading to this
distribution has been performed using a pure Schechter estimator of the LF. Employing the other LF models listed above,
however, gives very similar results and will leave our conclusions unchanged.\footnote{Despite having less degrees of
freedom, this is also true in the case of the fixed LF estimator, where only the spread in the $v_{z}$-component is
significantly reduced.} As can be seen from the figure, the observed distributions in both redshift bins are well described
by Gaussian profiles with (nearly) zero mean and standard deviations of approximately $170$ and $200$ km s$^{-1}$, respectively.
In contrast to the Fisher errors, the dispersion found from the mocks includes both the cosmic signal and contributions due
to the magnitude dipole introduced in section \ref{section3b}. Repeating the mock analysis after removing the latter leads to
a decrease in the dispersions of around $9$\% and $62$\% for the first and second bin, respectively. Systematic errors induced
by the magnitude dipole are expected to increase with the redshift since a bulk flow with amplitude $v_{B}$ is expected to
induce a magnitude offset around $\delta{m}=5\log_{10}(1-v_{B}/cz)$ \cite{Nusser2011}. On the contrary, cosmic variance is
expected to decrease with the volume, and thus with the redshift. The increase of the dispersion with the redshift indicates
that the errors induced by the magnitude dipole obliterate the contribution due to cosmic variance.

We find our measurements of $v_{K}$ to be fully compatible with the distribution obtained from the mocks and consistent with
zero at a $1\sigma$ (first redshift bin) and $2\sigma$ (second redshift bin) confidence level. Given that the estimated flow
components are not necessarily uncorrelated, however, it is more appropriate to consider the joint distribution of the bulk
flow components which is adequately characterized by a multivariate Gaussian. From our set of mocks and the actual data, we
find that correlations between the different components are relatively mild and the corresponding (linear) correlation
coefficients typically take values around $0.1$--$0.3$. Choosing the hybrid model of the LF, for instance, the bulk flow
measured in both bins is consistent with zero within $1.5\sigma$ if we adopt the usual confidence levels of multivariate
normally distributed data with three degrees of freedom. Surprisingly, the bulk flow amplitudes associated with the estimated
components in this case are $v_{B}\approx 490\pm 100$ and $580\pm 80$ expressed in units of km s$^{-1}$, where the errors are
purely statistical and relatively small. Despite these rather large values, the recovered flows correspond to detections at
the $1.5\sigma$ level. The reason is that the distribution of amplitudes does not follow a Gaussian, but has a long tail and
is closely related to the $\chi^{2}$-distribution. The estimated flows are pointing toward
$(l,b)\approx (310^{\circ},-25^{\circ})\pm (30^{\circ},10^{\circ})$ and $(310^{\circ},5^{\circ})\pm (10^{\circ},15^{\circ})$
in Galactic coordinates for the first and second redshift bin, respectively. Note that changing the cosmological parameters
to {\tt param{\_}planck} or using the other LF models yields basically the same results.

\begin{figure*}
\vspace{10pt}
\includegraphics[width=0.88\linewidth]{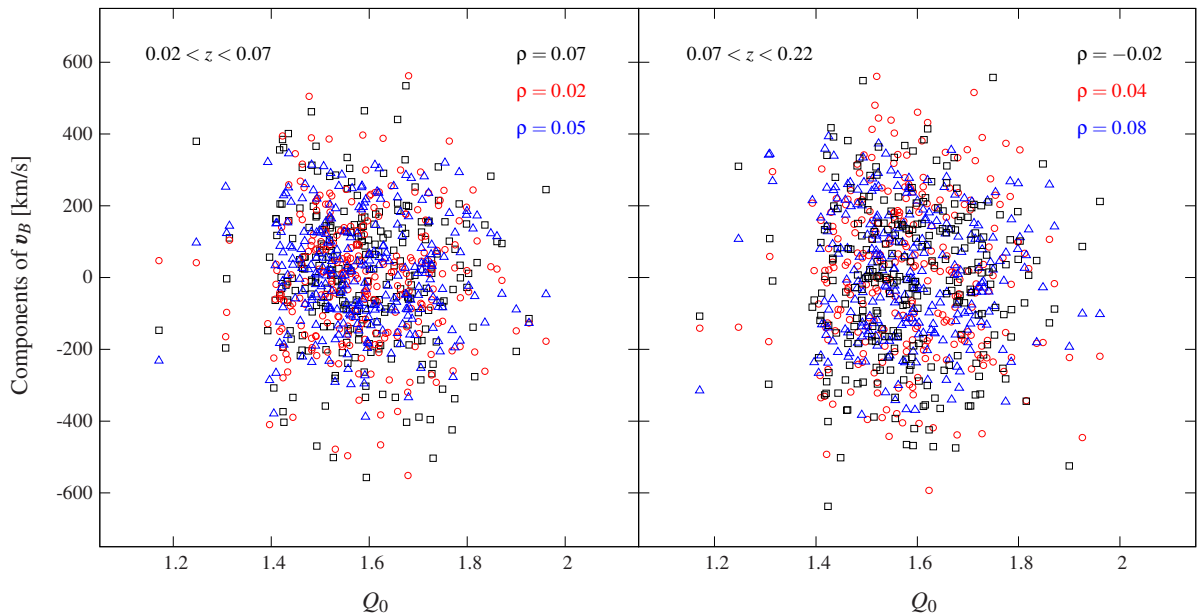}
\caption{Scatter plots of ``bulk flow'' components versus the linear evolution parameter $Q_{0}$ obtained from the
simple NYU-VAGC mocks: the panels illustrate the resulting distributions of $v_{x}$ (black squares), $v_{y}$ (red
circles), and $v_{z}$ (blue triangles) for the two redshift bins with $0.02<z<0.07$ (left) and $0.07<z<0.22$ (right).}
\label{fig5}
\end{figure*}

As for the comparison of our flow measurements with the mock catalog results, we point out that the simple NYU-VAGC
mocks are not only built and analyzed with a slightly different cosmological model, but also ignore any redshift
dependence of the peculiar velocity field which is assumed at $z=0$ (see section \ref{section3b}). For typical choices
of cosmological parameters and the redshift range of interest, this amounts to small differences $\ltsim 3$\% and can,
therefore, be ignored in our analysis. Another concern is that fixing the linear evolution as described in section
\ref{sectionnum} causes a bias in the flow components since the monopole-like term $Q(z)$ might leak in through the mask. 
To ensure that this is not the case, we plot the inferred components for both redshift bins against the estimate of
the parameter $Q_{0}$ in figure \ref{fig5}. A brief visual inspection of the scatter already indicates that
there is no evidence for a correlation between these quantities. This is confirmed by calculating the linear correlation
coefficients which turn out smaller than $0.1$ in all the cases. Together with the above findings, we thus conclude that
the SDSS galaxy data exhibit no hint toward anomalously large flows. Accounting for the known magnitude tilt in the
photometric calibration, our velocity measurements further appear fully consistent with the expectations of a
$\Lambda$CDM cosmology.

\subsection{Higher-order multipoles: constraining angular power and \boldmath{$\sigma_{8}$}}
\label{sectionvcosmo}
As we have outlined in section \ref{section2c}, the luminosity-based approach considered in this work is analogous to
the analysis of CMB anisotropies and, in principle, it is straightforward to constrain the angular velocity power spectrum
using basically the same techniques. Given the characteristics of the SDSS data and our previous findings from section
\ref{sec:bf}, however, we already expect such constraints to be rather weak and potentially biased because of the
systematic magnitude tilt described in section \ref{section3b}. Nevertheless, we shall explore the potential of this
approach and illustrate some examples involving simple velocity models.

\begin{figure*} 
\vspace{10pt}
\includegraphics[width=0.87\linewidth]{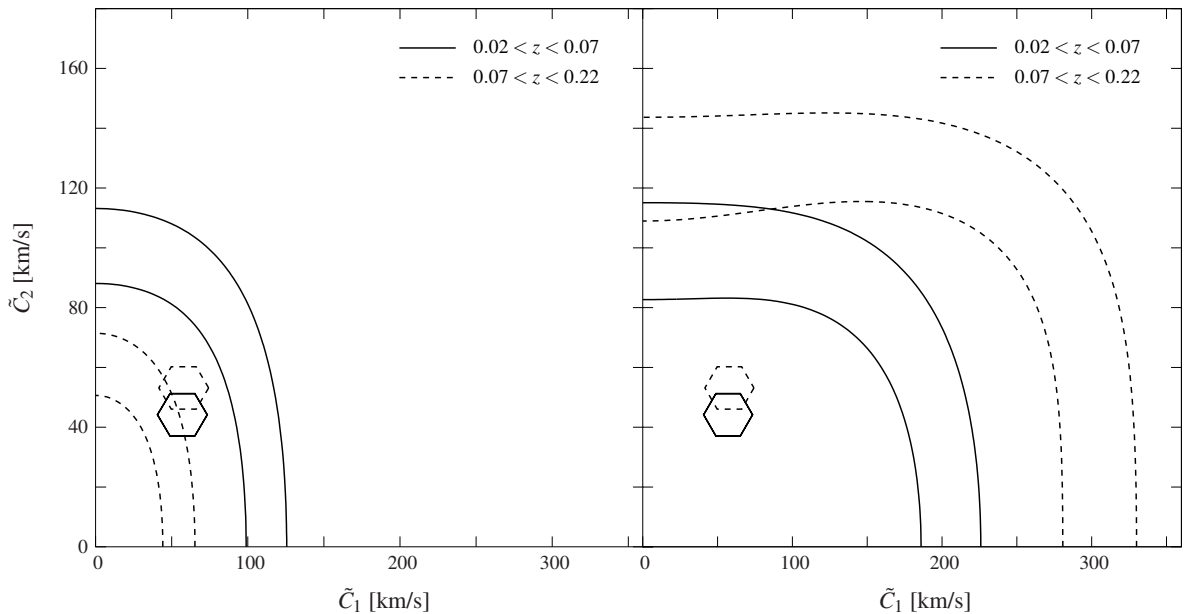}
\caption{Constraints on angular velocity power in a randomly selected mock for a model with $l_{\rm max}=2$: shown
are the joint $1\sigma$ and $2\sigma$ confidence regions of $\tilde{C}_{1}$ and $\tilde{C}_{2}$ (see text) for the
first (solid lines) and second (dashed lines) redshift bin, respectively, estimated with (right panel) and without
(left panel) a systematic dipole in the galaxy magnitudes. The hexagons indicate the corresponding results obtained
from directly using the known galaxy velocities.}
\label{fig6}
\end{figure*}

\subsubsection{Constraints with no cosmology priors}
Let us assume a velocity model with $l_{\rm max}=2$ and assess the impact of a tilt in the zero-point photometry
with the help of a mock galaxy catalog randomly chosen form the NYU-VAGC set. To facilitate a direct interpretation
in terms of velocities, we additionally define the dimensional quantity 
\begin{equation}
\tilde{C}_{l}\equiv \sqrt{\frac{2l+1}{4\pi}C_{l}}
\label{eq:4d}
\end{equation}
which will be used in what follows below. Again, we assume the latest pre-Planck $\Lambda$CDM cosmology determined
by {\tt param{\_}wmap} and also work with the hybrid estimator of the LF (see section \ref{sec:bf}). Figure \ref{fig6}
shows the joint $1\sigma$ and $2\sigma$ confidence regions of $\tilde{C}_{1}$ and $\tilde{C}_{2}$, estimated after
maximizing the likelihood $P(C_{l})$ in eq. \eqref{eq:2c4} for the same mock, with (right panel) and without (left
panel) mimicking the systematic magnitude dipole offset (see section \ref{section3b}). Here the posterior likelihood
is constructed separately for each redshift bin after marginalizing $P\left (\vd\vert a_{lm}\right )$ over the $a_{lm}$
of the respective other one, and the resulting contour lines are derived using the quadratic estimator presented in
ref. \cite{Bond1998}. The effect of a spurious magnitude dipole mostly affects the probability contour along the
$\tilde{C}_{1}$-axis, i.e. the power in the dipole, and as expected, the amplitude of the effect increases with the
redshift. To quantify the smearing introduced when one estimates velocities through luminosity variations, we further
compare the contours with the values of $\tilde{C}_{1}$ and $\tilde{C}_{2}$ inferred directly from the galaxy peculiar
velocities in the mocks (hexagons in the figure). The estimated constraints are consistent with these values within
the (large) $1$--2$\sigma$ bounds.

\begin{figure*}
\vspace{10pt} 
\includegraphics[width=0.45\linewidth]{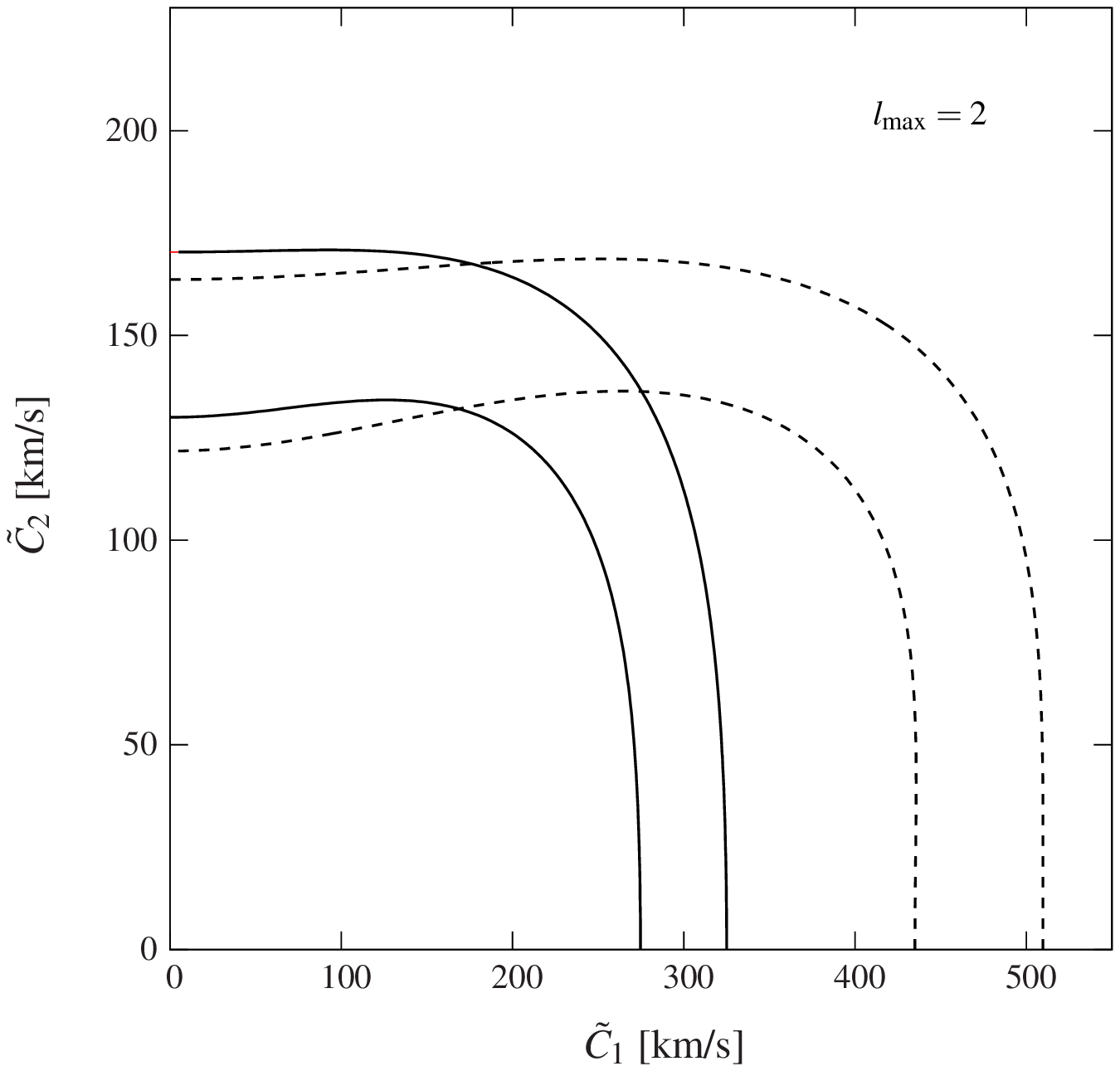}
\hfill
\includegraphics[width=0.45\linewidth]{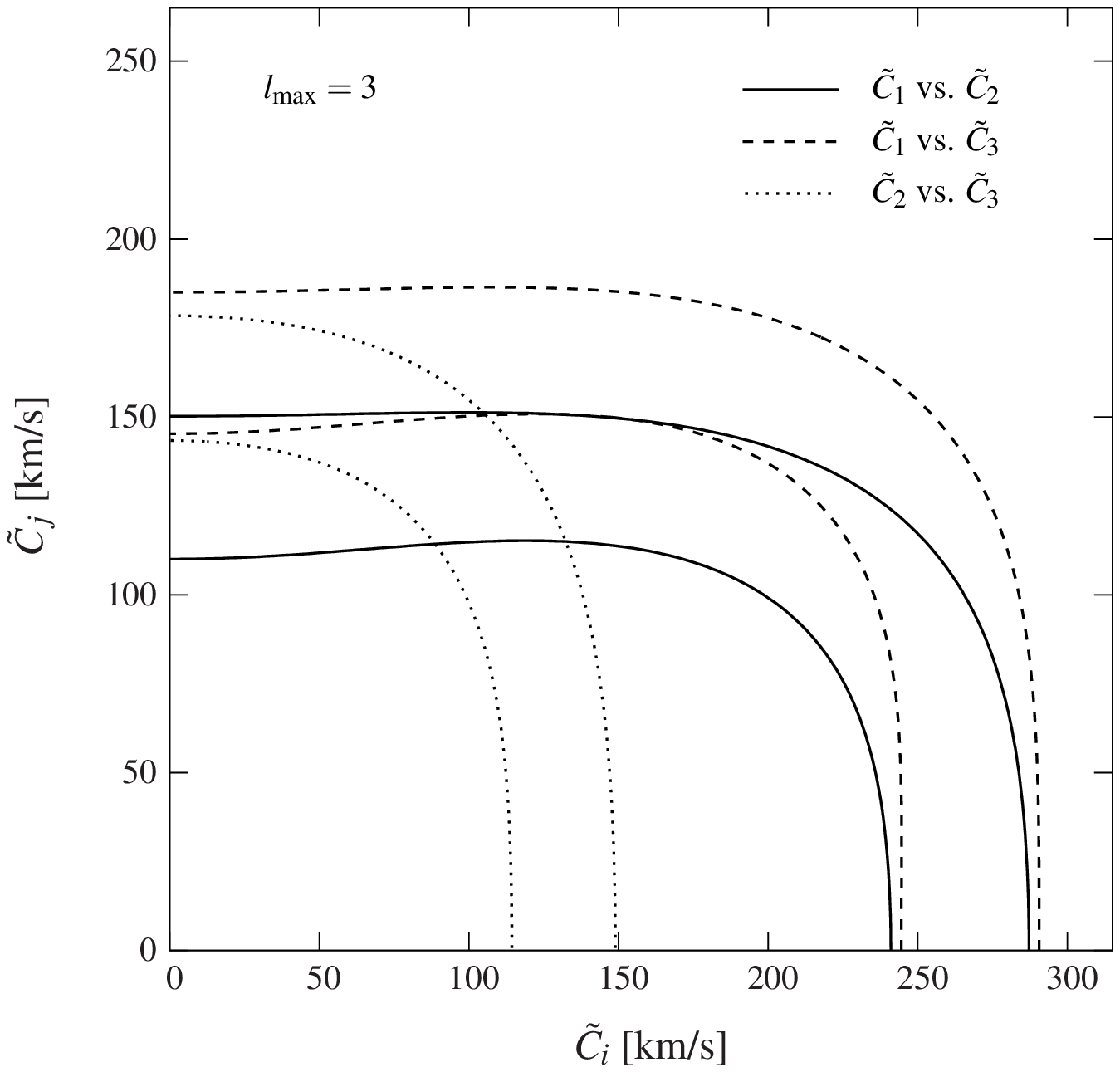}
\caption{Left panel: same as figure \ref{fig6}, but now for the real NYU-VAGC galaxy data which is missing direct
velocity estimates. Right panel: adopting a model with $l_{\rm max}=3$ and fixed LF to analyze the real galaxy
data, the plot shows the marginalized joint $1\sigma$ and $2\sigma$ confidence regions of $\tilde{C}_{1}$ and
$\tilde{C}_{2}$ (solid line), $\tilde{C}_{1}$ and $\tilde{C}_{3}$ (dashed line), and $\tilde{C}_{2}$ and
$\tilde{C}_{3}$ (dotted line) for the first redshift bin.}
\label{fig7}
\end{figure*}

Repeating the analysis for the real SDSS galaxies, we end up with the confidence regions depicted in the left panel
of figure \ref{fig7}. Although it is not very constraining in the present case, our analysis restricts the $\tilde{C}_{l}$
for $l_{\rm max}=2$ to several hundred km s$^{-1}$ and is consistent with zero power. This fully agrees with the
predictions of the $\Lambda$CDM model and does not suggest any anomalous properties. We also note a striking resemblance
in contour trends with the mock result in figure \ref{fig6} from which it is tempting to deduce the existence of a
formidable dipole contamination in the real data. Since, among other uncertainties, there is still leakage due to the
survey geometry, however, strong statements like that cannot be made.

Including higher velocity multipoles with $l_{\rm max}\geq 3$, the constraints become even weaker as the level of
degeneracy increases. To give a final example, we assume another model with fixed LF and set $l_{\rm max}=3$. The
corresponding confidence regions of the different $\tilde{C}_{l}$ are shown in right panel of figure \ref{fig7}.
Note that the tighter bounds are only a consequence of reducing the available degrees of freedom.

\begin{figure*} 
\vspace{5pt}
\includegraphics[width=0.9\linewidth]{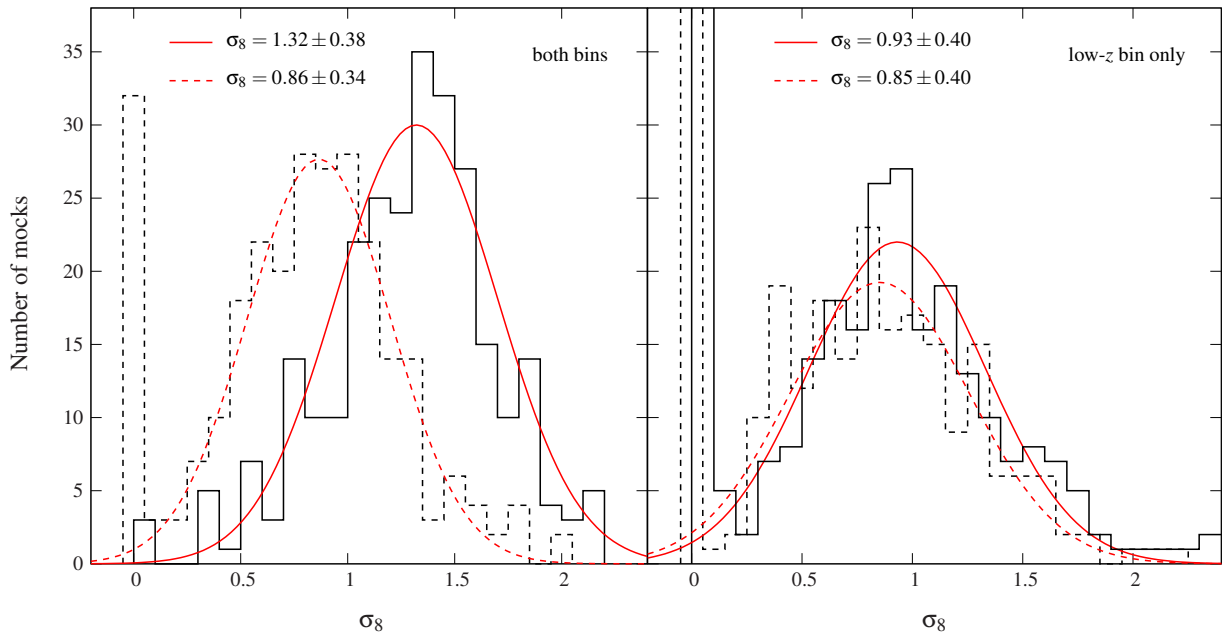}
\caption{Distribution of $\sigma_{8}$ estimated from the simple NYU-VAGC mocks: shown are the recovered
histograms (black lines) with (solid lines) and without (dashed lines) the inclusion of a systematic dipole
in the galaxy magnitudes, using the information in both redshift bins (left) and the first redshift bin
with $0.02<z<0.07$ only (right). Note that the Gaussian curves (red lines) are obtained by fitting the
observed cumulative distribution, excluding the cases where no large-scale power is detected.}
\label{fig8}
\end{figure*}

\subsubsection{Constraints with cosmology priors}
Next, we shall consider constraints on cosmological parameters by imposing a $\Lambda$CDM prior on the angular power
spectrum as detailed above in section \ref{section2c}. In doing so, it is convenient to divide the parameters that define
the cosmological models into two categories. The ones that characterize the background cosmology and that are used to 
estimate absolute magnitudes and compute distances for sample selection, and those that characterize the density fluctuations.
Here we focus on $\sigma_{8}$ which belongs to the latter category, and assume that all other parameters are fixed to their
values in {\tt param{\_}wmap}. At the end of this section, we shall briefly discuss other choices and comment on the
possibility of constraining background parameters such as $\Omega_{m}$.

\begin{figure*} 
\vspace{10pt}
\includegraphics[width=0.91\linewidth]{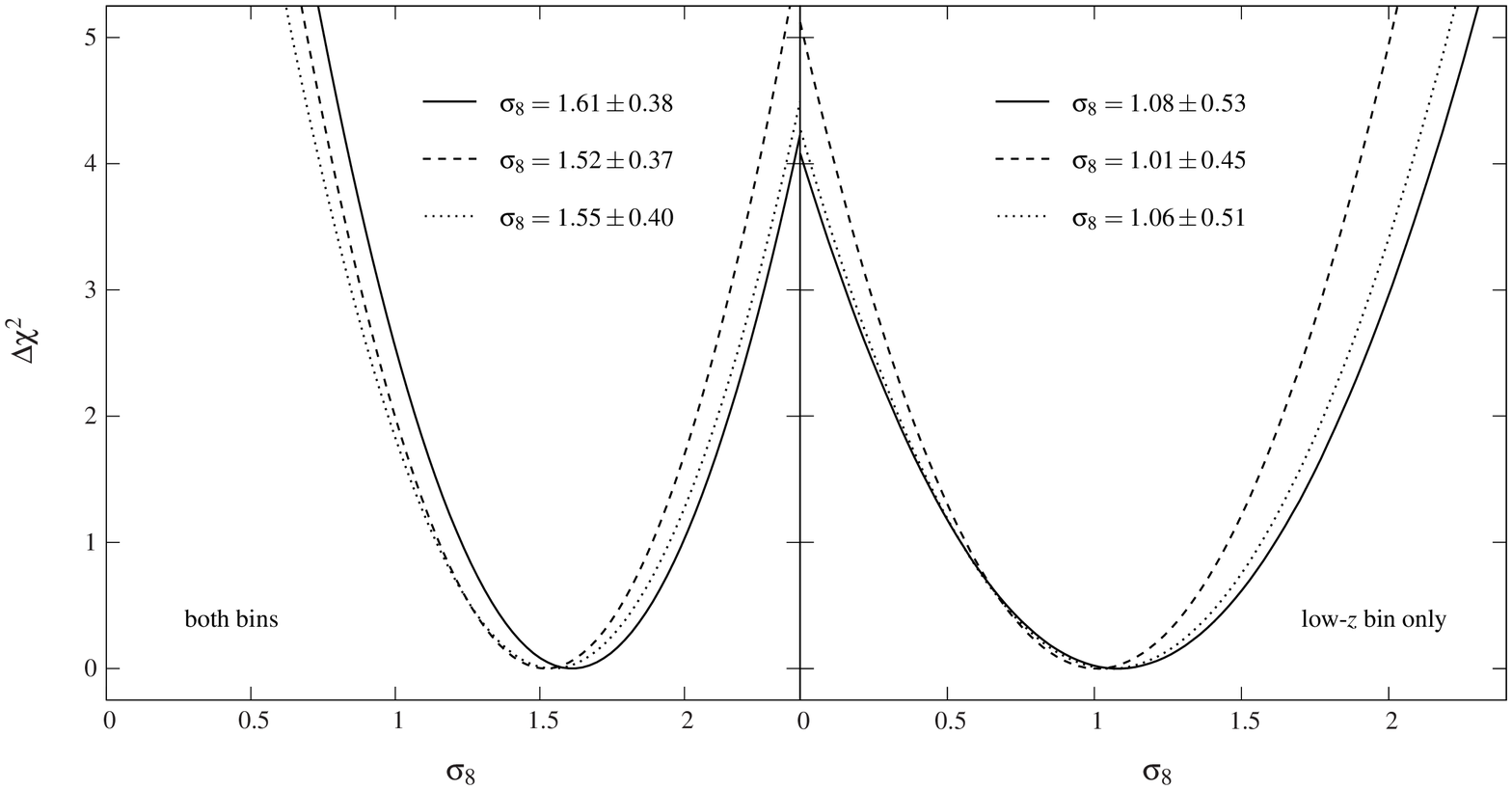}
\caption{Raw estimates of $\sigma_{8}$ obtained from the real NYU-VAGC galaxy data: shown is the derived
$\Delta\chi^{2}$ as a function of $\sigma_{8}$ for both redshift bins (left panel) and the first redshift bin
with $0.02<z<0.07$ only (right panel), adopting the hybrid (solid line), fixed (dashed line), and Schechter
(dotted line) estimator of the LF.}
\label{fig9}
\end{figure*}

Like in the previous section, we assess the impact of a spurious tilt in the estimated magnitudes. To guarantee the validity
of linear theory, we set $l_{\rm max}=5$ which corresponds to considering physical scales above $\sim 100h^{-1}$ Mpc.
Concerning the calculation of the theoretical $C_{l}$ which is required for the prior probability and summarized in appendix
\ref{app2}, we adopt the parameterized form of the matter power spectrum $P(k)$ given in ref. \cite{EH98}. Moreover, the
galaxy redshift distribution $p(z)$ used to compute the bin-averaged velocity field given by eq. \eqref{eq:app2a} is taken to
be of the form
\begin{equation}
p(z) \propto z^{a}\exp{\left\lbrack -\left (z/\overline{z}\right )^{b}\right\rbrack },
\label{eq:4e}
\end{equation}
where the parameters $a=1.31$, $b=1.94$, and $\overline{z}=0.1$ are found by directly fitting eq. \eqref{eq:4e} to the
observed distribution. As is customary, $\sigma_{8}$ is inferred from discretely sampling the posterior probability and
interpolating the corresponding result. In our calculations, we will choose a step size of $0.05$.

Applying this procedure to the full suite of mock catalogs with and without the inclusion of a systematic magnitude
dipole, we obtain the histograms shown in figure \ref{fig8}. While the results for the combination of both redshift
bins stem from maximizing $P(C_{l})$ in eq. \eqref{eq:2c4} using the full probability $P\left (\vd\vert a_{lm}\right )$,
those for the low-$z$ bin are computed by constructing the posterior probability after marginalizing
$P\left (\vd\vert a_{lm}\right )$ over all $a_{lm}$ in the high-$z$ bin. Note that the former approach is actually
inconsistent as it incorrectly assumes that the $a_{lm}$ are uncorrelated between different bins. Accounting for these
missing correlations, however, leads only to differences of a few percent in the estimated values of $\sigma_{8}$ and
its error, suggesting that they may be safely neglected. As described in section \ref{section3b}, the NYU-VAGC mocks
are based on the parameter set {\tt param{\_}mock} and assume an input value of $\sigma_{8}=0.8$. In addition, we assume
a Schechter LF for the mock analysis (just like in section \ref{sec:bf}), but using the other LF models does not significantly
change the results. The spikes in the histograms at $\sigma_8=0$ correspond to the cases in which we do not detect
any power. Once we exclude those, the histograms are reasonably well represented by Gaussian distributions with
standard deviations of $\sim 0.3$--$0.4$ (solid and dashed, red lines). As is readily seen from comparing the left
and right panels, the presence of a systematic magnitude dipole (solid lines) causes a bias in the estimate of
$\sigma_{8}$ which is rather severe for higher redshifts, i.e. $0.07<z<0.22$ (left panel). As expected, removing the
dipole (dashed lines) also eliminates the bias, thus leading to the same mean value of $\sigma_{8}$ in both cases.
Expressing the bias in numbers, the dipole contribution to galaxy magnitudes amounts to a systematic shift of
$\Delta\sigma_{8}\approx 0.13$ and $\Delta\sigma_{8}\approx 0.52$ for the low-$z$ bin and the combination of both
redshift bins, respectively.

Considering now the real SDSS galaxy sample, we perform exactly the same analysis to obtain measurements of $\sigma_{8}$
for the different LF estimators introduced in section \ref{sec:bf}. Our results are presented in figure \ref{fig9} which
shows the derived $\Delta\chi^{2}$ as a function of $\sigma_{8}$, obtained using the information from both redshift bins
(left panel) and that of the first redshift bin only (right panel). Similar to what we have discovered in
our investigation of ``bulk flows'', the values based on different LF models agree very well within their corresponding
$1\sigma$ errors, and we get $\sigma_{8}\sim 1.0$--$1.1$ in the low-$z$ bin and $\sigma_{8}\sim 1.5$--$1.6$ over the full
$z$-range. Remarkably, the measured values and uncertainties closely match the inferred biased distributions of the previous
mock analysis depicted in figure \ref{fig8}. If the magnitude tilt in the SDSS data is the only relevant source of
systematic errors and sufficiently characterized by a dipole-like modulation, we can use the bias estimated from the
mocks to correct our measurements. Taking the result of the hybrid LF estimator (solid line in figure \ref{fig9}), for
example, we obtain corrected values of $\sigma_{8}=1.09\pm 0.38$ (both bins) and $\sigma_{8}=0.95\pm 0.53$ (low-$z$ bin)
which are fully compatible with each other and also consistent with the expectation of the $\Lambda$CDM model. The quoted
errors are the statistical errors inferred from the NYU-VAGC data. Note that changing the cosmology to {\tt param{\_}planck}
or choosing a different LF estimator has only a minor impact on the results.

\begin{figure} 
\centering
\includegraphics[width=0.95\linewidth]{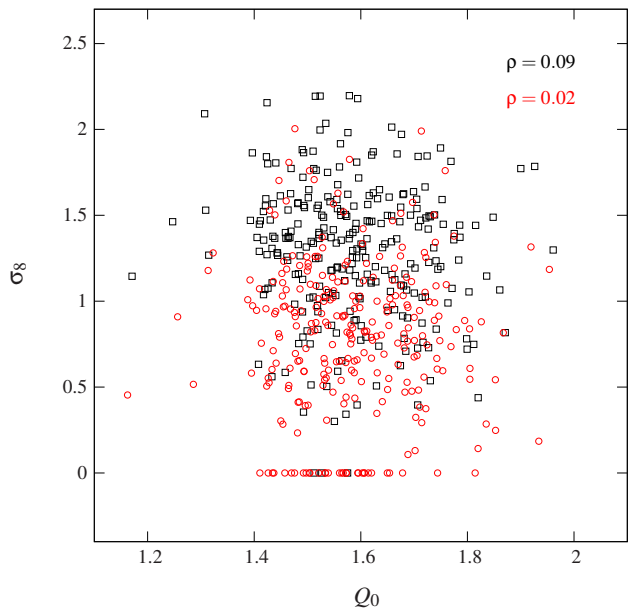}
%\vspace{0pt}
\caption{Scatter plots of $\sigma_{8}$ versus the evolution parameter $Q_{0}$ for the simple NYU-VAGC mocks:
using the information from both redshift bins, the plot illustrates the resulting distributions obtained with
(black squares) and without (red circles) a systematic dipole in the galaxy magnitudes.}
\label{fig10}
\end{figure}

Again, one may ask whether fixing the linear evolution as described in section \ref{sectionnum} causes an additional bias
in our measurements of $\sigma_{8}$. To answer this question, we plot the derived values of $\sigma_{8}$ for both redshift
bins against the estimate of the evolution parameter $Q_{0}$ in figure \ref{fig10}, using the simple NYU-VAGC mocks with
(black squares) and without (red circles) the magnitude dipole. As before (see section \ref{sec:bf}), the linear correlation
coefficients turn out $\ltsim 0.1$, and there is no indication for a correlation between these quantities. 

Of course, one is not restricted to $\sigma_{8}$, but also free to look at other cosmological parameters or various
combinations thereof. Considering the two parameters $h$ and $\Omega_{b}$ which, together, determine the baryonic matter
density, for instance, we find that the respective constraints turn out weaker than before, and are also highly
degenerate with $\sigma_{8}$. Similar statements should hold for the parameter $\Omega_{m}$. However, we do not explicitly
check this because changing $\Omega_{m}$ alters the background cosmology and implies a modification of the survey volume and
the selection of the galaxy sample. Taking this into account would substantially increase the workload of the analysis.

\subsection{Caveats}
\label{sectioncaveats}
\subsubsection{Coherent photometric errors and spurious signals}
 
Although our analysis of the SDSS galaxy data has tried to account for known systematics such as the zero-point photometric
tilt (see the mock description in section \ref{section3b}), there could exist additional errors in the photometric calibration.
It is important to note that the impact of such errors increases with the redshift and significantly affects the results
obtained at $z\gtsim 0.1$ in our analysis. An example is the possibility of zero-point offsets in magnitudes between the
whole northern and southern hemispheres due to observing the galaxies in disconnected regions. If this offset was around
$0.01$--$0.02$, it should basically contribute a spurious flow of $\sim 100$--$250$ km s$^{-1}$ to the actual bulk motion
along the connecting axis, assuming a redshift of $z=0.1$ for all galaxies. Since the SDSS footprint gives much more weight
to the northern sample, however, the effect is much less pronounced. As we have verified with the help of our galaxy mocks,
it has just a mild impact on velocity measurements along the previously defined $z$-axis (see section \ref{sec:bf}), thus
leaving our conclusions unchanged. In fact, the missing evidence for any large-scale flow anomaly found in sections \ref{sec:bf}
and \ref{sectionvcosmo} already indicates that there are no other relevant systematics which would otherwise require serious
fine-tuning.

As for the photometric tilt, which constitutes the main source of systematic errors in the present analysis, it
is worth pointing out that, although difficult in practice, such a photometric bias could in principle be characterized and
corrected for by using additional information from star or galaxy counts and clustering from the very same SDSS DR7 dataset.
Another possibility is to recalibrate the SDSS photometry with the help of independent datasets. For example, this could be
accomplished using observations of the Panoramic Survey Telescope and Rapid Response System (Pan-STARRS) which covers $3/4$
of the sky visible from Hawaii \cite{Kaiser2002, Kaiser2010} or, indeed, any future, wide galaxy survey with good photometric
stability. With this respect, a major step forward in control on the photometric calibration is expected from galaxy surveys
carried out from space like, for instance, the planned Euclid mission \cite{euclid2011}.

Imperfect corrections for Galactic extinction might also cause a systematic large-scale offset in the estimated magnitudes.
In the NYU-VAGC, this correction is based on the maps given in ref. \cite{Schlegel1998}. The recent comparison between these
and the reddening maps obtained from the Pan-STARRS1 stellar photometry \cite{Schlafly2014} does not hint at any large-scale
coherent residual with an amplitude comparable to that of the known magnitude tilt. The new dust maps that will be obtained
from the Planck data will settle the issue.

\subsubsection{Environmental dependence of the luminosity function}
There are several studies which strongly hint toward a dependence of the LF on the large-scale
environment of galaxies \cite[e.g.,][]{Balogh2001, Mo2004, Croton2005, Park2007, Merluzzi2010, Faltenbacher2010}.
However, investigations trying to shed light on the connection between luminosity and galaxy density are typically
limited to scales of a few Mpc, thus not probing the large scales relevant to our work. As a matter of fact, an
analysis addressing these environmental dependencies of galaxy luminosities on scales $\sim 100h^{-1}$ Mpc and
above is still unavailable. Nonetheless, we may get an idea from extrapolating the observed dependence into the
large-scale domain. Using that the rms of density fluctuations averaged over scales $\gtsim 100h^{-1}$ is less
than about $0.07$, this gives rise to a particularly small effect if the overdensity of the large-scale
environment turns out to be the most important factor. As we have already suggested in ref. \cite{Nusser2013},
it should further be feasible to take such effects into account by performing measurements over independent
volumes which are classified in terms of their average density.

\subsubsection{Luminosity evolution and $K$-corrections}
The analysis conducted in this paper assumes that the evolution of galaxy luminosity can be effectively described with
a linear model of the form $Q(z)= Q_{0}(z-z_{0})$ where $z_{0}=0.1$ and $Q_{0}$ is determined according to section
\ref{sectionnum}. To assess the robustness of our results with respect to this specific model, we have carried out
a few simple tests. In a first run, we have examined the influence of varying $Q_{0}$ within its estimated $3\sigma$
confidence interval. This typically leads to changes in the measured ``bulk flow'' components of several tens of
km s$^{-1}$ and causes deviations less than $10$\% in the found values of $\sigma_{8}$. Moreover, we have explored
nonlinear evolution models with second- and third-order terms. In this case, the corresponding changes in our
velocity measurements turn out even smaller and can be safely neglected. Similarly, we have studied the impact of
different $K$-corrections using the mean correction given by eq. \eqref{eq:4b} and two-dimensional
polynomial fits as a function of redshift and $g-r$ color \cite{Chilin2010}.\footnote{As advised by the SDSS
collaboration, we use model magnitudes to estimate the $g-r$ colors of galaxies.} Again, the resulting differences
are marginal and do not affect any of our conclusions.

\section{Conclusions}
\label{section5}
We have exploited the well-known fact that peculiar motion induces spatially-coherent variations in the observed 
luminosity distribution of galaxies to probe the cosmic velocity field at $z\sim 0.1$ from the luminosity distribution
of SDSS galaxies in the NYU-VAGC. The method adopted here extends the maximum-likelihood approach proposed in ref.
\cite{Nusser2011} to constrain the peculiar velocity field beyond the bulk flow component. Considering the bin-averaged
peculiar velocity field in two different redshift bins, $0.02<z<0.07$ and $0.07<z<0.22$, we have demonstrated how 
the method permits bounds on the corresponding angular velocity power spectrum and cosmological parameters. The main
results of our analysis can be summarized as follows:
%The results have been assessed and calibrated using  mock catalogs designed to match the sky coverage and magnitude limit of
%the actual catalog. The mock catalogs are based on the standard $\Lamvbda$CDM model.
\begin{itemize}
\item
To assess the robustness of our analysis against potential systematic errors, we have used a suite of mock galaxy catalogs
obtained both from numerical simulations and from the NYU-VAGC dataset itself to match the real data as close as possible.
We have identified three main obstacles which potentially hamper the analysis of the SDSS data: the survey geometry, which
causes mixing between different moments, the possible degeneracies between the velocity multipoles and the estimator of the
LF, and the presence of a coherent photometric tilt of about $0.01$ magnitude across the survey region. While the impact of
mode-to-mode mixing can be readily quantified by modeling the sky coverage and the influence of the LF has been evaluated
by applying different estimators to the mocks, the latter effect is less trivial to account for. Here we have modeled the
zero-point photometry offset by adding a randomly oriented dipole normalized such that the corresponding rms over all
galaxies is $\delta m_{\rm dipole}=0.01$ for each individual mock galaxy catalog. Our results suggest that the systematic
tilt in the observed galaxy magnitudes is sufficiently described by this dipole contribution. 

\item
Accounting for the known systematics in the SDSS photometry, the estimated ``bulk flows'' are consistent with the predictions
of the standard $\Lambda$CDM model at $\ltsim 1$--$2\sigma$ confidence in both redshift bins. The combined analysis of the
corresponding three Cartesian components further corroborates this result. Using an independent estimator, this confirms the
findings of the CMB studies in refs. \cite{Osborne2011, planck_bf} which provide an upper bulk flow limit of a few hundred
km s$^{-1}$ at a $95$\% confidence limit on similar scales.

\item
Our analysis yields direct constraints on the angular velocity power spectrum $C_{l}$ (considering terms up to the octupole)
defined in section \ref{section2c}, independent of a prior on the cosmological model. All of the estimated $C_{l}$ are
consistent with the corresponding theoretical power spectra of the $\Lambda$CDM cosmology.

\item 
Assuming a prior on the $C_{l}$ as dictated by the $\Lambda$CDM model with fixed density parameters and a Hubble constant,
we have used the method to infer the parameter $\sigma_{8}$ which determines the amplitude of the velocity field. After
correcting for known systematics, we obtain $\sigma_{8}\approx 1.1\pm 0.4$ for the combination of both redshift bins and
$\sigma_{8}\approx 1.0\pm 0.5$ for the low-redshift bin only. As anticipated, the found constraints on velocity moments and
$\sigma_{8}$ are not very tight. However, they show the validity of our approach in view of future analyses with different
datasets.

\item 
As for the encountered data-inherent issues, current and next-generation spectroscopic surveys are designed to alleviate
most of them, thanks to their large sky coverage (e.g., eBOSS\footnote{ http://www.sdss3.org/future/}, DESI \cite{bigboss2011,
Levi2013}) and improved photometric calibration in ground-based surveys (e.g., PAN-STARRS \cite{Kaiser2002, Kaiser2010}) and
especially in space-borne experiments like Euclid \cite{euclid2011}. Note that since uncertainties in the measured redshifts
play a little role in our error budgets, the method is also suitable for application to wide photometric redshift surveys
such as the 2MASS Photometric Redshift catalog (2MPZ) \citep{Bilicki2014} and, again, Euclid.

\item  
These excellent observational perspectives give us confidence that the method considered here will become a full-fledged
cosmological probe, independent and alternative to the more traditional ones based on galaxy clustering, gravitational
lensing and redshift space distortions. We expect that combining all these approaches will result in superior control
over potential systematic errors that might affect the estimate of cosmological quantities, chief among them the growth
rate $f(\Omega)$ of density fluctuations \cite{Nusser2012}. 

\item The main interest here is the methodological aspect since the novel approach to estimate the angular velocity power
spectrum or cosmological parameters, developed in analogy to the statistical treatment of CMB anisotropies, is to be regarded
as a proof of concept guiding future analyses. As a final remark, we point out that it should be conceivable to reverse the
ansatz taken in this work, allowing one to constrain luminosity evolution and to improve the photometric calibration of a
galaxy sample in a given cosmological framework.
\end{itemize}

\begin{acknowledgments}
This research was supported by the I-CORE Program of the Planning and Budgeting Committee, THE ISRAEL SCIENCE FOUNDATION (grants No. 1829/12
and No. 203/09), the German-Israeli Foundation for Research and Development, the Asher Space Research Institute, and in part by the Lady
Davis Foundation. M.F. is supported by a fellowship from the Minerva Foundation. E.B. is supported by INFN-PD51 INDARK, MIUR PRIN 2011
``The dark Universe and the cosmic evolution of baryons: from current surveys to Euclid'', and the Agenzia Spaziale Italiana  from the
agreement ASI/INAF/I/023/12/0.
\end{acknowledgments}
 
\appendix

\onecolumngrid

\section{Quadratic approximation of \boldmath{$\log P_{\rm tot}$}}
\label{app1}
As previously discussed in section \ref{section2c}, the total log-likelihood of observing galaxies with absolute magnitudes
$M_{i}$ given their redshifts and radial peculiar velocities in a real (or simulated) dataset is
\begin{equation}
\log P_{\rm tot} = \sum_{i}\log P_{i}\left (M_{i}\vert z_{i},V_{i}\right ) = \sum_{i}\log\frac{\phi(M_{i})}{\eta\left (M^{+}_{i},M^{-}_{i}\right )},
\label{eq:app1a}
\end{equation}
where the function $\eta\left (M^{+}_{i},M^{-}_{i}\right )$ and the limiting magnitudes $M_{i}^{\pm}$ are defined through
eqs. \eqref{eq:2b2} and \eqref{eq:app1c}. Assuming a set of LF and evolution parameters $q_{j}$ as well as a redshift-binned
velocity model $\tilde{V}(\hvr ) = \sum a_{lm}Y_{lm}^{*}(\hvr )$, we now seek an expansion of the form
\begin{equation}
\log P_{i} \approx \log P_{i}\vert_{\vx =\vx_{0}} + \sum\limits_{\alpha}\left.\frac{\partial\log P_{i}}{\partial x_{\alpha}}\right
\vert_{\vx =\vx_{0}}x_{\alpha} + \sum\limits_{\alpha ,\beta}\left.\frac{\partial^{2}\log P_{i}}{\partial x_{\alpha}\partial x_{\beta}}
\right\vert_{\vx =\vx_{0}}x_{\alpha}x_{\beta}.
\label{eq:app1d}
\end{equation}
Here we have introduced $\vx^{\rm T} = (q_{j},a_{lm})$ and $\vx_{0}$ corresponds to an arbitrary fixed parameter vector. Given a
model of $\phi (M)$, it is then straightforward (but tedious) to derive the specific form of the above expansion by using eqs.
(\ref{eq:2c2}--\ref{eq:app1c}) and the relations
\begin{equation}
\begin{split}
\frac{\partial{\rm DM}(z_{c})}{\partial a_{lm}} &=
-\frac{1+z}{c}\left.\frac{\partial{\rm DM}(z)}{\partial z}\right\vert_{z=z_{c}}Y_{lm}^{*}\\
\frac{\partial^{2}{\rm DM}(z_{c})}{\partial a_{lm}\partial a_{l^{\prime}m^{\prime}}} &=
\left (\frac{1+z}{c}\right )^{2}\left.\frac{\partial^{2}{\rm DM}(z)}{\partial z^{2}}\right\vert_{z=z_{c}}Y_{lm}^{*}
Y_{l^{\prime}m^{\prime}}^{*},
\end{split}
\end{equation}
where the $Y_{lm}$ are evaluated at the position of the galaxy in question. In the following, we shall give some details
regarding the calculation for the two models of $\phi(M)$ adopted in our analysis.

\subsection{Schechter form}
\label{app1a}
Assuming a Schechter LF, we start from \cite{schechter}
\begin{equation}
\phi (M) = 0.4\log{(10)}\phi^{\star}10^{0.4(1+\alpha^{\star})(M^{\star}-M)}\exp{\left (-10^{0.4(M^{\star}-M)}\right )},
\end{equation}
with the usual Schechter parameters given as $M^{\star}$, $\alpha^{\star}$, and $\phi^{\star}$. The only non-trivial derivatives
appearing in eq. \eqref{eq:app1d} are those of $\eta\left (M^{+},M^{-}\right )$ with respect to $\alpha^{\star}$. Since eq.
\eqref{eq:2b2} can be written in terms of the incomplete gamma function, i.e.
\begin{equation}
\eta\left (M^{+},M^{-}\right ) = \phi^{\star}\left\lbrack\Gamma\left (1+\alpha^{\star},\tilde{L}^{-}\right ) - \Gamma\left (1+
\alpha^{\star},\tilde{L}^{+}\right )\right\rbrack,
\end{equation}
the corresponding expressions can be obtained with the help of
\begin{equation}
\begin{split}
\frac{\partial\Gamma\left (1+\alpha^{\star},\tilde{L}\right )}{\partial\alpha^{\star}} =
&\left.\Gamma\left (1+\alpha^{\star},\tilde{L}\right )\log\tilde{L}\right. + \mathcal{A}\left (\alpha^{\star},\tilde{L}\right ),\\
\frac{\partial^{2}\Gamma\left (1+\alpha^{\star},\tilde{L}\right )}{\partial\alpha^{\star 2}} =
&\left\lbrack\Gamma\left (1+\alpha^{\star},\tilde{L}\right )\log\tilde{L} + 2\mathcal{A}\left (\alpha^{\star},\tilde{L}\right )
\right\rbrack\log\tilde{L} + \mathcal{B}\left (\alpha^{\star},\tilde{L}\right ),
\end{split}
\end{equation}
where $\tilde{L}=L/L^{\star}=10^{0.4(M^{\star}-M)}$,
\begin{equation}
\begin{split}
\mathcal{A} &= \frac{\psi^{(0)}(1+\alpha^{\star})-\log\tilde{L}}{\Gamma (-
\alpha^{\star})\sin\left\lbrack\pi(1+\alpha^{\star})\right\rbrack}\pi + \tilde{L}^{1+\alpha^{\star}}\sum\limits_{n=0}^{\infty}\frac{(-1)^{n}
\tilde{L}^{n}}{n!(1+\alpha^{\star}+n)^{2}},\\
\mathcal{B} &= \frac{\psi^{(1)}(1+\alpha^{\star})+\left (\psi^{(0)}(1+\alpha^{\star})-\log\tilde{L}\right )^{2}}{\Gamma (-
\alpha^{\star})\sin\left\lbrack\pi(1+\alpha^{\star})\right\rbrack}\pi - 2\tilde{L}^{1+\alpha^{\star}}\sum\limits_{n=0}^{\infty}\frac{(-1)^{n}
\tilde{L}^{n}}{n!(1+\alpha^{\star}+n)^{3}},
\end{split}
\end{equation}
and $\psi^{(s)}$ denotes the polygamma function of degree $s$. Note that the above relations remain strictly valid only as long
as $\tilde{L}<1$ and $1+\alpha^{\star}$ is neither zero nor a negative integer.

\subsection{Cubic spline}
\label{app1b}
Choosing an equidistant set of sampling points $(M_{i},\phi_{i})$ with $0\leq i<N$, the LF may be modeled as a natural cubic
spline with piecewise definition \cite{Press2002}
\begin{equation}
\phi(M) = (1-t_{i})\phi_{i-1} + t_{i}\phi_{i} + t_{i}(1-t_{i})\left\lbrack a_{i}(1-t_{i}) + b_{i}t_{i}\right\rbrack,\qquad M_{i-1}\leq M<M_{i},
\label{eq:cs}
\end{equation}
where $i$ now runs from $1$ to $N-1$, $t_{i}=(M-M_{i})/\Delta M$ and
\begin{equation}
a_{i} = k_{i-1}\Delta M - (\phi_{i} - \phi_{i-1}),\qquad
b_{i} = -k_{i}\Delta M + (\phi_{i} - \phi_{i-1}).
\end{equation}
The $k_{i}$ can be written in terms of an appropriate tridiagonal matrix $A_{ij}$ whose inverse solves the spline problem, and
introducing the Kronecker delta $\delta^{K}_{i,j}$, one has
\begin{equation}
k_{i} = \frac{3}{\Delta M^{2}}\sum\limits_{j}A_{ij}^{-1}\left\lbrack\left (1-\delta^{K}_{0,j}\right )(\phi_{j}-\phi_{j-1}) +
\left (1-\delta^{K}_{N-1,j}\right )(\phi_{j+1}-\phi_{j})\right\rbrack.
\end{equation}
Since we are dealing with polynomials, evaluating the expansion in \eqref{eq:app1d} only requires basic calculus. To obtain
the derivatives of $a_{i}$ and $b_{i}$, for instance, one easily verifies that
\begin{equation}
\frac{\partial k_{i}}{\partial\phi_{j}} = \frac{3}{\Delta M^{2}}\left\lbrack\left (1-\delta^{K}_{0,j}\right )A^{-1}_{i,j-1}+\left (
\delta^{K}_{N-1,j}-\delta^{K}_{0,j}\right )A^{-1}_{i,j}-\left (1-\delta^{K}_{N-1,j}\right )A^{-1}_{i,j+1}\right\rbrack.
\end{equation}
Similarly, piecewise integration of eq. \eqref{eq:cs} yields
\begin{equation}
\begin{split}
\frac{1}{\Delta M}\int\phi(M^{\prime})\dd M^{\prime} = &\left (t_{i}-\frac{1}{2}t_{i}^{2}\right )\phi_{i-1} + \frac{1}{2}t_{i}^{2}\phi_{i}
+ a_{i}\left (\frac{1}{2}-\frac{2}{3}t_{i}+\frac{1}{4}t_{i}^{2}\right )t_{i}^{2}\\
&\left. +\right. b_{i}\left (\frac{1}{3}-\frac{1}{4}t_{i}\right )t_{i}^{3},\qquad M_{i-1}\leq M^{\prime}<M_{i},
\end{split}
\end{equation}
which may be used to compute certain derivatives of $\eta\left (M^{+},M^{-}\right )$. Note that because of the spline equation's
linearity, all second derivatives with respect to the spline parameters $\phi_{j}$ trivially vanish.

\section{Constructing the probability \boldmath{$P(C_{l})$} for Gaussian fields}
\label{app3}
Starting from Bayes' theorem, we write the posterior probability as
\begin{equation}
P(a_{lm}\vert\vd )\propto P\left (\vd\vert a_{lm}\right )P\left (a_{lm}\vert C_{l}\right ).
\label{eq:app3a}
\end{equation}
Assuming that the probabilities $P(\vd\vert a_{lm})$ and $P(a_{lm}\vert C_{l})$ correspond to Gaussian distributions
$\mathcal{N}(\va,\bm{\Sigma}_{M})$ and $\mathcal{N}(0,\mathbf{D})$, respectively, evaluating the integral in eq. \eqref{eq:2c4}
yields
\begin{equation}
\log{P(C_{l})} = \log{P\left (\vd\vert C_{l}\right )} = -\frac{1}{2}\left\lbrack\va^{\rm T}\mathbf{Q}^{-1}\va
+ \tr{\left (\log\mathbf{Q}\right )}\right\rbrack + {\rm const},
\label{eq:app3b}
\end{equation}
where $\bm{\Sigma}_{M}$ denotes the marginal covariance matrix constructed from the original one (i.e. the full matrix $\bm{\Sigma}$
which has been introduced in section \ref{section2c}), $\mathbf{D}$ is a diagonal matrix which depends on the $C_{l}$, and
$\mathbf{Q}=\bm{\Sigma}_{M}+\mathbf{D}$ is assumed to be non-singular. Analog to the analysis of CMB anisotropies, the power
spectrum is now estimated by maximizing the probability in eq. \eqref{eq:app3b} with respect to the $C_{l}$. To obtain an error
for this estimate, we simply expand the logarithm of $P(C_{l})$ around the maximum (assumed as well-defined) to quadratic order
and compute the inverse of the corresponding curvature matrix. One may easily verify that the required derivatives with respect
to $C_{l}$ are explicitly given by
\begin{equation}
\frac{\partial\log{P(C_{l})}}{\partial C_{l}} = \frac{1}{2}\left\lbrack\va^{\rm T}\mathbf{Q}^{-1}
\frac{\partial\mathbf{D}}{\partial C_{l}}\mathbf{Q}^{-1}\va - \tr{\left (\mathbf{Q}^{-1}
\frac{\partial\mathbf{D}}{\partial C_{l}}\right )}\right\rbrack
\label{eq:app3c}
\end{equation}
and
\begin{equation}
\frac{\partial^{2}\log{P}}{\partial C_{l}\partial C_{l^{\prime}}} = -\va^{\rm T}\mathbf{Q}^{-1}
\frac{\partial\mathbf{D}}{\partial C_{l}}\mathbf{Q}^{-1}\frac{\partial\mathbf{D}}{\partial C_{l^{\prime}}}\mathbf{Q}^{-1}
\va + \frac{1}{2}\tr{\left (\mathbf{Q}^{-1}\frac{\partial\mathbf{D}}{\partial C_{l}}\mathbf{Q}^{-1}
\frac{\partial\mathbf{D}}{\partial C_{l^{\prime}}}\right )}.
\label{eq:app3d}
\end{equation}
Using that $\langle\va\va^{\rm T}\rangle = \mathbf{Q}$, it further follows that 
\begin{equation}
\left\langle\frac{\partial^{2}\log{P}}{\partial C_{l}\partial C_{l^{\prime}}}\right\rangle =
-\frac{1}{2}\tr{\left (\mathbf{Q}^{-1}\frac{\partial\mathbf{D}}{\partial C_{l}}\mathbf{Q}^{-1}
\frac{\partial\mathbf{D}}{\partial C_{l^{\prime}}}\right )}.
\end{equation}

\section{Theoretical \boldmath{$C_{l}$} for the \boldmath{$\Lambda$}CDM cosmology}
\label{app2}

In what follows, we will present predictions for the angular power spectrum of the peculiar velocity field introduced in section \ref{section2}
for the standard $\Lambda$CDM cosmology. Since the observed galaxies are divided into redshift bins, we consider the averaged velocity field
\begin{equation}
\tilde{V}(\hvr ) = \int_{r_{1}}^{r_{2}}V\left\lbrack\hvr r, t(r)\right\rbrack p(r)\dd r.
\label{eq:app2a}
\end{equation}
Here $r_{1}$ and $r_{2}$ are the comoving distances at the limiting redshifts of a given bin, and $p(r)dr$ is the probability of observing
a galaxy within the interval $[r,r+dr]$. We thus begin with
\begin{equation}
a_{lm} = \int\dd\Omega\tilde{V}(\hvr )Y_{lm}(\hvr ) = -\int\dd\Omega Y_{lm}(\hvr )\int_{r_{1}}^{r_{2}}W(r)\frac{\partial\Phi_{0}(\hvr r)}{\partial r}\dd r ,
\label{eq:app2b}
\end{equation}
where we have used $\tilde{V}(\hvr )=\sum a_{lm}Y_{lm}^{*}(\hvr )$, the definition $W(r) = 2a\dot{D}p(r)/3\Omega_{0}H_{0}^{2}$, and the
linear relation
\begin{equation}
V(\vr ,t) = -\frac{2}{3}\frac{a\dot{D}(t)}{\Omega_{0}H_{0}^{2}}\frac{\partial\Phi_{0}}{\partial r},
\label{eq:app2c}
\end{equation}
with $D(t)$ and $a(t)$ evaluated at $t=t(r)$. Expanding $\Phi_0(\vr )$ in Fourier space, i.e.
\begin{equation}
\Phi_0(\vr) = \frac{1}{(2\pi)^3}\int\dd^{3}k \Phi_{\vk}{\rm e}^{{\rm i}\vk\cdot{\vr}},
\label{eq:app2d}
\end{equation}
and exploiting the plane wave expansion
\begin{equation}
{\rm e}^{{\rm i}\vk \cdot{\vr}} = 4\pi\sum\limits_{l,m}{\rm i}^lj_l(kr) Y^{*}_{lm}(\hat{\vr})Y_{lm}(\hat{\vk}),
\label{eq:app2e}
\end{equation}
where $j_{l}$ are the usual spherical Bessel functions of the first kind, we get 
\begin{equation}
a_{lm} = -\frac{{\rm i}^l}{2\pi^{2}}\int_{r_{1}}^{r_{2}}\dd rW(r)\int\dd^{3}k\Phi_{\vk}\left (\frac{lj_{l}}{r}-kj_{l+1}\right )Y_{lm}(\hat{\vk}). 
\label{eq:app2f}
\end{equation}
Therefore, using that $\langle\Phi_{\vk}\Phi_{\vk^{\prime}}\rangle = (2\pi)^{3}\delta_{D}(\vk-\vk^{\prime})P_{\Phi}(k)$, we finally arrive
at
\begin{equation}
C_{l} = \left\langle\lvert a _{lm}\rvert^{2}\right\rangle =
\frac{2}{\pi}\int\dd kk^{2}P_{\Phi}(k)\left\lvert\int_{r_{1}}^{r_{2}}\dd rW(r)\left (\frac{lj_{l}}{r}-kj_{l+1}\right )\right\rvert^{2}.
\label{eq:app2g}
\end{equation}

\twocolumngrid

\bibliography{bulk_ref.bib}

%merlin.mbs 2010-03-15 4.21a (PWD, AO, DPC)
\begin{thebibliography}{80}%
\makeatletter
\providecommand \@ifxundefined [1]{%
 \@ifx{#1\undefined}
}%
\providecommand \@ifnum [1]{%
 \ifnum #1\expandafter \@firstoftwo
 \else \expandafter \@secondoftwo
 \fi
}%
\providecommand \@ifx [1]{%
 \ifx #1\expandafter \@firstoftwo
 \else \expandafter \@secondoftwo
 \fi
}%
\providecommand \natexlab [1]{#1}%
\providecommand \enquote  [1]{``#1''}%
\providecommand \bibnamefont  [1]{#1}%
\providecommand \bibfnamefont [1]{#1}%
\providecommand \citenamefont [1]{#1}%
\providecommand \href@noop [0]{\@secondoftwo}%
\providecommand \href [0]{\begingroup \@sanitize@url \@href}%
\providecommand \@href[1]{\@@startlink{#1}\@@href}%
\providecommand \@@href[1]{\endgroup#1\@@endlink}%
\providecommand \@sanitize@url [0]{\catcode `\\12\catcode `\$12\catcode
  `\&12\catcode `\#12\catcode `\^12\catcode `\_12\catcode `\%12\relax}%
\providecommand \@@startlink[1]{}%
\providecommand \@@endlink[0]{}%
\providecommand \url  [0]{\begingroup\@sanitize@url \@url }%
\providecommand \@url [1]{\endgroup\@href {#1}{\urlprefix }}%
\providecommand \urlprefix  [0]{URL }%
\providecommand \Eprint [0]{\href }%
\@ifxundefined \urlstyle {%
  \providecommand \doi  [0]{\begingroup \@sanitize@url \@doi}%
  \providecommand \@doi [1]{\endgroup \@@startlink {\doibase
  #1}doi:\discretionary {}{}{}#1\@@endlink }%
}{%
  \providecommand \doi  [0]{doi:\discretionary{}{}{}\begingroup
  \urlstyle{rm}\Url }%
}%
\providecommand \doibase [0]{http://dx.doi.org/}%
\providecommand \Doi [0]{\begingroup \@sanitize@url \@Doi }%
\providecommand \@Doi  [1]{\endgroup\@@startlink{\doibase#1}\@@Doi}%
\providecommand \@@Doi [1]{#1\@@endlink}%
\providecommand \selectlanguage [0]{\@gobble}%
\providecommand \bibinfo  [0]{\@secondoftwo}%
\providecommand \bibfield  [0]{\@secondoftwo}%
\providecommand \translation [1]{[#1]}%
\providecommand \BibitemOpen [0]{}%
\providecommand \bibitemStop [0]{}%
\providecommand \bibitemNoStop [0]{.\EOS\space}%
\providecommand \EOS [0]{\spacefactor3000\relax}%
\providecommand \BibitemShut  [1]{\csname bibitem#1\endcsname}%
%</preamble>
\bibitem [{\citenamefont {{Percival}}\ \emph {et~al.}(2010)\citenamefont
  {{Percival}}, \citenamefont {{Reid}}, \citenamefont {{Eisenstein}},
  \citenamefont {{Bahcall}}, \citenamefont {{Budavari}}, \citenamefont
  {{Frieman}}, \citenamefont {{Fukugita}}, \citenamefont {{Gunn}},
  \citenamefont {{Ivezi{\'c}}}, \citenamefont {{Knapp}},\ and\ \citenamefont
  {et~al.}}]{Percival2010}%
  \BibitemOpen
  \bibfield  {author} {\bibinfo {author} {\bibfnamefont {W.~J.}\ \bibnamefont
  {{Percival}}}, \bibinfo {author} {\bibfnamefont {B.~A.}\ \bibnamefont
  {{Reid}}}, \bibinfo {author} {\bibfnamefont {D.~J.}\ \bibnamefont
  {{Eisenstein}}}, \bibinfo {author} {\bibfnamefont {N.~A.}\ \bibnamefont
  {{Bahcall}}}, \bibinfo {author} {\bibfnamefont {T.}~\bibnamefont
  {{Budavari}}}, \bibinfo {author} {\bibfnamefont {J.~A.}\ \bibnamefont
  {{Frieman}}}, \bibinfo {author} {\bibfnamefont {M.}~\bibnamefont
  {{Fukugita}}}, \bibinfo {author} {\bibfnamefont {J.~E.}\ \bibnamefont
  {{Gunn}}}, \bibinfo {author} {\bibfnamefont {{\v Z}.}~\bibnamefont
  {{Ivezi{\'c}}}}, \bibinfo {author} {\bibfnamefont {G.~R.}\ \bibnamefont
  {{Knapp}}}, \ and\ \bibinfo {author} {\bibnamefont {et~al.}},\ }\Doi
  {10.1111/j.1365-2966.2009.15812.x} {\bibfield  {journal} {\bibinfo  {journal}
  {\mnras},\ }\textbf {\bibinfo {volume} {401}},\ \bibinfo {pages} {2148}
  (\bibinfo {year} {2010})},\ \Eprint {http://arxiv.org/abs/0907.1660}
  {arXiv:0907.1660 [astro-ph.CO]} \BibitemShut {NoStop}%
\bibitem [{\citenamefont {{Riess}}\ \emph {et~al.}(2011)\citenamefont
  {{Riess}}, \citenamefont {{Macri}}, \citenamefont {{Casertano}},
  \citenamefont {{Lampeitl}}, \citenamefont {{Ferguson}}, \citenamefont
  {{Filippenko}}, \citenamefont {{Jha}}, \citenamefont {{Li}},\ and\
  \citenamefont {{Chornock}}}]{Riess2011}%
  \BibitemOpen
  \bibfield  {author} {\bibinfo {author} {\bibfnamefont {A.~G.}\ \bibnamefont
  {{Riess}}}, \bibinfo {author} {\bibfnamefont {L.}~\bibnamefont {{Macri}}},
  \bibinfo {author} {\bibfnamefont {S.}~\bibnamefont {{Casertano}}}, \bibinfo
  {author} {\bibfnamefont {H.}~\bibnamefont {{Lampeitl}}}, \bibinfo {author}
  {\bibfnamefont {H.~C.}\ \bibnamefont {{Ferguson}}}, \bibinfo {author}
  {\bibfnamefont {A.~V.}\ \bibnamefont {{Filippenko}}}, \bibinfo {author}
  {\bibfnamefont {S.~W.}\ \bibnamefont {{Jha}}}, \bibinfo {author}
  {\bibfnamefont {W.}~\bibnamefont {{Li}}}, \ and\ \bibinfo {author}
  {\bibfnamefont {R.}~\bibnamefont {{Chornock}}},\ }\Doi
  {10.1088/0004-637X/730/2/119} {\bibfield  {journal} {\bibinfo  {journal}
  {\apj},\ }\textbf {\bibinfo {volume} {730}},\ \bibinfo {eid} {119} (\bibinfo
  {year} {2011})},\ \Eprint {http://arxiv.org/abs/1103.2976} {arXiv:1103.2976
  [astro-ph.CO]} \BibitemShut {NoStop}%
\bibitem [{\citenamefont {{Hinshaw}}\ \emph {et~al.}(2013)\citenamefont
  {{Hinshaw}}, \citenamefont {{Larson}}, \citenamefont {{Komatsu}},
  \citenamefont {{Spergel}}, \citenamefont {{Bennett}}, \citenamefont
  {{Dunkley}}, \citenamefont {{Nolta}}, \citenamefont {{Halpern}},
  \citenamefont {{Hill}}, \citenamefont {{Odegard}},\ and\ \citenamefont
  {et~al.}}]{Hinshaw2013}%
  \BibitemOpen
  \bibfield  {author} {\bibinfo {author} {\bibfnamefont {G.}~\bibnamefont
  {{Hinshaw}}}, \bibinfo {author} {\bibfnamefont {D.}~\bibnamefont {{Larson}}},
  \bibinfo {author} {\bibfnamefont {E.}~\bibnamefont {{Komatsu}}}, \bibinfo
  {author} {\bibfnamefont {D.~N.}\ \bibnamefont {{Spergel}}}, \bibinfo {author}
  {\bibfnamefont {C.~L.}\ \bibnamefont {{Bennett}}}, \bibinfo {author}
  {\bibfnamefont {J.}~\bibnamefont {{Dunkley}}}, \bibinfo {author}
  {\bibfnamefont {M.~R.}\ \bibnamefont {{Nolta}}}, \bibinfo {author}
  {\bibfnamefont {M.}~\bibnamefont {{Halpern}}}, \bibinfo {author}
  {\bibfnamefont {R.~S.}\ \bibnamefont {{Hill}}}, \bibinfo {author}
  {\bibfnamefont {N.}~\bibnamefont {{Odegard}}}, \ and\ \bibinfo {author}
  {\bibnamefont {et~al.}},\ }\Doi {10.1088/0067-0049/208/2/19} {\bibfield
  {journal} {\bibinfo  {journal} {\apjs},\ }\textbf {\bibinfo {volume} {208}},\
  \bibinfo {eid} {19} (\bibinfo {year} {2013})},\ \Eprint
  {http://arxiv.org/abs/1212.5226} {arXiv:1212.5226 [astro-ph.CO]} \BibitemShut
  {NoStop}%
\bibitem [{\citenamefont {{Planck Collaboration}}\ \emph
  {et~al.}(2013)\citenamefont {{Planck Collaboration}}, \citenamefont {{Ade}},
  \citenamefont {{Aghanim}}, \citenamefont {{Armitage-Caplan}}, \citenamefont
  {{Arnaud}}, \citenamefont {{Ashdown}}, \citenamefont {{Atrio-Barandela}},
  \citenamefont {{Aumont}}, \citenamefont {{Baccigalupi}}, \citenamefont
  {{Banday}},\ and\ \citenamefont {et~al.}}]{Planck2013}%
  \BibitemOpen
  \bibfield  {author} {\bibinfo {author} {\bibnamefont {{Planck
  Collaboration}}}, \bibinfo {author} {\bibfnamefont {P.~A.~R.}\ \bibnamefont
  {{Ade}}}, \bibinfo {author} {\bibfnamefont {N.}~\bibnamefont {{Aghanim}}},
  \bibinfo {author} {\bibfnamefont {C.}~\bibnamefont {{Armitage-Caplan}}},
  \bibinfo {author} {\bibfnamefont {M.}~\bibnamefont {{Arnaud}}}, \bibinfo
  {author} {\bibfnamefont {M.}~\bibnamefont {{Ashdown}}}, \bibinfo {author}
  {\bibfnamefont {F.}~\bibnamefont {{Atrio-Barandela}}}, \bibinfo {author}
  {\bibfnamefont {J.}~\bibnamefont {{Aumont}}}, \bibinfo {author}
  {\bibfnamefont {C.}~\bibnamefont {{Baccigalupi}}}, \bibinfo {author}
  {\bibfnamefont {A.~J.}\ \bibnamefont {{Banday}}}, \ and\ \bibinfo {author}
  {\bibnamefont {et~al.}},\ }\href@noop {} {\bibfield  {journal} {\bibinfo
  {journal} {ArXiv e-prints}} (\bibinfo {year} {2013})},\ \Eprint
  {http://arxiv.org/abs/1303.5076} {arXiv:1303.5076 [astro-ph.CO]} \BibitemShut
  {NoStop}%
\bibitem [{\citenamefont {{Riess}}\ \emph {et~al.}(1995)\citenamefont
  {{Riess}}, \citenamefont {{Press}},\ and\ \citenamefont
  {{Kirshner}}}]{Riess1995}%
  \BibitemOpen
  \bibfield  {author} {\bibinfo {author} {\bibfnamefont {A.~G.}\ \bibnamefont
  {{Riess}}}, \bibinfo {author} {\bibfnamefont {W.~H.}\ \bibnamefont
  {{Press}}}, \ and\ \bibinfo {author} {\bibfnamefont {R.~P.}\ \bibnamefont
  {{Kirshner}}},\ }\Doi {10.1086/187897} {\bibfield  {journal} {\bibinfo
  {journal} {\apjl},\ }\textbf {\bibinfo {volume} {445}},\ \bibinfo {pages}
  {L91} (\bibinfo {year} {1995})},\ \Eprint
  {http://arxiv.org/abs/astro-ph/9412017} {astro-ph/9412017} \BibitemShut
  {NoStop}%
\bibitem [{\citenamefont {{Dekel}}\ \emph {et~al.}(1999)\citenamefont
  {{Dekel}}, \citenamefont {{Eldar}}, \citenamefont {{Kolatt}}, \citenamefont
  {{Yahil}}, \citenamefont {{Willick}}, \citenamefont {{Faber}}, \citenamefont
  {{Courteau}},\ and\ \citenamefont {{Burstein}}}]{Dekel1999}%
  \BibitemOpen
  \bibfield  {author} {\bibinfo {author} {\bibfnamefont {A.}~\bibnamefont
  {{Dekel}}}, \bibinfo {author} {\bibfnamefont {A.}~\bibnamefont {{Eldar}}},
  \bibinfo {author} {\bibfnamefont {T.}~\bibnamefont {{Kolatt}}}, \bibinfo
  {author} {\bibfnamefont {A.}~\bibnamefont {{Yahil}}}, \bibinfo {author}
  {\bibfnamefont {J.~A.}\ \bibnamefont {{Willick}}}, \bibinfo {author}
  {\bibfnamefont {S.~M.}\ \bibnamefont {{Faber}}}, \bibinfo {author}
  {\bibfnamefont {S.}~\bibnamefont {{Courteau}}}, \ and\ \bibinfo {author}
  {\bibfnamefont {D.}~\bibnamefont {{Burstein}}},\ }\Doi {10.1086/307636}
  {\bibfield  {journal} {\bibinfo  {journal} {\apj},\ }\textbf {\bibinfo
  {volume} {522}},\ \bibinfo {pages} {1} (\bibinfo {year} {1999})},\ \Eprint
  {http://arxiv.org/abs/astro-ph/9812197} {astro-ph/9812197} \BibitemShut
  {NoStop}%
\bibitem [{\citenamefont {{Zaroubi}}\ \emph {et~al.}(2001)\citenamefont
  {{Zaroubi}}, \citenamefont {{Bernardi}}, \citenamefont {{da Costa}},
  \citenamefont {{Hoffman}}, \citenamefont {{Alonso}}, \citenamefont
  {{Wegner}}, \citenamefont {{Willmer}},\ and\ \citenamefont
  {{Pellegrini}}}]{Zaroubi2001}%
  \BibitemOpen
  \bibfield  {author} {\bibinfo {author} {\bibfnamefont {S.}~\bibnamefont
  {{Zaroubi}}}, \bibinfo {author} {\bibfnamefont {M.}~\bibnamefont
  {{Bernardi}}}, \bibinfo {author} {\bibfnamefont {L.~N.}\ \bibnamefont {{da
  Costa}}}, \bibinfo {author} {\bibfnamefont {Y.}~\bibnamefont {{Hoffman}}},
  \bibinfo {author} {\bibfnamefont {M.~V.}\ \bibnamefont {{Alonso}}}, \bibinfo
  {author} {\bibfnamefont {G.}~\bibnamefont {{Wegner}}}, \bibinfo {author}
  {\bibfnamefont {C.~N.~A.}\ \bibnamefont {{Willmer}}}, \ and\ \bibinfo
  {author} {\bibfnamefont {P.~S.}\ \bibnamefont {{Pellegrini}}},\ }\Doi
  {10.1046/j.1365-8711.2001.04605.x} {\bibfield  {journal} {\bibinfo  {journal}
  {\mnras},\ }\textbf {\bibinfo {volume} {326}},\ \bibinfo {pages} {375}
  (\bibinfo {year} {2001})},\ \Eprint {http://arxiv.org/abs/astro-ph/0005558}
  {astro-ph/0005558} \BibitemShut {NoStop}%
\bibitem [{\citenamefont {{Hudson}}\ \emph {et~al.}(2004)\citenamefont
  {{Hudson}}, \citenamefont {{Smith}}, \citenamefont {{Lucey}},\ and\
  \citenamefont {{Branchini}}}]{Hudson2004}%
  \BibitemOpen
  \bibfield  {author} {\bibinfo {author} {\bibfnamefont {M.~J.}\ \bibnamefont
  {{Hudson}}}, \bibinfo {author} {\bibfnamefont {R.~J.}\ \bibnamefont
  {{Smith}}}, \bibinfo {author} {\bibfnamefont {J.~R.}\ \bibnamefont
  {{Lucey}}}, \ and\ \bibinfo {author} {\bibfnamefont {E.}~\bibnamefont
  {{Branchini}}},\ }\Doi {10.1111/j.1365-2966.2004.07893.x} {\bibfield
  {journal} {\bibinfo  {journal} {\mnras},\ }\textbf {\bibinfo {volume}
  {352}},\ \bibinfo {pages} {61} (\bibinfo {year} {2004})},\ \Eprint
  {http://arxiv.org/abs/astro-ph/0404386} {astro-ph/0404386} \BibitemShut
  {NoStop}%
\bibitem [{\citenamefont {{Sarkar}}\ \emph {et~al.}(2007)\citenamefont
  {{Sarkar}}, \citenamefont {{Feldman}},\ and\ \citenamefont
  {{Watkins}}}]{Sarkar2007}%
  \BibitemOpen
  \bibfield  {author} {\bibinfo {author} {\bibfnamefont {D.}~\bibnamefont
  {{Sarkar}}}, \bibinfo {author} {\bibfnamefont {H.~A.}\ \bibnamefont
  {{Feldman}}}, \ and\ \bibinfo {author} {\bibfnamefont {R.}~\bibnamefont
  {{Watkins}}},\ }\Doi {10.1111/j.1365-2966.2006.11334.x} {\bibfield  {journal}
  {\bibinfo  {journal} {\mnras},\ }\textbf {\bibinfo {volume} {375}},\ \bibinfo
  {pages} {691} (\bibinfo {year} {2007})},\ \Eprint
  {http://arxiv.org/abs/astro-ph/0607426} {astro-ph/0607426} \BibitemShut
  {NoStop}%
\bibitem [{\citenamefont {{Lavaux}}\ \emph {et~al.}(2010)\citenamefont
  {{Lavaux}}, \citenamefont {{Tully}}, \citenamefont {{Mohayaee}},\ and\
  \citenamefont {{Colombi}}}]{Lavaux2010}%
  \BibitemOpen
  \bibfield  {author} {\bibinfo {author} {\bibfnamefont {G.}~\bibnamefont
  {{Lavaux}}}, \bibinfo {author} {\bibfnamefont {R.~B.}\ \bibnamefont
  {{Tully}}}, \bibinfo {author} {\bibfnamefont {R.}~\bibnamefont {{Mohayaee}}},
  \ and\ \bibinfo {author} {\bibfnamefont {S.}~\bibnamefont {{Colombi}}},\
  }\Doi {10.1088/0004-637X/709/1/483} {\bibfield  {journal} {\bibinfo
  {journal} {\apj},\ }\textbf {\bibinfo {volume} {709}},\ \bibinfo {pages}
  {483} (\bibinfo {year} {2010})},\ \Eprint {http://arxiv.org/abs/0810.3658}
  {arXiv:0810.3658} \BibitemShut {NoStop}%
\bibitem [{\citenamefont {{Feldman}}\ \emph {et~al.}(2010)\citenamefont
  {{Feldman}}, \citenamefont {{Watkins}},\ and\ \citenamefont
  {{Hudson}}}]{feldwh10}%
  \BibitemOpen
  \bibfield  {author} {\bibinfo {author} {\bibfnamefont {H.~A.}\ \bibnamefont
  {{Feldman}}}, \bibinfo {author} {\bibfnamefont {R.}~\bibnamefont
  {{Watkins}}}, \ and\ \bibinfo {author} {\bibfnamefont {M.~J.}\ \bibnamefont
  {{Hudson}}},\ }\Doi {10.1111/j.1365-2966.2010.17052.x} {\bibfield  {journal}
  {\bibinfo  {journal} {\mnras},\ }\textbf {\bibinfo {volume} {407}},\ \bibinfo
  {pages} {2328} (\bibinfo {year} {2010})},\ \Eprint
  {http://arxiv.org/abs/0911.5516} {arXiv:0911.5516 [astro-ph.CO]} \BibitemShut
  {NoStop}%
\bibitem [{\citenamefont {{Nusser}}\ and\ \citenamefont
  {{Davis}}(2011)}]{ND11}%
  \BibitemOpen
  \bibfield  {author} {\bibinfo {author} {\bibfnamefont {A.}~\bibnamefont
  {{Nusser}}}\ and\ \bibinfo {author} {\bibfnamefont {M.}~\bibnamefont
  {{Davis}}},\ }\Doi {10.1088/0004-637X/736/2/93} {\bibfield  {journal}
  {\bibinfo  {journal} {\apj},\ }\textbf {\bibinfo {volume} {736}},\ \bibinfo
  {eid} {93} (\bibinfo {year} {2011})},\ \Eprint
  {http://arxiv.org/abs/1101.1650} {arXiv:1101.1650 [astro-ph.CO]} \BibitemShut
  {NoStop}%
\bibitem [{\citenamefont {{Turnbull}}\ \emph {et~al.}(2012)\citenamefont
  {{Turnbull}}, \citenamefont {{Hudson}}, \citenamefont {{Feldman}},
  \citenamefont {{Hicken}}, \citenamefont {{Kirshner}},\ and\ \citenamefont
  {{Watkins}}}]{Turnbull2012}%
  \BibitemOpen
  \bibfield  {author} {\bibinfo {author} {\bibfnamefont {S.~J.}\ \bibnamefont
  {{Turnbull}}}, \bibinfo {author} {\bibfnamefont {M.~J.}\ \bibnamefont
  {{Hudson}}}, \bibinfo {author} {\bibfnamefont {H.~A.}\ \bibnamefont
  {{Feldman}}}, \bibinfo {author} {\bibfnamefont {M.}~\bibnamefont {{Hicken}}},
  \bibinfo {author} {\bibfnamefont {R.~P.}\ \bibnamefont {{Kirshner}}}, \ and\
  \bibinfo {author} {\bibfnamefont {R.}~\bibnamefont {{Watkins}}},\ }\Doi
  {10.1111/j.1365-2966.2011.20050.x} {\bibfield  {journal} {\bibinfo  {journal}
  {\mnras},\ }\textbf {\bibinfo {volume} {420}},\ \bibinfo {pages} {447}
  (\bibinfo {year} {2012})},\ \Eprint {http://arxiv.org/abs/1111.0631}
  {arXiv:1111.0631 [astro-ph.CO]} \BibitemShut {NoStop}%
\bibitem [{\citenamefont {{Feindt}}\ \emph {et~al.}(2013)\citenamefont
  {{Feindt}}, \citenamefont {{Kerschhaggl}}, \citenamefont {{Kowalski}},
  \citenamefont {{Aldering}}, \citenamefont {{Antilogus}}, \citenamefont
  {{Aragon}}, \citenamefont {{Bailey}}, \citenamefont {{Baltay}}, \citenamefont
  {{Bongard}}, \citenamefont {{Buton}},\ and\ \citenamefont
  {et~al.}}]{Feindt2013}%
  \BibitemOpen
  \bibfield  {author} {\bibinfo {author} {\bibfnamefont {U.}~\bibnamefont
  {{Feindt}}}, \bibinfo {author} {\bibfnamefont {M.}~\bibnamefont
  {{Kerschhaggl}}}, \bibinfo {author} {\bibfnamefont {M.}~\bibnamefont
  {{Kowalski}}}, \bibinfo {author} {\bibfnamefont {G.}~\bibnamefont
  {{Aldering}}}, \bibinfo {author} {\bibfnamefont {P.}~\bibnamefont
  {{Antilogus}}}, \bibinfo {author} {\bibfnamefont {C.}~\bibnamefont
  {{Aragon}}}, \bibinfo {author} {\bibfnamefont {S.}~\bibnamefont {{Bailey}}},
  \bibinfo {author} {\bibfnamefont {C.}~\bibnamefont {{Baltay}}}, \bibinfo
  {author} {\bibfnamefont {S.}~\bibnamefont {{Bongard}}}, \bibinfo {author}
  {\bibfnamefont {C.}~\bibnamefont {{Buton}}}, \ and\ \bibinfo {author}
  {\bibnamefont {et~al.}},\ }\Doi {10.1051/0004-6361/201321880} {\bibfield
  {journal} {\bibinfo  {journal} {\aap},\ }\textbf {\bibinfo {volume} {560}},\
  \bibinfo {eid} {A90} (\bibinfo {year} {2013})},\ \Eprint
  {http://arxiv.org/abs/1310.4184} {arXiv:1310.4184 [astro-ph.CO]} \BibitemShut
  {NoStop}%
\bibitem [{\citenamefont {{Tully}}\ and\ \citenamefont
  {{Fisher}}(1977)}]{TF77}%
  \BibitemOpen
  \bibfield  {author} {\bibinfo {author} {\bibfnamefont {R.~B.}\ \bibnamefont
  {{Tully}}}\ and\ \bibinfo {author} {\bibfnamefont {J.~R.}\ \bibnamefont
  {{Fisher}}},\ }\href@noop {} {\bibfield  {journal} {\bibinfo  {journal}
  {\aap},\ }\textbf {\bibinfo {volume} {54}},\ \bibinfo {pages} {661} (\bibinfo
  {year} {1977})}\BibitemShut {NoStop}%
\bibitem [{\citenamefont {{Lynden-Bell}}\ \emph {et~al.}(1988)\citenamefont
  {{Lynden-Bell}}, \citenamefont {{Faber}}, \citenamefont {{Burstein}},
  \citenamefont {{Davies}}, \citenamefont {{Dressler}}, \citenamefont
  {{Terlevich}},\ and\ \citenamefont {{Wegner}}}]{lyn88}%
  \BibitemOpen
  \bibfield  {author} {\bibinfo {author} {\bibfnamefont {D.}~\bibnamefont
  {{Lynden-Bell}}}, \bibinfo {author} {\bibfnamefont {S.~M.}\ \bibnamefont
  {{Faber}}}, \bibinfo {author} {\bibfnamefont {D.}~\bibnamefont {{Burstein}}},
  \bibinfo {author} {\bibfnamefont {R.~L.}\ \bibnamefont {{Davies}}}, \bibinfo
  {author} {\bibfnamefont {A.}~\bibnamefont {{Dressler}}}, \bibinfo {author}
  {\bibfnamefont {R.~J.}\ \bibnamefont {{Terlevich}}}, \ and\ \bibinfo {author}
  {\bibfnamefont {G.}~\bibnamefont {{Wegner}}},\ }\Doi {10.1086/166066}
  {\bibfield  {journal} {\bibinfo  {journal} {\apj},\ }\textbf {\bibinfo
  {volume} {326}},\ \bibinfo {pages} {19} (\bibinfo {year} {1988})}\BibitemShut
  {NoStop}%
\bibitem [{\citenamefont {{Strauss}}\ and\ \citenamefont
  {{Willick}}(1995)}]{Strauss1995}%
  \BibitemOpen
  \bibfield  {author} {\bibinfo {author} {\bibfnamefont {M.~A.}\ \bibnamefont
  {{Strauss}}}\ and\ \bibinfo {author} {\bibfnamefont {J.~A.}\ \bibnamefont
  {{Willick}}},\ }\Doi {10.1016/0370-1573(95)00013-7} {\bibfield  {journal}
  {\bibinfo  {journal} {\physrep},\ }\textbf {\bibinfo {volume} {261}},\
  \bibinfo {pages} {271} (\bibinfo {year} {1995})},\ \Eprint
  {http://arxiv.org/abs/astro-ph/9502079} {astro-ph/9502079} \BibitemShut
  {NoStop}%
\bibitem [{\citenamefont {{Haehnelt}}\ and\ \citenamefont
  {{Tegmark}}(1996)}]{Haehnelt1996}%
  \BibitemOpen
  \bibfield  {author} {\bibinfo {author} {\bibfnamefont {M.~G.}\ \bibnamefont
  {{Haehnelt}}}\ and\ \bibinfo {author} {\bibfnamefont {M.}~\bibnamefont
  {{Tegmark}}},\ }\href@noop {} {\bibfield  {journal} {\bibinfo  {journal}
  {\mnras},\ }\textbf {\bibinfo {volume} {279}},\ \bibinfo {pages} {545}
  (\bibinfo {year} {1996})},\ \Eprint {http://arxiv.org/abs/astro-ph/9507077}
  {astro-ph/9507077} \BibitemShut {NoStop}%
\bibitem [{\citenamefont {{Osborne}}\ \emph {et~al.}(2011)\citenamefont
  {{Osborne}}, \citenamefont {{Mak}}, \citenamefont {{Church}},\ and\
  \citenamefont {{Pierpaoli}}}]{Osborne2011}%
  \BibitemOpen
  \bibfield  {author} {\bibinfo {author} {\bibfnamefont {S.~J.}\ \bibnamefont
  {{Osborne}}}, \bibinfo {author} {\bibfnamefont {D.~S.~Y.}\ \bibnamefont
  {{Mak}}}, \bibinfo {author} {\bibfnamefont {S.~E.}\ \bibnamefont {{Church}}},
  \ and\ \bibinfo {author} {\bibfnamefont {E.}~\bibnamefont {{Pierpaoli}}},\
  }\Doi {10.1088/0004-637X/737/2/98} {\bibfield  {journal} {\bibinfo  {journal}
  {\apj},\ }\textbf {\bibinfo {volume} {737}},\ \bibinfo {eid} {98} (\bibinfo
  {year} {2011})},\ \Eprint {http://arxiv.org/abs/1011.2781} {arXiv:1011.2781
  [astro-ph.CO]} \BibitemShut {NoStop}%
\bibitem [{\citenamefont {{Lavaux}}\ \emph {et~al.}(2013)\citenamefont
  {{Lavaux}}, \citenamefont {{Afshordi}},\ and\ \citenamefont
  {{Hudson}}}]{Lavaux2013}%
  \BibitemOpen
  \bibfield  {author} {\bibinfo {author} {\bibfnamefont {G.}~\bibnamefont
  {{Lavaux}}}, \bibinfo {author} {\bibfnamefont {N.}~\bibnamefont
  {{Afshordi}}}, \ and\ \bibinfo {author} {\bibfnamefont {M.~J.}\ \bibnamefont
  {{Hudson}}},\ }\Doi {10.1093/mnras/sts698} {\bibfield  {journal} {\bibinfo
  {journal} {\mnras},\ }\textbf {\bibinfo {volume} {430}},\ \bibinfo {pages}
  {1617} (\bibinfo {year} {2013})},\ \Eprint {http://arxiv.org/abs/1207.1721}
  {arXiv:1207.1721 [astro-ph.CO]} \BibitemShut {NoStop}%
\bibitem [{\citenamefont {{The Planck Collaboration}}\ \emph
  {et~al.}(2014)\citenamefont {{The Planck Collaboration}}, \citenamefont
  {{Ade}}, \citenamefont {{Aghanim}}, \citenamefont {{Arnaud}}, \citenamefont
  {{Ashdown}}, \citenamefont {{Aumont}}, \citenamefont {{Baccigalupi}},
  \citenamefont {{Balbi}}, \citenamefont {{Banday}}, \citenamefont
  {{Barreiro}},\ and\ \citenamefont {et~al.}}]{planck_bf}%
  \BibitemOpen
  \bibfield  {author} {\bibinfo {author} {\bibnamefont {{The Planck
  Collaboration}}}, \bibinfo {author} {\bibfnamefont {P.~A.~R.}\ \bibnamefont
  {{Ade}}}, \bibinfo {author} {\bibfnamefont {N.}~\bibnamefont {{Aghanim}}},
  \bibinfo {author} {\bibfnamefont {M.}~\bibnamefont {{Arnaud}}}, \bibinfo
  {author} {\bibfnamefont {M.}~\bibnamefont {{Ashdown}}}, \bibinfo {author}
  {\bibfnamefont {J.}~\bibnamefont {{Aumont}}}, \bibinfo {author}
  {\bibfnamefont {C.}~\bibnamefont {{Baccigalupi}}}, \bibinfo {author}
  {\bibfnamefont {A.}~\bibnamefont {{Balbi}}}, \bibinfo {author} {\bibfnamefont
  {A.~J.}\ \bibnamefont {{Banday}}}, \bibinfo {author} {\bibfnamefont {R.~B.}\
  \bibnamefont {{Barreiro}}}, \ and\ \bibinfo {author} {\bibnamefont
  {et~al.}},\ }\Doi {10.1051/0004-6361/201321299} {\bibfield  {journal}
  {\bibinfo  {journal} {\aap},\ }\textbf {\bibinfo {volume} {561}},\ \bibinfo
  {eid} {A97} (\bibinfo {year} {2014})},\ \Eprint
  {http://arxiv.org/abs/1303.5090} {arXiv:1303.5090} \BibitemShut {NoStop}%
\bibitem [{\citenamefont {{Hamilton}}(1998)}]{Hamilton1998}%
  \BibitemOpen
  \bibfield  {author} {\bibinfo {author} {\bibfnamefont {A.~J.~S.}\
  \bibnamefont {{Hamilton}}},\ }in\ \href@noop {} {\emph {\bibinfo {booktitle}
  {The Evolving Universe}}},\ \bibinfo {series} {Astrophysics and Space Science
  Library}, Vol.\ \bibinfo {volume} {231},\ \bibinfo {editor} {edited by\
  \bibinfo {editor} {\bibfnamefont {D.}~\bibnamefont {{Hamilton}}}}\ (\bibinfo
  {year} {1998})\ p.\ \bibinfo {pages} {185}\BibitemShut {NoStop}%
\bibitem [{\citenamefont {{Peacock}}\ \emph {et~al.}(2001)\citenamefont
  {{Peacock}}, \citenamefont {{Cole}}, \citenamefont {{Norberg}}, \citenamefont
  {{Baugh}}, \citenamefont {{Bland-Hawthorn}}, \citenamefont {{Bridges}},
  \citenamefont {{Cannon}}, \citenamefont {{Colless}}, \citenamefont
  {{Collins}}, \citenamefont {{Couch}},\ and\ \citenamefont
  {et~al.}}]{Peacock2001}%
  \BibitemOpen
  \bibfield  {author} {\bibinfo {author} {\bibfnamefont {J.~A.}\ \bibnamefont
  {{Peacock}}}, \bibinfo {author} {\bibfnamefont {S.}~\bibnamefont {{Cole}}},
  \bibinfo {author} {\bibfnamefont {P.}~\bibnamefont {{Norberg}}}, \bibinfo
  {author} {\bibfnamefont {C.~M.}\ \bibnamefont {{Baugh}}}, \bibinfo {author}
  {\bibfnamefont {J.}~\bibnamefont {{Bland-Hawthorn}}}, \bibinfo {author}
  {\bibfnamefont {T.}~\bibnamefont {{Bridges}}}, \bibinfo {author}
  {\bibfnamefont {R.~D.}\ \bibnamefont {{Cannon}}}, \bibinfo {author}
  {\bibfnamefont {M.}~\bibnamefont {{Colless}}}, \bibinfo {author}
  {\bibfnamefont {C.}~\bibnamefont {{Collins}}}, \bibinfo {author}
  {\bibfnamefont {W.}~\bibnamefont {{Couch}}}, \ and\ \bibinfo {author}
  {\bibnamefont {et~al.}},\ }\href@noop {} {\bibfield  {journal} {\bibinfo
  {journal} {\nat},\ }\textbf {\bibinfo {volume} {410}},\ \bibinfo {pages}
  {169} (\bibinfo {year} {2001})},\ \Eprint
  {http://arxiv.org/abs/astro-ph/0103143} {astro-ph/0103143} \BibitemShut
  {NoStop}%
\bibitem [{\citenamefont {{Scoccimarro}}(2004)}]{Scoccimarro2004}%
  \BibitemOpen
  \bibfield  {author} {\bibinfo {author} {\bibfnamefont {R.}~\bibnamefont
  {{Scoccimarro}}},\ }\Doi {10.1103/PhysRevD.70.083007} {\bibfield  {journal}
  {\bibinfo  {journal} {\prd},\ }\textbf {\bibinfo {volume} {70}},\ \bibinfo
  {eid} {083007} (\bibinfo {year} {2004})},\ \Eprint
  {http://arxiv.org/abs/astro-ph/0407214} {astro-ph/0407214} \BibitemShut
  {NoStop}%
\bibitem [{\citenamefont {{Guzzo}}\ \emph {et~al.}(2008)\citenamefont
  {{Guzzo}}, \citenamefont {{Pierleoni}}, \citenamefont {{Meneux}},
  \citenamefont {{Branchini}}, \citenamefont {{Le F{\`e}vre}}, \citenamefont
  {{Marinoni}}, \citenamefont {{Garilli}}, \citenamefont {{Blaizot}},
  \citenamefont {{De Lucia}}, \citenamefont {{Pollo}},\ and\ \citenamefont
  {et~al.}}]{Guz08}%
  \BibitemOpen
  \bibfield  {author} {\bibinfo {author} {\bibfnamefont {L.}~\bibnamefont
  {{Guzzo}}}, \bibinfo {author} {\bibfnamefont {M.}~\bibnamefont
  {{Pierleoni}}}, \bibinfo {author} {\bibfnamefont {B.}~\bibnamefont
  {{Meneux}}}, \bibinfo {author} {\bibfnamefont {E.}~\bibnamefont
  {{Branchini}}}, \bibinfo {author} {\bibfnamefont {O.}~\bibnamefont {{Le
  F{\`e}vre}}}, \bibinfo {author} {\bibfnamefont {C.}~\bibnamefont
  {{Marinoni}}}, \bibinfo {author} {\bibfnamefont {B.}~\bibnamefont
  {{Garilli}}}, \bibinfo {author} {\bibfnamefont {J.}~\bibnamefont
  {{Blaizot}}}, \bibinfo {author} {\bibfnamefont {G.}~\bibnamefont {{De
  Lucia}}}, \bibinfo {author} {\bibfnamefont {A.}~\bibnamefont {{Pollo}}}, \
  and\ \bibinfo {author} {\bibnamefont {et~al.}},\ }\Doi {10.1038/nature06555}
  {\bibfield  {journal} {\bibinfo  {journal} {\nat},\ }\textbf {\bibinfo
  {volume} {451}},\ \bibinfo {pages} {541} (\bibinfo {year} {2008})},\ \Eprint
  {http://arxiv.org/abs/0802.1944} {arXiv:0802.1944} \BibitemShut {NoStop}%
\bibitem [{\citenamefont {{Tammann}}\ \emph {et~al.}(1979)\citenamefont
  {{Tammann}}, \citenamefont {{Yahil}},\ and\ \citenamefont {{Sandage}}}]{TYS}%
  \BibitemOpen
  \bibfield  {author} {\bibinfo {author} {\bibfnamefont {G.~A.}\ \bibnamefont
  {{Tammann}}}, \bibinfo {author} {\bibfnamefont {A.}~\bibnamefont {{Yahil}}},
  \ and\ \bibinfo {author} {\bibfnamefont {A.}~\bibnamefont {{Sandage}}},\
  }\Doi {10.1086/157556} {\bibfield  {journal} {\bibinfo  {journal} {\apj},\
  }\textbf {\bibinfo {volume} {234}},\ \bibinfo {pages} {775} (\bibinfo {year}
  {1979})}\BibitemShut {NoStop}%
\bibitem [{\citenamefont {{Huchra}}\ \emph {et~al.}(2012)\citenamefont
  {{Huchra}}, \citenamefont {{Macri}}, \citenamefont {{Masters}}, \citenamefont
  {{Jarrett}}, \citenamefont {{Berlind}}, \citenamefont {{Calkins}},
  \citenamefont {{Crook}}, \citenamefont {{Cutri}}, \citenamefont {{Erdo{\v
  g}du}}, \citenamefont {{Falco}},\ and\ \citenamefont {et~al.}}]{Huchra2012}%
  \BibitemOpen
  \bibfield  {author} {\bibinfo {author} {\bibfnamefont {J.~P.}\ \bibnamefont
  {{Huchra}}}, \bibinfo {author} {\bibfnamefont {L.~M.}\ \bibnamefont
  {{Macri}}}, \bibinfo {author} {\bibfnamefont {K.~L.}\ \bibnamefont
  {{Masters}}}, \bibinfo {author} {\bibfnamefont {T.~H.}\ \bibnamefont
  {{Jarrett}}}, \bibinfo {author} {\bibfnamefont {P.}~\bibnamefont
  {{Berlind}}}, \bibinfo {author} {\bibfnamefont {M.}~\bibnamefont
  {{Calkins}}}, \bibinfo {author} {\bibfnamefont {A.~C.}\ \bibnamefont
  {{Crook}}}, \bibinfo {author} {\bibfnamefont {R.}~\bibnamefont {{Cutri}}},
  \bibinfo {author} {\bibfnamefont {P.}~\bibnamefont {{Erdo{\v g}du}}},
  \bibinfo {author} {\bibfnamefont {E.}~\bibnamefont {{Falco}}}, \ and\
  \bibinfo {author} {\bibnamefont {et~al.}},\ }\Doi
  {10.1088/0067-0049/199/2/26} {\bibfield  {journal} {\bibinfo  {journal}
  {\apjs},\ }\textbf {\bibinfo {volume} {199}},\ \bibinfo {eid} {26} (\bibinfo
  {year} {2012})},\ \Eprint {http://arxiv.org/abs/1108.0669} {arXiv:1108.0669
  [astro-ph.CO]} \BibitemShut {NoStop}%
\bibitem [{\citenamefont {{Nusser}}\ \emph {et~al.}(2011)\citenamefont
  {{Nusser}}, \citenamefont {{Branchini}},\ and\ \citenamefont
  {{Davis}}}]{Nusser2011}%
  \BibitemOpen
  \bibfield  {author} {\bibinfo {author} {\bibfnamefont {A.}~\bibnamefont
  {{Nusser}}}, \bibinfo {author} {\bibfnamefont {E.}~\bibnamefont
  {{Branchini}}}, \ and\ \bibinfo {author} {\bibfnamefont {M.}~\bibnamefont
  {{Davis}}},\ }\Doi {10.1088/0004-637X/735/2/77} {\bibfield  {journal}
  {\bibinfo  {journal} {\apj},\ }\textbf {\bibinfo {volume} {735}},\ \bibinfo
  {eid} {77} (\bibinfo {year} {2011})},\ \Eprint
  {http://arxiv.org/abs/1102.4189} {arXiv:1102.4189 [astro-ph.CO]} \BibitemShut
  {NoStop}%
\bibitem [{\citenamefont {{Branchini}}\ \emph {et~al.}(2012)\citenamefont
  {{Branchini}}, \citenamefont {{Davis}},\ and\ \citenamefont
  {{Nusser}}}]{Branchini2012}%
  \BibitemOpen
  \bibfield  {author} {\bibinfo {author} {\bibfnamefont {E.}~\bibnamefont
  {{Branchini}}}, \bibinfo {author} {\bibfnamefont {M.}~\bibnamefont
  {{Davis}}}, \ and\ \bibinfo {author} {\bibfnamefont {A.}~\bibnamefont
  {{Nusser}}},\ }\Doi {10.1111/j.1365-2966.2012.21210.x} {\bibfield  {journal}
  {\bibinfo  {journal} {\mnras},\ }\textbf {\bibinfo {volume} {424}},\ \bibinfo
  {pages} {472} (\bibinfo {year} {2012})},\ \Eprint
  {http://arxiv.org/abs/1202.5206} {arXiv:1202.5206 [astro-ph.CO]} \BibitemShut
  {NoStop}%
\bibitem [{\citenamefont {{Nusser}}\ and\ \citenamefont
  {{Davis}}(1994)}]{Nusser1994}%
  \BibitemOpen
  \bibfield  {author} {\bibinfo {author} {\bibfnamefont {A.}~\bibnamefont
  {{Nusser}}}\ and\ \bibinfo {author} {\bibfnamefont {M.}~\bibnamefont
  {{Davis}}},\ }\Doi {10.1086/187172} {\bibfield  {journal} {\bibinfo
  {journal} {\apjl},\ }\textbf {\bibinfo {volume} {421}},\ \bibinfo {pages}
  {L1} (\bibinfo {year} {1994})},\ \Eprint
  {http://arxiv.org/abs/astro-ph/9309009} {astro-ph/9309009} \BibitemShut
  {NoStop}%
\bibitem [{\citenamefont {{Nusser}}\ \emph {et~al.}(2012)\citenamefont
  {{Nusser}}, \citenamefont {{Branchini}},\ and\ \citenamefont
  {{Davis}}}]{Nusser2012}%
  \BibitemOpen
  \bibfield  {author} {\bibinfo {author} {\bibfnamefont {A.}~\bibnamefont
  {{Nusser}}}, \bibinfo {author} {\bibfnamefont {E.}~\bibnamefont
  {{Branchini}}}, \ and\ \bibinfo {author} {\bibfnamefont {M.}~\bibnamefont
  {{Davis}}},\ }\Doi {10.1088/0004-637X/744/2/193} {\bibfield  {journal}
  {\bibinfo  {journal} {\apj},\ }\textbf {\bibinfo {volume} {744}},\ \bibinfo
  {eid} {193} (\bibinfo {year} {2012})},\ \Eprint
  {http://arxiv.org/abs/1106.6145} {arXiv:1106.6145 [astro-ph.CO]} \BibitemShut
  {NoStop}%
\bibitem [{\citenamefont {{York}}\ \emph {et~al.}(2000)\citenamefont {{York}},
  \citenamefont {{Adelman}}, \citenamefont {{Anderson}}, \citenamefont
  {{Anderson}}, \citenamefont {{Annis}}, \citenamefont {{Bahcall}},
  \citenamefont {{Bakken}}, \citenamefont {{Barkhouser}}, \citenamefont
  {{Bastian}}, \citenamefont {{Berman}},\ and\ \citenamefont {{SDSS
  Collaboration}}}]{York2000}%
  \BibitemOpen
  \bibfield  {author} {\bibinfo {author} {\bibfnamefont {D.~G.}\ \bibnamefont
  {{York}}}, \bibinfo {author} {\bibfnamefont {J.}~\bibnamefont {{Adelman}}},
  \bibinfo {author} {\bibfnamefont {J.~E.}\ \bibnamefont {{Anderson}},
  \bibfnamefont {Jr.}}, \bibinfo {author} {\bibfnamefont {S.~F.}\ \bibnamefont
  {{Anderson}}}, \bibinfo {author} {\bibfnamefont {J.}~\bibnamefont {{Annis}}},
  \bibinfo {author} {\bibfnamefont {N.~A.}\ \bibnamefont {{Bahcall}}}, \bibinfo
  {author} {\bibfnamefont {J.~A.}\ \bibnamefont {{Bakken}}}, \bibinfo {author}
  {\bibfnamefont {R.}~\bibnamefont {{Barkhouser}}}, \bibinfo {author}
  {\bibfnamefont {S.}~\bibnamefont {{Bastian}}}, \bibinfo {author}
  {\bibfnamefont {E.}~\bibnamefont {{Berman}}}, \ and\ \bibinfo {author}
  {\bibnamefont {{SDSS Collaboration}}},\ }\Doi {10.1086/301513} {\bibfield
  {journal} {\bibinfo  {journal} {\aj},\ }\textbf {\bibinfo {volume} {120}},\
  \bibinfo {pages} {1579} (\bibinfo {year} {2000})},\ \Eprint
  {http://arxiv.org/abs/astro-ph/0006396} {astro-ph/0006396} \BibitemShut
  {NoStop}%
\bibitem [{\citenamefont {{Fixsen}}\ \emph {et~al.}(1996)\citenamefont
  {{Fixsen}}, \citenamefont {{Cheng}}, \citenamefont {{Gales}}, \citenamefont
  {{Mather}}, \citenamefont {{Shafer}},\ and\ \citenamefont
  {{Wright}}}]{Fixsen1996}%
  \BibitemOpen
  \bibfield  {author} {\bibinfo {author} {\bibfnamefont {D.~J.}\ \bibnamefont
  {{Fixsen}}}, \bibinfo {author} {\bibfnamefont {E.~S.}\ \bibnamefont
  {{Cheng}}}, \bibinfo {author} {\bibfnamefont {J.~M.}\ \bibnamefont
  {{Gales}}}, \bibinfo {author} {\bibfnamefont {J.~C.}\ \bibnamefont
  {{Mather}}}, \bibinfo {author} {\bibfnamefont {R.~A.}\ \bibnamefont
  {{Shafer}}}, \ and\ \bibinfo {author} {\bibfnamefont {E.~L.}\ \bibnamefont
  {{Wright}}},\ }\Doi {10.1086/178173} {\bibfield  {journal} {\bibinfo
  {journal} {\apj},\ }\textbf {\bibinfo {volume} {473}},\ \bibinfo {pages}
  {576} (\bibinfo {year} {1996})},\ \Eprint
  {http://arxiv.org/abs/astro-ph/9605054} {astro-ph/9605054} \BibitemShut
  {NoStop}%
\bibitem [{\citenamefont {{Sachs}}\ and\ \citenamefont {{Wolfe}}(1967)}]{SW}%
  \BibitemOpen
  \bibfield  {author} {\bibinfo {author} {\bibfnamefont {R.~K.}\ \bibnamefont
  {{Sachs}}}\ and\ \bibinfo {author} {\bibfnamefont {A.~M.}\ \bibnamefont
  {{Wolfe}}},\ }\Doi {10.1086/148982} {\bibfield  {journal} {\bibinfo
  {journal} {\apj},\ }\textbf {\bibinfo {volume} {147}},\ \bibinfo {pages} {73}
  (\bibinfo {year} {1967})}\BibitemShut {NoStop}%
\bibitem [{\citenamefont {{Blanton}}\ and\ \citenamefont
  {{Roweis}}(2007)}]{Blanton2007}%
  \BibitemOpen
  \bibfield  {author} {\bibinfo {author} {\bibfnamefont {M.~R.}\ \bibnamefont
  {{Blanton}}}\ and\ \bibinfo {author} {\bibfnamefont {S.}~\bibnamefont
  {{Roweis}}},\ }\Doi {10.1086/510127} {\bibfield  {journal} {\bibinfo
  {journal} {\aj},\ }\textbf {\bibinfo {volume} {133}},\ \bibinfo {pages} {734}
  (\bibinfo {year} {2007})},\ \Eprint {http://arxiv.org/abs/astro-ph/0606170}
  {astro-ph/0606170} \BibitemShut {NoStop}%
\bibitem [{\citenamefont {{Tegmark}}(1997)}]{Tegmark1997}%
  \BibitemOpen
  \bibfield  {author} {\bibinfo {author} {\bibfnamefont {M.}~\bibnamefont
  {{Tegmark}}},\ }\Doi {10.1103/PhysRevD.55.5895} {\bibfield  {journal}
  {\bibinfo  {journal} {\prd},\ }\textbf {\bibinfo {volume} {55}},\ \bibinfo
  {pages} {5895} (\bibinfo {year} {1997})},\ \Eprint
  {http://arxiv.org/abs/astro-ph/9611174} {astro-ph/9611174} \BibitemShut
  {NoStop}%
\bibitem [{\citenamefont {{Bond}}\ \emph {et~al.}(1998)\citenamefont {{Bond}},
  \citenamefont {{Jaffe}},\ and\ \citenamefont {{Knox}}}]{Bond1998}%
  \BibitemOpen
  \bibfield  {author} {\bibinfo {author} {\bibfnamefont {J.~R.}\ \bibnamefont
  {{Bond}}}, \bibinfo {author} {\bibfnamefont {A.~H.}\ \bibnamefont {{Jaffe}}},
  \ and\ \bibinfo {author} {\bibfnamefont {L.}~\bibnamefont {{Knox}}},\ }\Doi
  {10.1103/PhysRevD.57.2117} {\bibfield  {journal} {\bibinfo  {journal}
  {\prd},\ }\textbf {\bibinfo {volume} {57}},\ \bibinfo {pages} {2117}
  (\bibinfo {year} {1998})},\ \Eprint {http://arxiv.org/abs/astro-ph/9708203}
  {astro-ph/9708203} \BibitemShut {NoStop}%
\bibitem [{\citenamefont {{Itoh}}\ \emph {et~al.}(2010)\citenamefont {{Itoh}},
  \citenamefont {{Yahata}},\ and\ \citenamefont {{Takada}}}]{Itoh2010}%
  \BibitemOpen
  \bibfield  {author} {\bibinfo {author} {\bibfnamefont {Y.}~\bibnamefont
  {{Itoh}}}, \bibinfo {author} {\bibfnamefont {K.}~\bibnamefont {{Yahata}}}, \
  and\ \bibinfo {author} {\bibfnamefont {M.}~\bibnamefont {{Takada}}},\ }\Doi
  {10.1103/PhysRevD.82.043530} {\bibfield  {journal} {\bibinfo  {journal}
  {\prd},\ }\textbf {\bibinfo {volume} {82}},\ \bibinfo {eid} {043530}
  (\bibinfo {year} {2010})},\ \Eprint {http://arxiv.org/abs/0912.1460}
  {arXiv:0912.1460 [astro-ph.CO]} \BibitemShut {NoStop}%
\bibitem [{\citenamefont {{Abate}}\ and\ \citenamefont
  {{Feldman}}(2012)}]{Abate2012}%
  \BibitemOpen
  \bibfield  {author} {\bibinfo {author} {\bibfnamefont {A.}~\bibnamefont
  {{Abate}}}\ and\ \bibinfo {author} {\bibfnamefont {H.~A.}\ \bibnamefont
  {{Feldman}}},\ }\Doi {10.1111/j.1365-2966.2011.19988.x} {\bibfield  {journal}
  {\bibinfo  {journal} {\mnras},\ }\textbf {\bibinfo {volume} {419}},\ \bibinfo
  {pages} {3482} (\bibinfo {year} {2012})},\ \Eprint
  {http://arxiv.org/abs/1106.5791} {arXiv:1106.5791 [astro-ph.CO]} \BibitemShut
  {NoStop}%
\bibitem [{\citenamefont {{Efstathiou}}\ \emph {et~al.}(1988)\citenamefont
  {{Efstathiou}}, \citenamefont {{Ellis}},\ and\ \citenamefont
  {{Peterson}}}]{efs88}%
  \BibitemOpen
  \bibfield  {author} {\bibinfo {author} {\bibfnamefont {G.}~\bibnamefont
  {{Efstathiou}}}, \bibinfo {author} {\bibfnamefont {R.~S.}\ \bibnamefont
  {{Ellis}}}, \ and\ \bibinfo {author} {\bibfnamefont {B.~A.}\ \bibnamefont
  {{Peterson}}},\ }\href@noop {} {\bibfield  {journal} {\bibinfo  {journal}
  {\mnras},\ }\textbf {\bibinfo {volume} {232}},\ \bibinfo {pages} {431}
  (\bibinfo {year} {1988})}\BibitemShut {NoStop}%
\bibitem [{\citenamefont {{Press}}\ \emph {et~al.}(2002)\citenamefont
  {{Press}}, \citenamefont {{Teukolsky}}, \citenamefont {{Vetterling}},\ and\
  \citenamefont {{Flannery}}}]{Press2002}%
  \BibitemOpen
  \bibfield  {author} {\bibinfo {author} {\bibfnamefont {W.~H.}\ \bibnamefont
  {{Press}}}, \bibinfo {author} {\bibfnamefont {S.~A.}\ \bibnamefont
  {{Teukolsky}}}, \bibinfo {author} {\bibfnamefont {W.~T.}\ \bibnamefont
  {{Vetterling}}}, \ and\ \bibinfo {author} {\bibfnamefont {B.~P.}\
  \bibnamefont {{Flannery}}},\ }\href@noop {} {\emph {\bibinfo {title}
  {Numerical recipes in C++ : the art of scientific computing by William
  H.~Press.~xxviii, 1,002 p.~: ill.~; 26 cm.~ Includes bibliographical
  references and index.~ISBN : 0521750334}}}\ (\bibinfo  {publisher} {Cambridge
  University Press, 2nd ed.},\ \bibinfo {year} {2002})\BibitemShut {NoStop}%
\bibitem [{\citenamefont {{Sandage}}\ \emph {et~al.}(1979)\citenamefont
  {{Sandage}}, \citenamefont {{Tammann}},\ and\ \citenamefont
  {{Yahil}}}]{Sandage1979}%
  \BibitemOpen
  \bibfield  {author} {\bibinfo {author} {\bibfnamefont {A.}~\bibnamefont
  {{Sandage}}}, \bibinfo {author} {\bibfnamefont {G.~A.}\ \bibnamefont
  {{Tammann}}}, \ and\ \bibinfo {author} {\bibfnamefont {A.}~\bibnamefont
  {{Yahil}}},\ }\Doi {10.1086/157295} {\bibfield  {journal} {\bibinfo
  {journal} {\apj},\ }\textbf {\bibinfo {volume} {232}},\ \bibinfo {pages}
  {352} (\bibinfo {year} {1979})}\BibitemShut {NoStop}%
\bibitem [{\citenamefont {{Schechter}}(1980)}]{schechter}%
  \BibitemOpen
  \bibfield  {author} {\bibinfo {author} {\bibfnamefont {P.~L.}\ \bibnamefont
  {{Schechter}}},\ }\Doi {10.1086/112742} {\bibfield  {journal} {\bibinfo
  {journal} {\astjo},\ }\textbf {\bibinfo {volume} {85}},\ \bibinfo {pages}
  {801} (\bibinfo {year} {1980})}\BibitemShut {NoStop}%
\bibitem [{\citenamefont {{Blanton}}\ \emph {et~al.}(2005)\citenamefont
  {{Blanton}}, \citenamefont {{Schlegel}}, \citenamefont {{Strauss}},
  \citenamefont {{Brinkmann}}, \citenamefont {{Finkbeiner}}, \citenamefont
  {{Fukugita}}, \citenamefont {{Gunn}}, \citenamefont {{Hogg}}, \citenamefont
  {{Ivezi{\'c}}}, \citenamefont {{Knapp}},\ and\ \citenamefont
  {et~al.}}]{Blanton2005}%
  \BibitemOpen
  \bibfield  {author} {\bibinfo {author} {\bibfnamefont {M.~R.}\ \bibnamefont
  {{Blanton}}}, \bibinfo {author} {\bibfnamefont {D.~J.}\ \bibnamefont
  {{Schlegel}}}, \bibinfo {author} {\bibfnamefont {M.~A.}\ \bibnamefont
  {{Strauss}}}, \bibinfo {author} {\bibfnamefont {J.}~\bibnamefont
  {{Brinkmann}}}, \bibinfo {author} {\bibfnamefont {D.}~\bibnamefont
  {{Finkbeiner}}}, \bibinfo {author} {\bibfnamefont {M.}~\bibnamefont
  {{Fukugita}}}, \bibinfo {author} {\bibfnamefont {J.~E.}\ \bibnamefont
  {{Gunn}}}, \bibinfo {author} {\bibfnamefont {D.~W.}\ \bibnamefont {{Hogg}}},
  \bibinfo {author} {\bibfnamefont {{\v Z}.}~\bibnamefont {{Ivezi{\'c}}}},
  \bibinfo {author} {\bibfnamefont {G.~R.}\ \bibnamefont {{Knapp}}}, \ and\
  \bibinfo {author} {\bibnamefont {et~al.}},\ }\Doi {10.1086/429803} {\bibfield
   {journal} {\bibinfo  {journal} {\aj},\ }\textbf {\bibinfo {volume} {129}},\
  \bibinfo {pages} {2562} (\bibinfo {year} {2005})},\ \Eprint
  {http://arxiv.org/abs/astro-ph/0410166} {astro-ph/0410166} \BibitemShut
  {NoStop}%
\bibitem [{\citenamefont {{Abazajian}}\ \emph {et~al.}(2009)\citenamefont
  {{Abazajian}}, \citenamefont {{Adelman-McCarthy}}, \citenamefont
  {{Ag{\"u}eros}}, \citenamefont {{Allam}}, \citenamefont {{Allende Prieto}},
  \citenamefont {{An}}, \citenamefont {{Anderson}}, \citenamefont {{Anderson}},
  \citenamefont {{Annis}}, \citenamefont {{Bahcall}},\ and\ \citenamefont
  {et~al.}}]{abaz}%
  \BibitemOpen
  \bibfield  {author} {\bibinfo {author} {\bibfnamefont {K.~N.}\ \bibnamefont
  {{Abazajian}}}, \bibinfo {author} {\bibfnamefont {J.~K.}\ \bibnamefont
  {{Adelman-McCarthy}}}, \bibinfo {author} {\bibfnamefont {M.~A.}\ \bibnamefont
  {{Ag{\"u}eros}}}, \bibinfo {author} {\bibfnamefont {S.~S.}\ \bibnamefont
  {{Allam}}}, \bibinfo {author} {\bibfnamefont {C.}~\bibnamefont {{Allende
  Prieto}}}, \bibinfo {author} {\bibfnamefont {D.}~\bibnamefont {{An}}},
  \bibinfo {author} {\bibfnamefont {K.~S.~J.}\ \bibnamefont {{Anderson}}},
  \bibinfo {author} {\bibfnamefont {S.~F.}\ \bibnamefont {{Anderson}}},
  \bibinfo {author} {\bibfnamefont {J.}~\bibnamefont {{Annis}}}, \bibinfo
  {author} {\bibfnamefont {N.~A.}\ \bibnamefont {{Bahcall}}}, \ and\ \bibinfo
  {author} {\bibnamefont {et~al.}},\ }\Doi {10.1088/0067-0049/182/2/543}
  {\bibfield  {journal} {\bibinfo  {journal} {\apjs},\ }\textbf {\bibinfo
  {volume} {182}},\ \bibinfo {pages} {543} (\bibinfo {year} {2009})},\ \Eprint
  {http://arxiv.org/abs/0812.0649} {arXiv:0812.0649} \BibitemShut {NoStop}%
\bibitem [{\citenamefont {{Schlegel}}\ \emph {et~al.}(1998)\citenamefont
  {{Schlegel}}, \citenamefont {{Finkbeiner}},\ and\ \citenamefont
  {{Davis}}}]{Schlegel1998}%
  \BibitemOpen
  \bibfield  {author} {\bibinfo {author} {\bibfnamefont {D.~J.}\ \bibnamefont
  {{Schlegel}}}, \bibinfo {author} {\bibfnamefont {D.~P.}\ \bibnamefont
  {{Finkbeiner}}}, \ and\ \bibinfo {author} {\bibfnamefont {M.}~\bibnamefont
  {{Davis}}},\ }\Doi {10.1086/305772} {\bibfield  {journal} {\bibinfo
  {journal} {\apj},\ }\textbf {\bibinfo {volume} {500}},\ \bibinfo {pages}
  {525} (\bibinfo {year} {1998})},\ \Eprint
  {http://arxiv.org/abs/astro-ph/9710327} {astro-ph/9710327} \BibitemShut
  {NoStop}%
\bibitem [{\citenamefont {{Blanton}}\ \emph {et~al.}(2001)\citenamefont
  {{Blanton}}, \citenamefont {{Dalcanton}}, \citenamefont {{Eisenstein}},
  \citenamefont {{Loveday}}, \citenamefont {{Strauss}}, \citenamefont
  {{SubbaRao}}, \citenamefont {{Weinberg}}, \citenamefont {{Anderson}},
  \citenamefont {{Annis}}, \citenamefont {{Bahcall}},\ and\ \citenamefont
  {et~al.}}]{Blanton2001}%
  \BibitemOpen
  \bibfield  {author} {\bibinfo {author} {\bibfnamefont {M.~R.}\ \bibnamefont
  {{Blanton}}}, \bibinfo {author} {\bibfnamefont {J.}~\bibnamefont
  {{Dalcanton}}}, \bibinfo {author} {\bibfnamefont {D.}~\bibnamefont
  {{Eisenstein}}}, \bibinfo {author} {\bibfnamefont {J.}~\bibnamefont
  {{Loveday}}}, \bibinfo {author} {\bibfnamefont {M.~A.}\ \bibnamefont
  {{Strauss}}}, \bibinfo {author} {\bibfnamefont {M.}~\bibnamefont
  {{SubbaRao}}}, \bibinfo {author} {\bibfnamefont {D.~H.}\ \bibnamefont
  {{Weinberg}}}, \bibinfo {author} {\bibfnamefont {J.~E.}\ \bibnamefont
  {{Anderson}}, \bibfnamefont {Jr.}}, \bibinfo {author} {\bibfnamefont
  {J.}~\bibnamefont {{Annis}}}, \bibinfo {author} {\bibfnamefont {N.~A.}\
  \bibnamefont {{Bahcall}}}, \ and\ \bibinfo {author} {\bibnamefont {et~al.}},\
  }\Doi {10.1086/320405} {\bibfield  {journal} {\bibinfo  {journal} {\aj},\
  }\textbf {\bibinfo {volume} {121}},\ \bibinfo {pages} {2358} (\bibinfo {year}
  {2001})},\ \Eprint {http://arxiv.org/abs/astro-ph/0012085} {astro-ph/0012085}
  \BibitemShut {NoStop}%
\bibitem [{\citenamefont {{Strauss}}\ \emph {et~al.}(2002)\citenamefont
  {{Strauss}}, \citenamefont {{Weinberg}}, \citenamefont {{Lupton}},
  \citenamefont {{Narayanan}}, \citenamefont {{Annis}}, \citenamefont
  {{Bernardi}}, \citenamefont {{Blanton}}, \citenamefont {{Burles}},
  \citenamefont {{Connolly}}, \citenamefont {{Dalcanton}},\ and\ \citenamefont
  {et~al.}}]{Strauss2002}%
  \BibitemOpen
  \bibfield  {author} {\bibinfo {author} {\bibfnamefont {M.~A.}\ \bibnamefont
  {{Strauss}}}, \bibinfo {author} {\bibfnamefont {D.~H.}\ \bibnamefont
  {{Weinberg}}}, \bibinfo {author} {\bibfnamefont {R.~H.}\ \bibnamefont
  {{Lupton}}}, \bibinfo {author} {\bibfnamefont {V.~K.}\ \bibnamefont
  {{Narayanan}}}, \bibinfo {author} {\bibfnamefont {J.}~\bibnamefont
  {{Annis}}}, \bibinfo {author} {\bibfnamefont {M.}~\bibnamefont {{Bernardi}}},
  \bibinfo {author} {\bibfnamefont {M.}~\bibnamefont {{Blanton}}}, \bibinfo
  {author} {\bibfnamefont {S.}~\bibnamefont {{Burles}}}, \bibinfo {author}
  {\bibfnamefont {A.~J.}\ \bibnamefont {{Connolly}}}, \bibinfo {author}
  {\bibfnamefont {J.}~\bibnamefont {{Dalcanton}}}, \ and\ \bibinfo {author}
  {\bibnamefont {et~al.}},\ }\Doi {10.1086/342343} {\bibfield  {journal}
  {\bibinfo  {journal} {\aj},\ }\textbf {\bibinfo {volume} {124}},\ \bibinfo
  {pages} {1810} (\bibinfo {year} {2002})},\ \Eprint
  {http://arxiv.org/abs/astro-ph/0206225} {astro-ph/0206225} \BibitemShut
  {NoStop}%
\bibitem [{\citenamefont {{Oke}}\ and\ \citenamefont {{Gunn}}(1983)}]{Oke1983}%
  \BibitemOpen
  \bibfield  {author} {\bibinfo {author} {\bibfnamefont {J.~B.}\ \bibnamefont
  {{Oke}}}\ and\ \bibinfo {author} {\bibfnamefont {J.~E.}\ \bibnamefont
  {{Gunn}}},\ }\Doi {10.1086/160817} {\bibfield  {journal} {\bibinfo  {journal}
  {\apj},\ }\textbf {\bibinfo {volume} {266}},\ \bibinfo {pages} {713}
  (\bibinfo {year} {1983})}\BibitemShut {NoStop}%
\bibitem [{\citenamefont {{Eisenstein}}\ \emph {et~al.}(2006)\citenamefont
  {{Eisenstein}}, \citenamefont {{Liebert}}, \citenamefont {{Harris}},
  \citenamefont {{Kleinman}}, \citenamefont {{Nitta}}, \citenamefont
  {{Silvestri}}, \citenamefont {{Anderson}}, \citenamefont {{Barentine}},
  \citenamefont {{Brewington}}, \citenamefont {{Brinkmann}},\ and\
  \citenamefont {et~al.}}]{Eisenstein2006}%
  \BibitemOpen
  \bibfield  {author} {\bibinfo {author} {\bibfnamefont {D.~J.}\ \bibnamefont
  {{Eisenstein}}}, \bibinfo {author} {\bibfnamefont {J.}~\bibnamefont
  {{Liebert}}}, \bibinfo {author} {\bibfnamefont {H.~C.}\ \bibnamefont
  {{Harris}}}, \bibinfo {author} {\bibfnamefont {S.~J.}\ \bibnamefont
  {{Kleinman}}}, \bibinfo {author} {\bibfnamefont {A.}~\bibnamefont {{Nitta}}},
  \bibinfo {author} {\bibfnamefont {N.}~\bibnamefont {{Silvestri}}}, \bibinfo
  {author} {\bibfnamefont {S.~A.}\ \bibnamefont {{Anderson}}}, \bibinfo
  {author} {\bibfnamefont {J.~C.}\ \bibnamefont {{Barentine}}}, \bibinfo
  {author} {\bibfnamefont {H.~J.}\ \bibnamefont {{Brewington}}}, \bibinfo
  {author} {\bibfnamefont {J.}~\bibnamefont {{Brinkmann}}}, \ and\ \bibinfo
  {author} {\bibnamefont {et~al.}},\ }\Doi {10.1086/507110} {\bibfield
  {journal} {\bibinfo  {journal} {\apjs},\ }\textbf {\bibinfo {volume} {167}},\
  \bibinfo {pages} {40} (\bibinfo {year} {2006})},\ \Eprint
  {http://arxiv.org/abs/astro-ph/0606700} {astro-ph/0606700} \BibitemShut
  {NoStop}%
\bibitem [{\citenamefont {{Blanton}}\ \emph
  {et~al.}(2003){\natexlab{a}}\citenamefont {{Blanton}}, \citenamefont
  {{Brinkmann}}, \citenamefont {{Csabai}}, \citenamefont {{Doi}}, \citenamefont
  {{Eisenstein}}, \citenamefont {{Fukugita}}, \citenamefont {{Gunn}},
  \citenamefont {{Hogg}},\ and\ \citenamefont {{Schlegel}}}]{Blanton2003B}%
  \BibitemOpen
  \bibfield  {author} {\bibinfo {author} {\bibfnamefont {M.~R.}\ \bibnamefont
  {{Blanton}}}, \bibinfo {author} {\bibfnamefont {J.}~\bibnamefont
  {{Brinkmann}}}, \bibinfo {author} {\bibfnamefont {I.}~\bibnamefont
  {{Csabai}}}, \bibinfo {author} {\bibfnamefont {M.}~\bibnamefont {{Doi}}},
  \bibinfo {author} {\bibfnamefont {D.}~\bibnamefont {{Eisenstein}}}, \bibinfo
  {author} {\bibfnamefont {M.}~\bibnamefont {{Fukugita}}}, \bibinfo {author}
  {\bibfnamefont {J.~E.}\ \bibnamefont {{Gunn}}}, \bibinfo {author}
  {\bibfnamefont {D.~W.}\ \bibnamefont {{Hogg}}}, \ and\ \bibinfo {author}
  {\bibfnamefont {D.~J.}\ \bibnamefont {{Schlegel}}},\ }\Doi {10.1086/342935}
  {\bibfield  {journal} {\bibinfo  {journal} {\aj},\ }\textbf {\bibinfo
  {volume} {125}},\ \bibinfo {pages} {2348} (\bibinfo {year}
  {2003}{\natexlab{a}})},\ \Eprint {http://arxiv.org/abs/astro-ph/0205243}
  {astro-ph/0205243} \BibitemShut {NoStop}%
\bibitem [{\citenamefont {{McBride}}\ \emph {et~al.}(2009)\citenamefont
  {{McBride}}, \citenamefont {{Berlind}}, \citenamefont {{Scoccimarro}},
  \citenamefont {{Wechsler}}, \citenamefont {{Busha}}, \citenamefont
  {{Gardner}},\ and\ \citenamefont {{van den Bosch}}}]{McBride2009}%
  \BibitemOpen
  \bibfield  {author} {\bibinfo {author} {\bibfnamefont {C.}~\bibnamefont
  {{McBride}}}, \bibinfo {author} {\bibfnamefont {A.}~\bibnamefont
  {{Berlind}}}, \bibinfo {author} {\bibfnamefont {R.}~\bibnamefont
  {{Scoccimarro}}}, \bibinfo {author} {\bibfnamefont {R.}~\bibnamefont
  {{Wechsler}}}, \bibinfo {author} {\bibfnamefont {M.}~\bibnamefont {{Busha}}},
  \bibinfo {author} {\bibfnamefont {J.}~\bibnamefont {{Gardner}}}, \ and\
  \bibinfo {author} {\bibfnamefont {F.}~\bibnamefont {{van den Bosch}}},\ }in\
  \href@noop {} {\emph {\bibinfo {booktitle} {American Astronomical Society
  Meeting Abstracts \#213}}},\ \bibinfo {series} {Bulletin of the American
  Astronomical Society}, Vol.~\bibinfo {volume} {41}\ (\bibinfo {year} {2009})\
  p.\ \bibinfo {pages} {425.06}\BibitemShut {NoStop}%
\bibitem [{\citenamefont {{Peacock}}\ and\ \citenamefont
  {{Smith}}(2000)}]{pesm}%
  \BibitemOpen
  \bibfield  {author} {\bibinfo {author} {\bibfnamefont {J.~A.}\ \bibnamefont
  {{Peacock}}}\ and\ \bibinfo {author} {\bibfnamefont {R.~E.}\ \bibnamefont
  {{Smith}}},\ }\Doi {10.1046/j.1365-8711.2000.03779.x} {\bibfield  {journal}
  {\bibinfo  {journal} {\mnras},\ }\textbf {\bibinfo {volume} {318}},\ \bibinfo
  {pages} {1144} (\bibinfo {year} {2000})},\ \Eprint
  {http://arxiv.org/abs/astro-ph/0005010} {astro-ph/0005010} \BibitemShut
  {NoStop}%
\bibitem [{\citenamefont {{Seljak}}(2000)}]{Seljak2000}%
  \BibitemOpen
  \bibfield  {author} {\bibinfo {author} {\bibfnamefont {U.}~\bibnamefont
  {{Seljak}}},\ }\Doi {10.1046/j.1365-8711.2000.03715.x} {\bibfield  {journal}
  {\bibinfo  {journal} {\mnras},\ }\textbf {\bibinfo {volume} {318}},\ \bibinfo
  {pages} {203} (\bibinfo {year} {2000})},\ \Eprint
  {http://arxiv.org/abs/astro-ph/0001493} {astro-ph/0001493} \BibitemShut
  {NoStop}%
\bibitem [{\citenamefont {{Berlind}}\ and\ \citenamefont
  {{Weinberg}}(2002)}]{Berlind2002}%
  \BibitemOpen
  \bibfield  {author} {\bibinfo {author} {\bibfnamefont {A.~A.}\ \bibnamefont
  {{Berlind}}}\ and\ \bibinfo {author} {\bibfnamefont {D.~H.}\ \bibnamefont
  {{Weinberg}}},\ }\Doi {10.1086/341469} {\bibfield  {journal} {\bibinfo
  {journal} {\apj},\ }\textbf {\bibinfo {volume} {575}},\ \bibinfo {pages}
  {587} (\bibinfo {year} {2002})},\ \Eprint
  {http://arxiv.org/abs/astro-ph/0109001} {astro-ph/0109001} \BibitemShut
  {NoStop}%
\bibitem [{\citenamefont {{Blanton}}\ \emph
  {et~al.}(2003){\natexlab{b}}\citenamefont {{Blanton}}, \citenamefont
  {{Hogg}}, \citenamefont {{Bahcall}}, \citenamefont {{Brinkmann}},
  \citenamefont {{Britton}}, \citenamefont {{Connolly}}, \citenamefont
  {{Csabai}}, \citenamefont {{Fukugita}}, \citenamefont {{Loveday}},
  \citenamefont {{Meiksin}},\ and\ \citenamefont {et~al.}}]{Blanton2003}%
  \BibitemOpen
  \bibfield  {author} {\bibinfo {author} {\bibfnamefont {M.~R.}\ \bibnamefont
  {{Blanton}}}, \bibinfo {author} {\bibfnamefont {D.~W.}\ \bibnamefont
  {{Hogg}}}, \bibinfo {author} {\bibfnamefont {N.~A.}\ \bibnamefont
  {{Bahcall}}}, \bibinfo {author} {\bibfnamefont {J.}~\bibnamefont
  {{Brinkmann}}}, \bibinfo {author} {\bibfnamefont {M.}~\bibnamefont
  {{Britton}}}, \bibinfo {author} {\bibfnamefont {A.~J.}\ \bibnamefont
  {{Connolly}}}, \bibinfo {author} {\bibfnamefont {I.}~\bibnamefont
  {{Csabai}}}, \bibinfo {author} {\bibfnamefont {M.}~\bibnamefont
  {{Fukugita}}}, \bibinfo {author} {\bibfnamefont {J.}~\bibnamefont
  {{Loveday}}}, \bibinfo {author} {\bibfnamefont {A.}~\bibnamefont
  {{Meiksin}}}, \ and\ \bibinfo {author} {\bibnamefont {et~al.}},\ }\Doi
  {10.1086/375776} {\bibfield  {journal} {\bibinfo  {journal} {\apj},\ }\textbf
  {\bibinfo {volume} {592}},\ \bibinfo {pages} {819} (\bibinfo {year}
  {2003}{\natexlab{b}})},\ \Eprint {http://arxiv.org/abs/astro-ph/0210215}
  {astro-ph/0210215} \BibitemShut {NoStop}%
\bibitem [{\citenamefont {{Padmanabhan}}\ \emph {et~al.}(2008)\citenamefont
  {{Padmanabhan}}, \citenamefont {{Schlegel}}, \citenamefont {{Finkbeiner}},
  \citenamefont {{Barentine}}, \citenamefont {{Blanton}}, \citenamefont
  {{Brewington}}, \citenamefont {{Gunn}}, \citenamefont {{Harvanek}},
  \citenamefont {{Hogg}}, \citenamefont {{Ivezi{\'c}}},\ and\ \citenamefont
  {et~al.}}]{Pad2008}%
  \BibitemOpen
  \bibfield  {author} {\bibinfo {author} {\bibfnamefont {N.}~\bibnamefont
  {{Padmanabhan}}}, \bibinfo {author} {\bibfnamefont {D.~J.}\ \bibnamefont
  {{Schlegel}}}, \bibinfo {author} {\bibfnamefont {D.~P.}\ \bibnamefont
  {{Finkbeiner}}}, \bibinfo {author} {\bibfnamefont {J.~C.}\ \bibnamefont
  {{Barentine}}}, \bibinfo {author} {\bibfnamefont {M.~R.}\ \bibnamefont
  {{Blanton}}}, \bibinfo {author} {\bibfnamefont {H.~J.}\ \bibnamefont
  {{Brewington}}}, \bibinfo {author} {\bibfnamefont {J.~E.}\ \bibnamefont
  {{Gunn}}}, \bibinfo {author} {\bibfnamefont {M.}~\bibnamefont {{Harvanek}}},
  \bibinfo {author} {\bibfnamefont {D.~W.}\ \bibnamefont {{Hogg}}}, \bibinfo
  {author} {\bibfnamefont {{\v Z}.}~\bibnamefont {{Ivezi{\'c}}}}, \ and\
  \bibinfo {author} {\bibnamefont {et~al.}},\ }\Doi {10.1086/524677} {\bibfield
   {journal} {\bibinfo  {journal} {\apj},\ }\textbf {\bibinfo {volume} {674}},\
  \bibinfo {pages} {1217} (\bibinfo {year} {2008})},\ \Eprint
  {http://arxiv.org/abs/astro-ph/0703454} {astro-ph/0703454} \BibitemShut
  {NoStop}%
\bibitem [{\citenamefont {{Tegmark}}\ \emph {et~al.}(2004)\citenamefont
  {{Tegmark}}, \citenamefont {{Blanton}}, \citenamefont {{Strauss}},
  \citenamefont {{Hoyle}}, \citenamefont {{Schlegel}}, \citenamefont
  {{Scoccimarro}}, \citenamefont {{Vogeley}}, \citenamefont {{Weinberg}},
  \citenamefont {{Zehavi}}, \citenamefont {{Berlind}},\ and\ \citenamefont
  {et~al.}}]{Tegmark2004}%
  \BibitemOpen
  \bibfield  {author} {\bibinfo {author} {\bibfnamefont {M.}~\bibnamefont
  {{Tegmark}}}, \bibinfo {author} {\bibfnamefont {M.~R.}\ \bibnamefont
  {{Blanton}}}, \bibinfo {author} {\bibfnamefont {M.~A.}\ \bibnamefont
  {{Strauss}}}, \bibinfo {author} {\bibfnamefont {F.}~\bibnamefont {{Hoyle}}},
  \bibinfo {author} {\bibfnamefont {D.}~\bibnamefont {{Schlegel}}}, \bibinfo
  {author} {\bibfnamefont {R.}~\bibnamefont {{Scoccimarro}}}, \bibinfo {author}
  {\bibfnamefont {M.~S.}\ \bibnamefont {{Vogeley}}}, \bibinfo {author}
  {\bibfnamefont {D.~H.}\ \bibnamefont {{Weinberg}}}, \bibinfo {author}
  {\bibfnamefont {I.}~\bibnamefont {{Zehavi}}}, \bibinfo {author}
  {\bibfnamefont {A.}~\bibnamefont {{Berlind}}}, \ and\ \bibinfo {author}
  {\bibnamefont {et~al.}},\ }\Doi {10.1086/382125} {\bibfield  {journal}
  {\bibinfo  {journal} {\apj},\ }\textbf {\bibinfo {volume} {606}},\ \bibinfo
  {pages} {702} (\bibinfo {year} {2004})},\ \Eprint
  {http://arxiv.org/abs/astro-ph/0310725} {astro-ph/0310725} \BibitemShut
  {NoStop}%
\bibitem [{\citenamefont {{James}}(2006)}]{James2006}%
  \BibitemOpen
  \bibfield  {author} {\bibinfo {author} {\bibfnamefont {F.}~\bibnamefont
  {{James}}},\ }\href@noop {} {\emph {\bibinfo {title} {Statistical Methods in
  Experimental Physics: 2nd Edition, by Frederick James.~ISBN-10 981-256-795-X;
  ISBN-13 981-270-527-9.~Published by World Scientific Publishing Co.,
  Pte.~Ltd., Singapore, 2006.}}}\ (\bibinfo  {publisher} {World Scientific
  Publishing Co},\ \bibinfo {year} {2006})\BibitemShut {NoStop}%
\bibitem [{\citenamefont {{Calabrese}}\ \emph {et~al.}(2013)\citenamefont
  {{Calabrese}}, \citenamefont {{Hlozek}}, \citenamefont {{Battaglia}},
  \citenamefont {{Battistelli}}, \citenamefont {{Bond}}, \citenamefont
  {{Chluba}}, \citenamefont {{Crichton}}, \citenamefont {{Das}}, \citenamefont
  {{Devlin}}, \citenamefont {{Dunkley}},\ and\ \citenamefont
  {et~al.}}]{Calabrese2013}%
  \BibitemOpen
  \bibfield  {author} {\bibinfo {author} {\bibfnamefont {E.}~\bibnamefont
  {{Calabrese}}}, \bibinfo {author} {\bibfnamefont {R.~A.}\ \bibnamefont
  {{Hlozek}}}, \bibinfo {author} {\bibfnamefont {N.}~\bibnamefont
  {{Battaglia}}}, \bibinfo {author} {\bibfnamefont {E.~S.}\ \bibnamefont
  {{Battistelli}}}, \bibinfo {author} {\bibfnamefont {J.~R.}\ \bibnamefont
  {{Bond}}}, \bibinfo {author} {\bibfnamefont {J.}~\bibnamefont {{Chluba}}},
  \bibinfo {author} {\bibfnamefont {D.}~\bibnamefont {{Crichton}}}, \bibinfo
  {author} {\bibfnamefont {S.}~\bibnamefont {{Das}}}, \bibinfo {author}
  {\bibfnamefont {M.~J.}\ \bibnamefont {{Devlin}}}, \bibinfo {author}
  {\bibfnamefont {J.}~\bibnamefont {{Dunkley}}}, \ and\ \bibinfo {author}
  {\bibnamefont {et~al.}},\ }\Doi {10.1103/PhysRevD.87.103012} {\bibfield
  {journal} {\bibinfo  {journal} {\prd},\ }\textbf {\bibinfo {volume} {87}},\
  \bibinfo {eid} {103012} (\bibinfo {year} {2013})},\ \Eprint
  {http://arxiv.org/abs/1302.1841} {arXiv:1302.1841 [astro-ph.CO]} \BibitemShut
  {NoStop}%
\bibitem [{\citenamefont {{Montero-Dorta}}\ and\ \citenamefont
  {{Prada}}(2009)}]{Montero2009}%
  \BibitemOpen
  \bibfield  {author} {\bibinfo {author} {\bibfnamefont {A.~D.}\ \bibnamefont
  {{Montero-Dorta}}}\ and\ \bibinfo {author} {\bibfnamefont {F.}~\bibnamefont
  {{Prada}}},\ }\Doi {10.1111/j.1365-2966.2009.15197.x} {\bibfield  {journal}
  {\bibinfo  {journal} {\mnras},\ }\textbf {\bibinfo {volume} {399}},\ \bibinfo
  {pages} {1106} (\bibinfo {year} {2009})},\ \Eprint
  {http://arxiv.org/abs/0806.4930} {arXiv:0806.4930} \BibitemShut {NoStop}%
\bibitem [{\citenamefont {{Kashlinsky}}\ \emph {et~al.}(2008)\citenamefont
  {{Kashlinsky}}, \citenamefont {{Atrio-Barandela}}, \citenamefont
  {{Kocevski}},\ and\ \citenamefont {{Ebeling}}}]{Kash2008}%
  \BibitemOpen
  \bibfield  {author} {\bibinfo {author} {\bibfnamefont {A.}~\bibnamefont
  {{Kashlinsky}}}, \bibinfo {author} {\bibfnamefont {F.}~\bibnamefont
  {{Atrio-Barandela}}}, \bibinfo {author} {\bibfnamefont {D.}~\bibnamefont
  {{Kocevski}}}, \ and\ \bibinfo {author} {\bibfnamefont {H.}~\bibnamefont
  {{Ebeling}}},\ }\Doi {10.1086/592947} {\bibfield  {journal} {\bibinfo
  {journal} {\apjl},\ }\textbf {\bibinfo {volume} {686}},\ \bibinfo {pages}
  {L49} (\bibinfo {year} {2008})},\ \Eprint {http://arxiv.org/abs/0809.3734}
  {arXiv:0809.3734} \BibitemShut {NoStop}%
\bibitem [{\citenamefont {{Keisler}}(2009)}]{Keisler2009}%
  \BibitemOpen
  \bibfield  {author} {\bibinfo {author} {\bibfnamefont {R.}~\bibnamefont
  {{Keisler}}},\ }\Doi {10.1088/0004-637X/707/1/L42} {\bibfield  {journal}
  {\bibinfo  {journal} {\apjl},\ }\textbf {\bibinfo {volume} {707}},\ \bibinfo
  {pages} {L42} (\bibinfo {year} {2009})},\ \Eprint
  {http://arxiv.org/abs/0910.4233} {arXiv:0910.4233 [astro-ph.CO]} \BibitemShut
  {NoStop}%
\bibitem [{\citenamefont {{Kashlinsky}}\ \emph {et~al.}(2010)\citenamefont
  {{Kashlinsky}}, \citenamefont {{Atrio-Barandela}}, \citenamefont {{Ebeling}},
  \citenamefont {{Edge}},\ and\ \citenamefont {{Kocevski}}}]{Kash2010}%
  \BibitemOpen
  \bibfield  {author} {\bibinfo {author} {\bibfnamefont {A.}~\bibnamefont
  {{Kashlinsky}}}, \bibinfo {author} {\bibfnamefont {F.}~\bibnamefont
  {{Atrio-Barandela}}}, \bibinfo {author} {\bibfnamefont {H.}~\bibnamefont
  {{Ebeling}}}, \bibinfo {author} {\bibfnamefont {A.}~\bibnamefont {{Edge}}}, \
  and\ \bibinfo {author} {\bibfnamefont {D.}~\bibnamefont {{Kocevski}}},\ }\Doi
  {10.1088/2041-8205/712/1/L81} {\bibfield  {journal} {\bibinfo  {journal}
  {\apjl},\ }\textbf {\bibinfo {volume} {712}},\ \bibinfo {pages} {L81}
  (\bibinfo {year} {2010})},\ \Eprint {http://arxiv.org/abs/0910.4958}
  {arXiv:0910.4958 [astro-ph.CO]} \BibitemShut {NoStop}%
\bibitem [{\citenamefont {{Eisenstein}}\ and\ \citenamefont
  {{Hu}}(1998)}]{EH98}%
  \BibitemOpen
  \bibfield  {author} {\bibinfo {author} {\bibfnamefont {D.~J.}\ \bibnamefont
  {{Eisenstein}}}\ and\ \bibinfo {author} {\bibfnamefont {W.}~\bibnamefont
  {{Hu}}},\ }\Doi {10.1086/305424} {\bibfield  {journal} {\bibinfo  {journal}
  {\apj},\ }\textbf {\bibinfo {volume} {496}},\ \bibinfo {pages} {605}
  (\bibinfo {year} {1998})},\ \Eprint {http://arxiv.org/abs/astro-ph/9709112}
  {astro-ph/9709112} \BibitemShut {NoStop}%
\bibitem [{\citenamefont {{Kaiser}}\ \emph {et~al.}(2002)\citenamefont
  {{Kaiser}}, \citenamefont {{Aussel}}, \citenamefont {{Burke}}, \citenamefont
  {{Boesgaard}}, \citenamefont {{Chambers}}, \citenamefont {{Chun}},
  \citenamefont {{Heasley}}, \citenamefont {{Hodapp}}, \citenamefont {{Hunt}},
  \citenamefont {{Jedicke}},\ and\ \citenamefont {et~al.}}]{Kaiser2002}%
  \BibitemOpen
  \bibfield  {author} {\bibinfo {author} {\bibfnamefont {N.}~\bibnamefont
  {{Kaiser}}}, \bibinfo {author} {\bibfnamefont {H.}~\bibnamefont {{Aussel}}},
  \bibinfo {author} {\bibfnamefont {B.~E.}\ \bibnamefont {{Burke}}}, \bibinfo
  {author} {\bibfnamefont {H.}~\bibnamefont {{Boesgaard}}}, \bibinfo {author}
  {\bibfnamefont {K.}~\bibnamefont {{Chambers}}}, \bibinfo {author}
  {\bibfnamefont {M.~R.}\ \bibnamefont {{Chun}}}, \bibinfo {author}
  {\bibfnamefont {J.~N.}\ \bibnamefont {{Heasley}}}, \bibinfo {author}
  {\bibfnamefont {K.-W.}\ \bibnamefont {{Hodapp}}}, \bibinfo {author}
  {\bibfnamefont {B.}~\bibnamefont {{Hunt}}}, \bibinfo {author} {\bibfnamefont
  {R.}~\bibnamefont {{Jedicke}}}, \ and\ \bibinfo {author} {\bibnamefont
  {et~al.}},\ }in\ \Doi {10.1117/12.457365} {\emph {\bibinfo {booktitle}
  {Survey and Other Telescope Technologies and Discoveries}}},\ \bibinfo
  {series} {Society of Photo-Optical Instrumentation Engineers (SPIE)
  Conference Series}, Vol.\ \bibinfo {volume} {4836},\ \bibinfo {editor}
  {edited by\ \bibinfo {editor} {\bibfnamefont {J.~A.}\ \bibnamefont
  {{Tyson}}}\ and\ \bibinfo {editor} {\bibfnamefont {S.}~\bibnamefont
  {{Wolff}}}}\ (\bibinfo {year} {2002})\ pp.\ \bibinfo {pages}
  {154--164}\BibitemShut {NoStop}%
\bibitem [{\citenamefont {{Kaiser}}\ \emph {et~al.}(2010)\citenamefont
  {{Kaiser}}, \citenamefont {{Burgett}}, \citenamefont {{Chambers}},
  \citenamefont {{Denneau}}, \citenamefont {{Heasley}}, \citenamefont
  {{Jedicke}}, \citenamefont {{Magnier}}, \citenamefont {{Morgan}},
  \citenamefont {{Onaka}},\ and\ \citenamefont {{Tonry}}}]{Kaiser2010}%
  \BibitemOpen
  \bibfield  {author} {\bibinfo {author} {\bibfnamefont {N.}~\bibnamefont
  {{Kaiser}}}, \bibinfo {author} {\bibfnamefont {W.}~\bibnamefont {{Burgett}}},
  \bibinfo {author} {\bibfnamefont {K.}~\bibnamefont {{Chambers}}}, \bibinfo
  {author} {\bibfnamefont {L.}~\bibnamefont {{Denneau}}}, \bibinfo {author}
  {\bibfnamefont {J.}~\bibnamefont {{Heasley}}}, \bibinfo {author}
  {\bibfnamefont {R.}~\bibnamefont {{Jedicke}}}, \bibinfo {author}
  {\bibfnamefont {E.}~\bibnamefont {{Magnier}}}, \bibinfo {author}
  {\bibfnamefont {J.}~\bibnamefont {{Morgan}}}, \bibinfo {author}
  {\bibfnamefont {P.}~\bibnamefont {{Onaka}}}, \ and\ \bibinfo {author}
  {\bibfnamefont {J.}~\bibnamefont {{Tonry}}},\ }in\ \Doi {10.1117/12.859188}
  {\emph {\bibinfo {booktitle} {Ground-based and Airborne Telescopes III}}},\
  \bibinfo {series} {Society of Photo-Optical Instrumentation Engineers (SPIE)
  Conference Series}, Vol.\ \bibinfo {volume} {7733}\ (\bibinfo {year}
  {2010})\BibitemShut {NoStop}%
\bibitem [{\citenamefont {{Laureijs}}\ \emph {et~al.}(2011)\citenamefont
  {{Laureijs}}, \citenamefont {{Amiaux}}, \citenamefont {{Arduini}},
  \citenamefont {{Augu{\`e}res}}, \citenamefont {{Brinchmann}}, \citenamefont
  {{Cole}}, \citenamefont {{Cropper}}, \citenamefont {{Dabin}}, \citenamefont
  {{Duvet}}, \citenamefont {{Ealet}},\ and\ \citenamefont
  {et~al.}}]{euclid2011}%
  \BibitemOpen
  \bibfield  {author} {\bibinfo {author} {\bibfnamefont {R.}~\bibnamefont
  {{Laureijs}}}, \bibinfo {author} {\bibfnamefont {J.}~\bibnamefont
  {{Amiaux}}}, \bibinfo {author} {\bibfnamefont {S.}~\bibnamefont {{Arduini}}},
  \bibinfo {author} {\bibfnamefont {J.~.}\ \bibnamefont {{Augu{\`e}res}}},
  \bibinfo {author} {\bibfnamefont {J.}~\bibnamefont {{Brinchmann}}}, \bibinfo
  {author} {\bibfnamefont {R.}~\bibnamefont {{Cole}}}, \bibinfo {author}
  {\bibfnamefont {M.}~\bibnamefont {{Cropper}}}, \bibinfo {author}
  {\bibfnamefont {C.}~\bibnamefont {{Dabin}}}, \bibinfo {author} {\bibfnamefont
  {L.}~\bibnamefont {{Duvet}}}, \bibinfo {author} {\bibfnamefont
  {A.}~\bibnamefont {{Ealet}}}, \ and\ \bibinfo {author} {\bibnamefont
  {et~al.}},\ }\href@noop {} {\bibfield  {journal} {\bibinfo  {journal} {ArXiv
  e-prints}} (\bibinfo {year} {2011})},\ \Eprint
  {http://arxiv.org/abs/1110.3193} {arXiv:1110.3193 [astro-ph.CO]} \BibitemShut
  {NoStop}%
\bibitem [{\citenamefont {{Schlafly}}\ \emph {et~al.}(2014)\citenamefont
  {{Schlafly}}, \citenamefont {{Green}}, \citenamefont {{Finkbeiner}},
  \citenamefont {{Juric}}, \citenamefont {{Rix}}, \citenamefont {{Martin}},
  \citenamefont {{Burgett}}, \citenamefont {{Chambers}}, \citenamefont
  {{Draper}}, \citenamefont {{Hodapp}},\ and\ \citenamefont
  {et~al.}}]{Schlafly2014}%
  \BibitemOpen
  \bibfield  {author} {\bibinfo {author} {\bibfnamefont {E.~F.}\ \bibnamefont
  {{Schlafly}}}, \bibinfo {author} {\bibfnamefont {G.}~\bibnamefont {{Green}}},
  \bibinfo {author} {\bibfnamefont {D.~P.}\ \bibnamefont {{Finkbeiner}}},
  \bibinfo {author} {\bibfnamefont {M.}~\bibnamefont {{Juric}}}, \bibinfo
  {author} {\bibfnamefont {H.-W.}\ \bibnamefont {{Rix}}}, \bibinfo {author}
  {\bibfnamefont {N.~F.}\ \bibnamefont {{Martin}}}, \bibinfo {author}
  {\bibfnamefont {W.~S.}\ \bibnamefont {{Burgett}}}, \bibinfo {author}
  {\bibfnamefont {K.~C.}\ \bibnamefont {{Chambers}}}, \bibinfo {author}
  {\bibfnamefont {P.~W.}\ \bibnamefont {{Draper}}}, \bibinfo {author}
  {\bibfnamefont {K.~W.}\ \bibnamefont {{Hodapp}}}, \ and\ \bibinfo {author}
  {\bibnamefont {et~al.}},\ }\href@noop {} {\bibfield  {journal} {\bibinfo
  {journal} {ArXiv e-prints}} (\bibinfo {year} {2014})},\ \Eprint
  {http://arxiv.org/abs/1405.2922} {arXiv:1405.2922} \BibitemShut {NoStop}%
\bibitem [{\citenamefont {{Balogh}}\ \emph {et~al.}(2001)\citenamefont
  {{Balogh}}, \citenamefont {{Christlein}}, \citenamefont {{Zabludoff}},\ and\
  \citenamefont {{Zaritsky}}}]{Balogh2001}%
  \BibitemOpen
  \bibfield  {author} {\bibinfo {author} {\bibfnamefont {M.~L.}\ \bibnamefont
  {{Balogh}}}, \bibinfo {author} {\bibfnamefont {D.}~\bibnamefont
  {{Christlein}}}, \bibinfo {author} {\bibfnamefont {A.~I.}\ \bibnamefont
  {{Zabludoff}}}, \ and\ \bibinfo {author} {\bibfnamefont {D.}~\bibnamefont
  {{Zaritsky}}},\ }\Doi {10.1086/321670} {\bibfield  {journal} {\bibinfo
  {journal} {\apj},\ }\textbf {\bibinfo {volume} {557}},\ \bibinfo {pages}
  {117} (\bibinfo {year} {2001})},\ \Eprint
  {http://arxiv.org/abs/astro-ph/0104042} {astro-ph/0104042} \BibitemShut
  {NoStop}%
\bibitem [{\citenamefont {{Mo}}\ \emph {et~al.}(2004)\citenamefont {{Mo}},
  \citenamefont {{Yang}}, \citenamefont {{van den Bosch}},\ and\ \citenamefont
  {{Jing}}}]{Mo2004}%
  \BibitemOpen
  \bibfield  {author} {\bibinfo {author} {\bibfnamefont {H.~J.}\ \bibnamefont
  {{Mo}}}, \bibinfo {author} {\bibfnamefont {X.}~\bibnamefont {{Yang}}},
  \bibinfo {author} {\bibfnamefont {F.~C.}\ \bibnamefont {{van den Bosch}}}, \
  and\ \bibinfo {author} {\bibfnamefont {Y.~P.}\ \bibnamefont {{Jing}}},\ }\Doi
  {10.1111/j.1365-2966.2004.07485.x} {\bibfield  {journal} {\bibinfo  {journal}
  {\mnras},\ }\textbf {\bibinfo {volume} {349}},\ \bibinfo {pages} {205}
  (\bibinfo {year} {2004})},\ \Eprint {http://arxiv.org/abs/astro-ph/0310147}
  {astro-ph/0310147} \BibitemShut {NoStop}%
\bibitem [{\citenamefont {{Croton}}\ \emph {et~al.}(2005)\citenamefont
  {{Croton}}, \citenamefont {{Farrar}}, \citenamefont {{Norberg}},
  \citenamefont {{Colless}}, \citenamefont {{Peacock}}, \citenamefont
  {{Baldry}}, \citenamefont {{Baugh}}, \citenamefont {{Bland-Hawthorn}},
  \citenamefont {{Bridges}}, \citenamefont {{Cannon}},\ and\ \citenamefont
  {et~al.}}]{Croton2005}%
  \BibitemOpen
  \bibfield  {author} {\bibinfo {author} {\bibfnamefont {D.~J.}\ \bibnamefont
  {{Croton}}}, \bibinfo {author} {\bibfnamefont {G.~R.}\ \bibnamefont
  {{Farrar}}}, \bibinfo {author} {\bibfnamefont {P.}~\bibnamefont {{Norberg}}},
  \bibinfo {author} {\bibfnamefont {M.}~\bibnamefont {{Colless}}}, \bibinfo
  {author} {\bibfnamefont {J.~A.}\ \bibnamefont {{Peacock}}}, \bibinfo {author}
  {\bibfnamefont {I.~K.}\ \bibnamefont {{Baldry}}}, \bibinfo {author}
  {\bibfnamefont {C.~M.}\ \bibnamefont {{Baugh}}}, \bibinfo {author}
  {\bibfnamefont {J.}~\bibnamefont {{Bland-Hawthorn}}}, \bibinfo {author}
  {\bibfnamefont {T.}~\bibnamefont {{Bridges}}}, \bibinfo {author}
  {\bibfnamefont {R.}~\bibnamefont {{Cannon}}}, \ and\ \bibinfo {author}
  {\bibnamefont {et~al.}},\ }\Doi {10.1111/j.1365-2966.2004.08546.x} {\bibfield
   {journal} {\bibinfo  {journal} {\mnras},\ }\textbf {\bibinfo {volume}
  {356}},\ \bibinfo {pages} {1155} (\bibinfo {year} {2005})},\ \Eprint
  {http://arxiv.org/abs/astro-ph/0407537} {astro-ph/0407537} \BibitemShut
  {NoStop}%
\bibitem [{\citenamefont {{Park}}\ \emph {et~al.}(2007)\citenamefont {{Park}},
  \citenamefont {{Choi}}, \citenamefont {{Vogeley}}, \citenamefont {{Gott}},
  \citenamefont {{Blanton}},\ and\ \citenamefont {{SDSS
  Collaboration}}}]{Park2007}%
  \BibitemOpen
  \bibfield  {author} {\bibinfo {author} {\bibfnamefont {C.}~\bibnamefont
  {{Park}}}, \bibinfo {author} {\bibfnamefont {Y.-Y.}\ \bibnamefont {{Choi}}},
  \bibinfo {author} {\bibfnamefont {M.~S.}\ \bibnamefont {{Vogeley}}}, \bibinfo
  {author} {\bibfnamefont {J.~R.}\ \bibnamefont {{Gott}}, \bibfnamefont {III}},
  \bibinfo {author} {\bibfnamefont {M.~R.}\ \bibnamefont {{Blanton}}}, \ and\
  \bibinfo {author} {\bibnamefont {{SDSS Collaboration}}},\ }\Doi
  {10.1086/511059} {\bibfield  {journal} {\bibinfo  {journal} {\apj},\ }\textbf
  {\bibinfo {volume} {658}},\ \bibinfo {pages} {898} (\bibinfo {year}
  {2007})},\ \Eprint {http://arxiv.org/abs/astro-ph/0611610} {astro-ph/0611610}
  \BibitemShut {NoStop}%
\bibitem [{\citenamefont {{Merluzzi}}\ \emph {et~al.}(2010)\citenamefont
  {{Merluzzi}}, \citenamefont {{Mercurio}}, \citenamefont {{Haines}},
  \citenamefont {{Smith}}, \citenamefont {{Busarello}},\ and\ \citenamefont
  {{Lucey}}}]{Merluzzi2010}%
  \BibitemOpen
  \bibfield  {author} {\bibinfo {author} {\bibfnamefont {P.}~\bibnamefont
  {{Merluzzi}}}, \bibinfo {author} {\bibfnamefont {A.}~\bibnamefont
  {{Mercurio}}}, \bibinfo {author} {\bibfnamefont {C.~P.}\ \bibnamefont
  {{Haines}}}, \bibinfo {author} {\bibfnamefont {R.~J.}\ \bibnamefont
  {{Smith}}}, \bibinfo {author} {\bibfnamefont {G.}~\bibnamefont
  {{Busarello}}}, \ and\ \bibinfo {author} {\bibfnamefont {J.~R.}\ \bibnamefont
  {{Lucey}}},\ }\Doi {10.1111/j.1365-2966.2009.15929.x} {\bibfield  {journal}
  {\bibinfo  {journal} {\mnras},\ }\textbf {\bibinfo {volume} {402}},\ \bibinfo
  {pages} {753} (\bibinfo {year} {2010})},\ \Eprint
  {http://arxiv.org/abs/0910.3877} {arXiv:0910.3877 [astro-ph.CO]} \BibitemShut
  {NoStop}%
\bibitem [{\citenamefont {{Faltenbacher}}\ \emph {et~al.}(2010)\citenamefont
  {{Faltenbacher}}, \citenamefont {{Finoguenov}},\ and\ \citenamefont
  {{Drory}}}]{Faltenbacher2010}%
  \BibitemOpen
  \bibfield  {author} {\bibinfo {author} {\bibfnamefont {A.}~\bibnamefont
  {{Faltenbacher}}}, \bibinfo {author} {\bibfnamefont {A.}~\bibnamefont
  {{Finoguenov}}}, \ and\ \bibinfo {author} {\bibfnamefont {N.}~\bibnamefont
  {{Drory}}},\ }\Doi {10.1088/0004-637X/712/1/484} {\bibfield  {journal}
  {\bibinfo  {journal} {\apj},\ }\textbf {\bibinfo {volume} {712}},\ \bibinfo
  {pages} {484} (\bibinfo {year} {2010})},\ \Eprint
  {http://arxiv.org/abs/1002.0844} {arXiv:1002.0844 [astro-ph.CO]} \BibitemShut
  {NoStop}%
\bibitem [{\citenamefont {{Nusser}}\ \emph {et~al.}(2013)\citenamefont
  {{Nusser}}, \citenamefont {{Branchini}},\ and\ \citenamefont
  {{Feix}}}]{Nusser2013}%
  \BibitemOpen
  \bibfield  {author} {\bibinfo {author} {\bibfnamefont {A.}~\bibnamefont
  {{Nusser}}}, \bibinfo {author} {\bibfnamefont {E.}~\bibnamefont
  {{Branchini}}}, \ and\ \bibinfo {author} {\bibfnamefont {M.}~\bibnamefont
  {{Feix}}},\ }\Doi {10.1088/1475-7516/2013/01/018} {\bibfield  {journal}
  {\bibinfo  {journal} {\jcap},\ }\textbf {\bibinfo {volume} {1}},\ \bibinfo
  {eid} {018} (\bibinfo {year} {2013})},\ \Eprint
  {http://arxiv.org/abs/1207.5800} {arXiv:1207.5800 [astro-ph.CO]} \BibitemShut
  {NoStop}%
\bibitem [{\citenamefont {{Chilingarian}}\ \emph {et~al.}(2010)\citenamefont
  {{Chilingarian}}, \citenamefont {{Melchior}},\ and\ \citenamefont
  {{Zolotukhin}}}]{Chilin2010}%
  \BibitemOpen
  \bibfield  {author} {\bibinfo {author} {\bibfnamefont {I.~V.}\ \bibnamefont
  {{Chilingarian}}}, \bibinfo {author} {\bibfnamefont {A.-L.}\ \bibnamefont
  {{Melchior}}}, \ and\ \bibinfo {author} {\bibfnamefont {I.~Y.}\ \bibnamefont
  {{Zolotukhin}}},\ }\Doi {10.1111/j.1365-2966.2010.16506.x} {\bibfield
  {journal} {\bibinfo  {journal} {\mnras},\ }\textbf {\bibinfo {volume}
  {405}},\ \bibinfo {pages} {1409} (\bibinfo {year} {2010})},\ \Eprint
  {http://arxiv.org/abs/1002.2360} {arXiv:1002.2360 [astro-ph.IM]} \BibitemShut
  {NoStop}%
\bibitem [{\citenamefont {{Schlegel}}\ \emph {et~al.}(2011)\citenamefont
  {{Schlegel}}, \citenamefont {{Abdalla}}, \citenamefont {{Abraham}},
  \citenamefont {{Ahn}}, \citenamefont {{Allende Prieto}}, \citenamefont
  {{Annis}}, \citenamefont {{Aubourg}}, \citenamefont {{Azzaro}}, \citenamefont
  {{Baltay}}, \citenamefont {{Baugh}},\ and\ \citenamefont
  {et~al.}}]{bigboss2011}%
  \BibitemOpen
  \bibfield  {author} {\bibinfo {author} {\bibfnamefont {D.}~\bibnamefont
  {{Schlegel}}}, \bibinfo {author} {\bibfnamefont {F.}~\bibnamefont
  {{Abdalla}}}, \bibinfo {author} {\bibfnamefont {T.}~\bibnamefont
  {{Abraham}}}, \bibinfo {author} {\bibfnamefont {C.}~\bibnamefont {{Ahn}}},
  \bibinfo {author} {\bibfnamefont {C.}~\bibnamefont {{Allende Prieto}}},
  \bibinfo {author} {\bibfnamefont {J.}~\bibnamefont {{Annis}}}, \bibinfo
  {author} {\bibfnamefont {E.}~\bibnamefont {{Aubourg}}}, \bibinfo {author}
  {\bibfnamefont {M.}~\bibnamefont {{Azzaro}}}, \bibinfo {author}
  {\bibfnamefont {S.~B.~C.}\ \bibnamefont {{Baltay}}}, \bibinfo {author}
  {\bibfnamefont {C.}~\bibnamefont {{Baugh}}}, \ and\ \bibinfo {author}
  {\bibnamefont {et~al.}},\ }\href@noop {} {\bibfield  {journal} {\bibinfo
  {journal} {ArXiv e-prints}} (\bibinfo {year} {2011})},\ \Eprint
  {http://arxiv.org/abs/1106.1706} {arXiv:1106.1706 [astro-ph.IM]} \BibitemShut
  {NoStop}%
\bibitem [{\citenamefont {{Levi}}\ \emph {et~al.}(2013)\citenamefont {{Levi}},
  \citenamefont {{Bebek}}, \citenamefont {{Beers}}, \citenamefont {{Blum}},
  \citenamefont {{Cahn}}, \citenamefont {{Eisenstein}}, \citenamefont
  {{Flaugher}}, \citenamefont {{Honscheid}}, \citenamefont {{Kron}},
  \citenamefont {{Lahav}},\ and\ \citenamefont {{representing the DESI
  collaboration}}}]{Levi2013}%
  \BibitemOpen
  \bibfield  {author} {\bibinfo {author} {\bibfnamefont {M.}~\bibnamefont
  {{Levi}}}, \bibinfo {author} {\bibfnamefont {C.}~\bibnamefont {{Bebek}}},
  \bibinfo {author} {\bibfnamefont {T.}~\bibnamefont {{Beers}}}, \bibinfo
  {author} {\bibfnamefont {R.}~\bibnamefont {{Blum}}}, \bibinfo {author}
  {\bibfnamefont {R.}~\bibnamefont {{Cahn}}}, \bibinfo {author} {\bibfnamefont
  {D.}~\bibnamefont {{Eisenstein}}}, \bibinfo {author} {\bibfnamefont
  {B.}~\bibnamefont {{Flaugher}}}, \bibinfo {author} {\bibfnamefont
  {K.}~\bibnamefont {{Honscheid}}}, \bibinfo {author} {\bibfnamefont
  {R.}~\bibnamefont {{Kron}}}, \bibinfo {author} {\bibfnamefont
  {O.}~\bibnamefont {{Lahav}}}, \ and\ \bibinfo {author} {\bibnamefont
  {{representing the DESI collaboration}}},\ }\href@noop {} {\bibfield
  {journal} {\bibinfo  {journal} {ArXiv e-prints}} (\bibinfo {year} {2013})},\
  \Eprint {http://arxiv.org/abs/1308.0847} {arXiv:1308.0847 [astro-ph.CO]}
  \BibitemShut {NoStop}%
\bibitem [{\citenamefont {{Bilicki}}\ \emph {et~al.}(2014)\citenamefont
  {{Bilicki}}, \citenamefont {{Jarrett}}, \citenamefont {{Peacock}},
  \citenamefont {{Cluver}},\ and\ \citenamefont {{Steward}}}]{Bilicki2014}%
  \BibitemOpen
  \bibfield  {author} {\bibinfo {author} {\bibfnamefont {M.}~\bibnamefont
  {{Bilicki}}}, \bibinfo {author} {\bibfnamefont {T.~H.}\ \bibnamefont
  {{Jarrett}}}, \bibinfo {author} {\bibfnamefont {J.~A.}\ \bibnamefont
  {{Peacock}}}, \bibinfo {author} {\bibfnamefont {M.~E.}\ \bibnamefont
  {{Cluver}}}, \ and\ \bibinfo {author} {\bibfnamefont {L.}~\bibnamefont
  {{Steward}}},\ }\Doi {10.1088/0067-0049/210/1/9} {\bibfield  {journal}
  {\bibinfo  {journal} {\apjs},\ }\textbf {\bibinfo {volume} {210}},\ \bibinfo
  {eid} {9} (\bibinfo {year} {2014})},\ \Eprint
  {http://arxiv.org/abs/1311.5246} {arXiv:1311.5246 [astro-ph.CO]} \BibitemShut
  {NoStop}%
\end{thebibliography}%
\end{document}